\documentclass[11pt,a4paper,twoside]{article}

\usepackage{graphicx}
\usepackage{dcolumn}
\usepackage{bm}
\usepackage{epsfig}
\usepackage{epstopdf}
\usepackage{color}
\usepackage{pdflscape}

\usepackage{amsmath}    
\usepackage{array}      
\usepackage{amssymb}    
\usepackage[compress]{cite}

\usepackage{booktabs} 
\usepackage{float}

\usepackage{multicol} 

\usepackage[hmarginratio=1:1,left=30mm,top=17mm,columnsep=15pt,centering]{geometry} 

\usepackage{lipsum} 

\usepackage[style]{abstract} 
\renewcommand{\abstitlestyle}[1]{}

\usepackage[hang, small,labelfont=bf,up,textfont=rm,up]{caption} 

\usepackage{titlesec} 
\titleformat{\section}[block]{\normalsize\normalfont\bfseries}{\thesection.}{1em}{} 
\titleformat{\subsection}[block]{\normalsize}{\thesubsection.}{1em}{} 

\title{\vspace{-10mm}\large\bf Kinetic-energy driven superconductivity in cuprate superconductors}

\date{}

\begin{document}

\baselineskip12pt

\maketitle

\begin{center}
\author{\vspace{-15mm}\normalsize {\small Shiping Feng$^{\rm a}$\renewcommand{\thefootnote}{*}\footnote{Corresponding author. Email: spfeng@bnu.edu.cn},
Yu Lan$^{\rm b}$, Huaisong Zhao$^{\rm c}$, L\"ulin Kuang$^{\rm a}$, Ling Qin$^{\rm a}$, and Xixiao Ma$^{\rm a}$} \\[2mm]
\it{
$^{\rm a}${\small Department of Physics, Beijing Normal University, Beijing 100875, China}}\\
\it{
$^{\rm b}${\small Department of Physics and Electronic Information Science, Hengyang Normal University, Hengyang 421002, China}}\\
\it{
$^{\rm c}${\small College of Physics, Qingdao University, Qingdao 266071, China}}}
\end{center}

\begin{abstract}\noindent
Superconductivity in cuprate superconductors occurs upon charge-carrier doping Mott insulators, where a central question is what mechanism causes the loss of electrical resistance below the superconducting transition temperature? In this review, we attempt to summarize the basic idea of the kinetic-energy driven superconducting mechanism in the description of superconductivity in cuprate superconductors. The mechanism of the kinetic-energy driven superconductivity is purely electronic without phonons, where the charge-carrier pairing interaction in the particle-particle channel arises directly from the kinetic energy by the exchange of spin excitations in the higher powers of the doping concentration. This kinetic-energy driven d-wave superconducting-state is controlled by both the superconducting gap and quasiparticle coherence, which leads to that the maximal superconducting transition temperature occurs around the optimal doping, and then decreases in both the underdoped and overdoped regimes. In particular, the same charge-carrier interaction mediated by spin excitations that induces the superconducting-state in the particle-particle channel also generates the normal-state pseudogap state in the particle-hole channel. The normal-state pseudogap crossover temperature is much larger than the superconducting transition temperature in the underdoped and optimally doped regimes, and then monotonically decreases upon the increase of doping, eventually disappearing together with superconductivity at the end of the superconducting dome. This kinetic-energy driven superconducting mechanism also indicates that the strong electron correlation favors superconductivity, since the main ingredient is identified into a charge-carrier pairing mechanism not from the external degree of freedom such as the phonon but rather solely from the internal spin degree of freedom of the electron. The typical properties of cuprate superconductors discussed within the framework of the kinetic-energy driven superconducting mechanism are also reviewed.
\end{abstract}

\section{Introduction}\label{Introduction}

\begin{figure}[h!]
\centering
\includegraphics[scale=0.7]{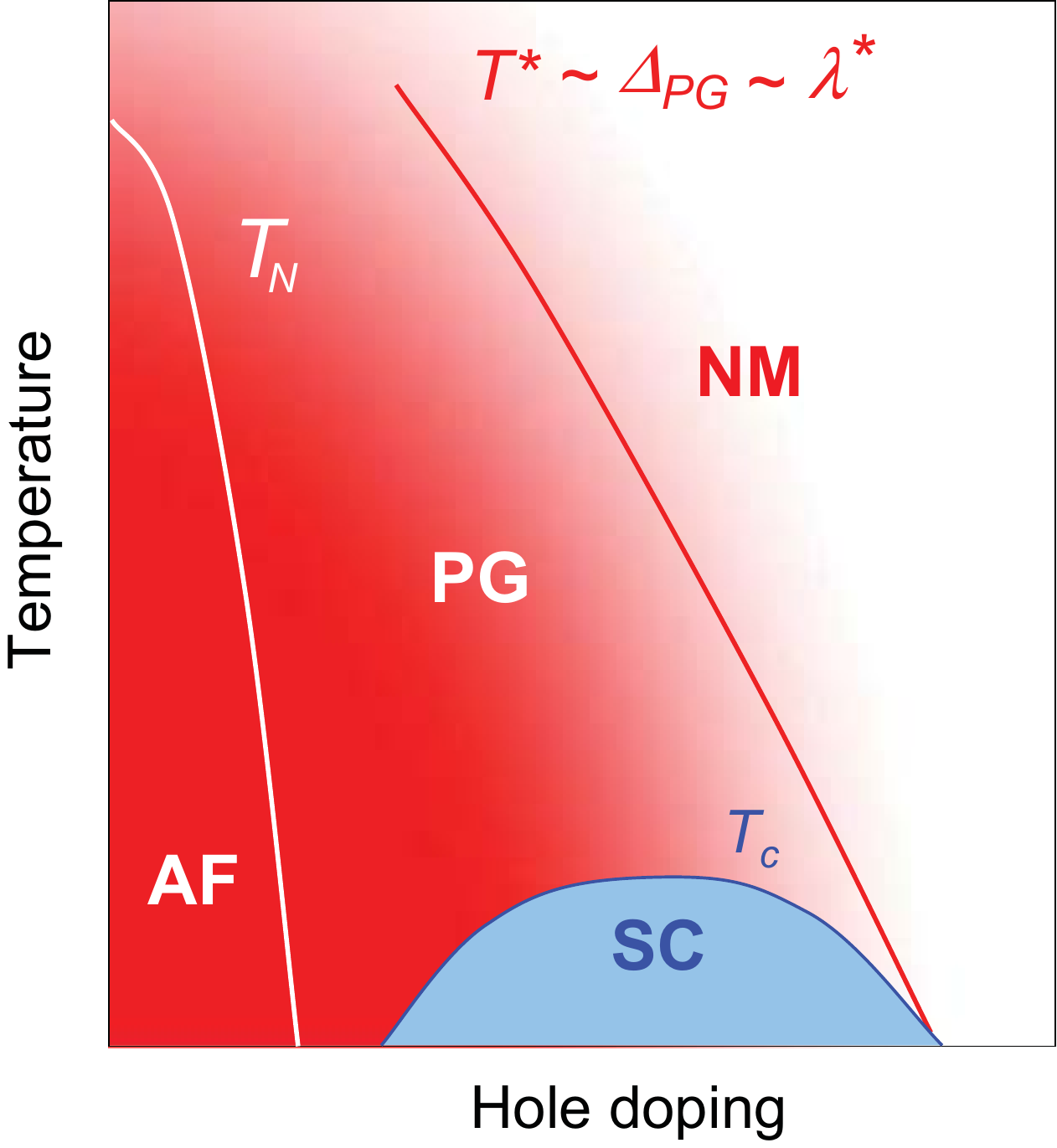}
\caption{(Color) Schematic phase diagram of cuprate superconductors (AFLRO region AF, SC-state SC, pseudogap phase PG, normal-metal phase NM, magnitude of the pseudogap $\Delta_{\rm PG}$, coupling strength $\lambda^{*}$). $T_{\rm N}$ is the N\'eel temperature, while the temperature below which superconductivity (a pseudogap) is observed is denoted by $T_{\rm c}$ ($T^{*}$). [From Ref. \cite{Kordyuk10}.] \label{phase-diagram-exp}}
\end{figure}

After intensive investigations over more than two decades, it has become clear that cuprate superconductors are among the most complicated systems studied in condensed matter physics \cite{Bednorz86,Wu87,Schilling93}. The complications arise mainly from that the parent compounds of cuprate superconductors are a form of non-conductor called a Mott insulator with an antiferromagnetic (AF) long-range order (AFLRO) \cite{Anderson87,Phillips10,Kastner98,Fujita12}, where a single common feature in the layered crystal structure is the presence of one to several CuO$_{2}$ planes in the unit cell \cite{Bednorz86,Wu87,Schilling93}. Inelastic neutron scattering (INS) experiments show that the low-energy spin excitations in these parent compounds are well described by an AF Heisenberg model \cite{Fujita12,Vaknin87,Hayden91,Hayden96,Coldea01} with the magnetic exchange coupling constant $J\sim 0.1$ eV. When these CuO$_{2}$ planes are doped with charge carriers, the AFLRO phase subsides and superconductivity emerges leaving the AF short-range order (AFSRO) correlation still intact \cite{Kastner98,Fujita12}. Although there are hundreds of cuprate superconducting (SC) compounds, they all fit into a universal phase diagram \cite{Kordyuk10} as schematically illustrated in Fig. \ref{phase-diagram-exp}, where the physical properties mainly depend on the extent of doping, and the regimes have been classified into the underdoped, optimally doped, and overdoped, respectively. After AFLRO is destroyed rapidly by doping, there are three apparent regions of the phase diagram: (a) a d-wave SC phase, where the maximal SC transition temperature $T_{\rm c}$ occurs around the optimal doping, and then decreases in both the underdoped and overdoped regimes \cite{Tallon95}; (b) a normal-state pseudogap metallic phase, where an energy gap called the normal-state pseudogap $\bar{\Delta}_{\rm pg}$ exists \cite{Timusk99,Norman05,Hufner08,Hufner08b,Batlogg94,Loeser96,Warren89,Johnston89,Alloul89} above $T_{\rm c}$ but below the normal-state pseudogap crossover temperature $T^{*}$. However, in contrast to the domelike shape of the doping dependence of $T_{\rm c}$, $T^{*}$ is much larger than $T_{\rm c}$ in the underdoped and optimally doped regimes \cite{Timusk99,Norman05,Hufner08,Hufner08b,Batlogg94,Loeser96,Warren89,Johnston89,Alloul89}, and then monotonically decreases upon the increase of doping. In particular, measurements taken by using a wide variety of techniques demonstrate that the normal-state pseudogap is present in both the spin and charge channels \cite{Timusk99}; (c) a {\it normal-metal} phase with largely transport properties. In the doped regime, charge carriers couple to spin excitations \cite{Fujita12,Eschrig06}. The combined INS and resonant inelastic X-ray scattering (RIXS) experimental data have identified spin excitations with high intensity over a large part of moment space, and shown that spin excitations exist across the entire range of the SC dome \cite{Fujita12,Dean14}. However, the charge-carrier doping causes substantial changes to the low-energy spin excitation spectrum \cite{Fujita12}, while it has a more modest effect on the high-energy spin excitations \cite{Dean14}. In particular, RIXS experiments \cite{Dean14} show that the high-energy spin excitations persist well into the overdoped regime and bear a striking resemblance to those found in the parent compounds, indicating that a local-moment picture accounts for the observed spin excitations at elevated energies even up to the overdoped regime \cite{Dean14}. Experimentally, a large body of data available from a wide variety of measurement techniques have provided rather detailed information on cuprate superconductors, where some essential agreements have emerged. We refer the readers to the more detailed summaries of experimental results available in the literatures \cite{Kastner98,Fujita12,Kordyuk10,Timusk99,Hufner08,Dean14,Bonn96,Damascelli03,Campuzano04,Deutscher05,Devereaux07,Fischer07}.

Superconductivity, the dissipationless flow of electrical current, is a striking manifestation of a subtle form of quantum rigidity on the  macroscopic scale \cite{Monthoux07}, where a central question is what mechanism causes the loss of electrical resistance below $T_{\rm c}$? It is commonly believed that the existence of electron Cooper pairs is the hallmark of superconductivity \cite{Monthoux07,Cooper56,Anderson07}, since these electron Cooper pairs behave as effective bosons, and can form something analogous to a Bose condensate that flows without resistance. This follows from a fact that although electrons repel each other because of the Coulomb interaction, at low energies there can be an effective attraction that originates by the exchange of bosons \cite{Monthoux07}. In conventional superconductors, as explained by the Bardeen-Cooper-Schrieffer (BCS) theory \cite{Bardeen57,Schrieffer64}, these exchanged bosons are phonons that act like a bosonic {\it glue} to hold the electron pairs together, and then these electron pairs condense into a coherent macroscopic quantum state that is insensitive to impurities and imperfections and hence conducts electricity without resistance \cite{Bardeen57,Schrieffer64}. The excitation in the SC-state has an energy gap $\bar{\Delta}$, which determines both the quasiparticle energy spectrum and the energy of the condensate \cite{Schrieffer64,Bardeen57,Bogoliubov58}. In this conventional electron-phonon SC mechanism \cite{Bardeen57}, the resulting wave function for the pairs turns out to be peaked at zero separation of the electrons, which leads to that the SC-state has an s-wave symmetry \cite{Norman11}. As a consequence, the pairs in conventional superconductors are always related to an increase in kinetic energy which is overcompensated by the lowering of potential energy \cite{Chester65}. At the temperature above $T_{\rm c}$, the electron is in the standard Landau Fermi-liquid state, which is generally referred to as a {\it normal-state}, where the density of states near the Fermi level is smooth and generally treated as featureless \cite{Schrieffer64}. As in conventional superconductors, superconductivity in cuprate superconductors results when charge carriers pair up into charge-carrier pairs, which is supported by many experimental evidences, including the factor of $2e$ occurring in the flux quantum and in the Josephson effect, as well as the electrodynamic and thermodynamic properties \cite{Tsuei00,Gough87,Gammel87,Hardy93,Wollman93,Tsuei94,Wollman95}. However, the normal-state of cuprate superconductors in the pseudogap phase is not normal at all, since the normal-state of cuprate superconductors in the pseudogap phase exhibits a number of the anomalous properties \cite{Timusk99,Norman05,Hufner08,Hufner08b,Batlogg94,Loeser96,Warren89,Johnston89,Alloul89} in the sense that they do not fit in with the standard Landau Fermi-liquid theory. Superconductivity is an instability of the normal-state. However, one of the most striking dilemmas is that the SC coherence of quasiparticle peaks in cuprate superconductors is described by a standard BCS formalism, although the normal-state is undoubtedly not the standard Landau Fermi-liquid on which the conventional BCS electron-phonon SC mechanism is based. Angle-resolved photoemission spectroscopy (ARPES) experiments reveal sharp spectral peaks in the excitation spectrum \cite{Damascelli03,Campuzano04}, indicating the presence of quasiparticle-like states, which is also consistent with the long lifetime of electronic state as it has been determined by the conductivity measurements \cite{Bonn96}. Moreover, ARPES experiments also observe the Bogoliubov-type dispersion of the SC-state \cite{Campuzano96,Matsui03} predicted by the standard BCS formalism. However, as a natural consequence of the unconventional SC mechanism that is responsible for the high $T_{\rm c}$, the charge-carrier pairs in cuprate superconductors have a dominant d-wave symmetry \cite{Tsuei00,Hardy93,Wollman93,Tsuei94}. This d-wave SC-state also implies that there is a strongly momentum-dependent attraction between charge carriers without phonons \cite{Monthoux07}. After over more than two decades of the painstaking effort, people are still debating the very mechanism of superconductivity in cuprate superconductors, where the crucial issues in cuprate superconductors are (a) what is the nature of the {\it glue} binding charge carriers into charge-carrier pairs, such that they can travel macroscopic distances without resistance? (b) whether the pseudogap has a competitive or collaborative role in engendering superconductivity?

Very soon after the discovery of superconductivity in cuprate superconductors \cite{Bednorz86}, Anderson \cite{Anderson87} proposed that in the parent compounds of cuprate superconductors, the spins form a superposition of singlets. This spin liquid of singlets is so-called the resonating valence bond (RVB) state. Upon the charge-carrier doping, these RVB singlets would become charged, resulting in a SC-state. The RVB state is fundamentally different from the conventional N\'eel state in which the doped charge carrier can move freely among the RVB spin liquid and then a better compromise between the charge-carrier kinetic energy and spin exchange energy may be achieved. Since then many elaborations of this idea followed \cite{Gros88,Lee88,Zhang88,Paramekanti01,Sorella02,Randeria04,Anderson04,Edegger06,Nave06,Chou06,Edegger07}. In particular, it was realized that essential aspects of the RVB concept can be formulated within the charge-spin separation (CSS) slave-particle approach \cite{Baskaran87,Zou88,Kotliar88,Yu92}, since the essential physics of cuprate superconductors is dominated by the short-range repulsive interaction which remains relevant and causes CSS \cite{Anderson00}. This CSS slave-particle approach also led to the development of the gauge theory for cuprate superconductors \cite{Yu92,Baskaran88,Ioffe89,Lee92,Lee06}. However, in the framework of the original RVB theory \cite{Anderson87}, the AF exchange coupling $J$ attracts electrons of opposite spins to be on neighboring sites. This is the result of states of very high energy with a spin gap, and the corresponding interaction has only high-energy dynamics \cite{Anderson07}. The normal-state pseudogap is identified as the spin gap in the RVB state with an energy scale set by $J$, and therefore is associated with the breaking of the RVB singlets \cite{Lee06,Anderson06}. In this case, Anderson \cite{Anderson00} suggested that the SC-state in cuprate superconductors is determined by the need to reduce the frustrated kinetic energy of the system, where the strong frustration of the kinetic energy in the normal-state is partially relieved upon entering the SC-state, indicating that the kinetic energy causes superconductivity \cite{Anderson02}. On the other hand, it has been argued phenomenologically that superconductivity in cuprate superconductors could arise from a lowering of the kinetic energy rather than the potential energy \cite{Hirsch92}. In this scenario, superconductors would exhibit qualitatively new features in their optical properties -- a violation of the low-energy optical sum rule, and a change in the high-energy optical absorption when the system becomes SC \cite{Hirsch92,Hirsch02}. Later, the high-precision optical measurements on cuprate superconductors in the near-infrared and visible region indicate small changes in the spectral weight associated with the onset of superconductivity \cite{Basov99,Molegraaf02}, therefore supporting this argument \cite{Hirsch92} that changes in the kinetic energy are indeed occurring. In particular, the similar experimental results have by now been obtained by differently experimental groups \cite{Syro03,Boris04,Homes04,Kuzmenko05,Carbone06}. More importantly, the recent experimental results \cite{Drachuck14} from the ARPES measurements on cuprate superconductors indicate that $T_{\rm c}$ is correlated with the charge-carrier kinetic energy, which supports the notion of the kinetic-energy driven superconductivity. Motivated by these experimental results \cite{Basov99,Molegraaf02,Syro03,Boris04,Homes04,Kuzmenko05,Carbone06}, several calculations based on the strongly correlated models have been done to show that superconductivity may be driven by a lowering of the kinetic energy upon the formation of the SC-state \cite{Wrobel03,Eckl03,Maier04,Yokoyama04,Yanase05,Haule07}. By constructing an effective Hamiltonian for spin polarons forming in weakly doped antiferromagnets, it has been demonstrated that the driving mechanism which gives rise to superconductivity in such system is the reduction of the kinetic energy \cite{Wrobel03}. Moreover, a numerical study of the two-dimensional Hubbard model within the dynamical cluster approximation has shown the lowering of the kinetic energy below $T_{\rm c}$ for different doping levels \cite{Maier04}. These theoretical calculation \cite{Wrobel03,Eckl03,Maier04,Yokoyama04,Yanase05,Haule07} shows that the paired charge carriers in the AFSRO background can be more mobile than the single charge carriers, and this can overcome the normal increase in the kinetic energy upon the pair formation \cite{Singh07}.

Superconductivity in cuprate superconductors is something entirely new, a manifestation of the strong electron correlation, or Mottness \cite{Anderson87,Phillips10}. In the early days of superconductivity, we \cite{Feng94,Feng04} have developed a fermion-spin theory to confront the strong electron correlation, where the constrained electron operator is decoupled as a product of a charge carrier and a localized spin, with the charge carrier represented the charge degree of freedom of the electron together with some effects of spin configuration rearrangements due to the presence of the doped charge carrier itself, while the spin operator represented the spin degree of freedom of the electron, and then the strong electron correlation can be treated properly in actual calculations. In particular, these charge carriers and spin are gauge invariant, and in this sense, the collective modes for these charge carriers and spins are real and can be interpreted as the physical excitations of the system. In this fermion-spin theory, the basic low-energy excitations are charge-carrier quasiparticles, the spin excitations, and the electron quasiparticles. In this case, the charge transport is mainly governed by the scattering of charge carriers due to spin fluctuations, and the scattering of spins due to charge-carrier fluctuations dominates the spin dynamics, while as a result of the charge-spin recombination, the electron quasiparticles are responsible for the electronic properties. Within the framework of the fermion-spin theory \cite{Feng94,Feng04}, we have established a kinetic-energy driven SC mechanism \cite{Feng03,Feng06,Feng06a}, where charge carriers are held together in d-wave pairs at low temperatures by the attractive interaction in the particle-particle channel that originates directly from the kinetic energy by the exchange of spin excitations in the higher powers of the doping concentration, and then these charge-carrier pairs (then electron Cooper pairs) condense to the d-wave SC-state. Although the physical properties of cuprate superconductors in the normal-state are fundamentally different from these in the standard Landau Fermi-liquid state, the kinetic-energy driven SC-state still is conventional BCS-like with the d-wave symmetry, and then the obtained formalism for the charge-carrier pairing can be used to compute $T_{\rm c}$ and the related SC coherence of the low-energy excitations in cuprate superconductors on the first-principles basis much as can be done for conventional superconductors. Moreover, this kinetic-energy driven SC-state is controlled by both the charge-carrier pair gap and quasiparticle coherence, which leads to that $T_{\rm c}$ takes a domelike shape with the underdoped and overdoped regimes on each side of the optimal doping $\delta_{\rm optimal}\approx 0.15$, where $T_{\rm c}$ reaches its maximum. On the other hand, the same charge-carrier interaction mediated by spin excitations that induces the SC-state in the particle-particle channel also generates the normal-state pseudogap state in the particle-hole channel \cite{Feng12}. As a consequence, the SC gap and normal-state pseudogap coexist but compete in the whole SC dome. However, the normal-state pseudogap crossover temperature $T^{*}$ is much larger than $T_{\rm c}$ in the underdoped and optimally doped regimes, and then monotonically decreases upon the increase of doping, eventually disappearing together with superconductivity at the end of the SC dome. This kinetic-energy driven SC mechanism therefore provides a natural explanation of both the origin of the normal-state pseudogap state and pairing mechanism for superconductivity.

It is beyond the scope of this article to provide an overview of various theories of superconductivity in cuprate superconductors that have been put forward in the literatures, and some earlier reviews and different perspectives appear in Refs. \cite{Phillips10,Norman05,Eschrig06,Monthoux07,Norman11,Anderson04,Edegger07,Yu92,Lee06,Scalapino12,Das14}. In this article, we only attempt to review comprehensively the general framework of the kinetic-energy driven SC mechanism in the context of our work \cite{Feng94,Feng04,Feng03,Feng06,Feng06a,Feng12} and to summarize several calculated results of physical quantities obtained based on the kinetic-energy driven SC mechanism. The number of topics for this review article is listed as follows. In section \ref{CSSFST}, we \cite{Feng94,Feng04} give an overview of the fermion-spin theory, and show that within the {\it decoupling} scheme, the fermion-spin representation is a natural representation of the constrained electron defined in a restricted Hilbert space without double electron occupancy. The kinetic-energy driven SC mechanism \cite{Feng03,Feng06,Feng06a} is introduced in section \ref{KEDM}. In the rest of sections, we show how this kinetic-energy driven SC mechanism yields many results that are in broad agreement with various key experimental facts observed on cuprate superconductors. Superconductors are not only perfect conductors, but also exhibit the so-called Meissner effect, where they expel magnetic fields. In section \ref{Meissner-effect}, we consider main features of the doping dependence of the electromagnetic response observed on cuprate superconductors by using the muon-spin-rotation measurement technique, and show that in analogy to the domelike shape of the doping dependence of $T_{\rm c}$, the maximal superfluid density $\rho_{\rm s}$ occurs around the critical doping $\delta_{\rm critical}\approx 0.195$, and then decreases in both lower doped and higher doped regimes. In section \ref{spin-response}, we turn to the comparison of the calculated result of the dynamical spin response with RIXS-INS experimental data. It is shown that the low-energy spin excitations in the SC-state have an hour-glass-shaped dispersion, with commensurate resonance that appears in the SC-state {\it only}, while the low-energy incommensurate (IC) spin fluctuations can persist into the normal-state. The high-energy spin excitations in the SC-state on the other hand retain roughly constant energy as a function of doping, with spectral weights and dispersion relations comparable to those found in the parent compounds. A brief description of the interplay between superconductivity and normal-state pseudogap state is given in section \ref{SC-Pseudogap}, where we \cite{Feng12} identify the normal-state pseudogap as being a region of the self-energy effect in the particle-hole channel in which the normal-state pseudogap suppresses the spectral weight of the low-energy excitation spectrum. This normal-state pseudogap disappears at $T^{*}$, and then system crossovers to the normal-metal phase with largely transport properties at the temperatures $T>T^{*}$. In section \ref{charge-transport}, we discuss the effect of the normal-state pseudogap on the infrared response of cuprate superconductors, and show that in the underdoped and optimally doped regimes, the transfer of the part of the low-energy spectral weight of the conductivity spectrum to the higher energy region to form a midinfrared band is intrinsically associated with the emergence of the normal-state pseudogap. Finally, the article concludes with the suggestions, in section \ref{conclusion}, for future work.

\section{Fermion-spin theory}\label{CSSFST}

\subsection{Model}\label{model}

In cuprate superconductors, the single common feature in the layered crystal structure is the presence of one to several CuO$_{2}$ planes in the unit cell \cite{Bednorz86,Wu87,Schilling93}, and it seems evident that the nonconventional behaviors of cuprate superconductors are dominated by the CuO$_{2}$ plane. In this case, as originally emphasized by Anderson \cite{Anderson87}, the essential physics of the doped CuO$_{2}$ plane is contained in the one-band large-$U$ Hubbard model \cite{Hubbard63} on a square lattice,
\begin{eqnarray}\label{Hubbard}
H=-\sum_{<ll'>\sigma}t_{ll'}C^{\dagger}_{l\sigma}C_{l'\sigma}+\mu\sum_{l\sigma}C^{\dagger}_{l\sigma}C_{l\sigma}+U\sum_{l}n_{l\uparrow}n_{l\downarrow},
\end{eqnarray}
where the summation is over all sites $l$, and the hopping integrals $t_{ll'}$ connect sites $l$ and $l'$. We will restrict to our attention to the nearest ($t$) and next nearest ($-t'$) neighbor hopping. $C^{\dagger}_{l\sigma}$ and $C_{l\sigma}$ are electron operators that respectively create and annihilate electrons with spin $\sigma$, $n_{l\sigma}=C^{\dagger}_{l\sigma}C_{l\sigma}$, and $\mu$ is the chemical potential. This large-$U$ Hubbard model (\ref{Hubbard}) indicates that the interactions in cuprate superconductors are dominated by the on-site Mott-Hubbard term $U$, which is very large as compared with the electron hopping integrals $t$ and $t'$, i.e., $U\gg t$, $t'$, and therefore leads to that electrons become strongly correlated to avoid double occupancy. In this case, the on-site Mott-Hubbard term must be dealt properly before bothering with relatively minor terms \cite{Anderson02,Anderson08}. It has been shown \cite{Gros87} that the correct way to deal with this large-$U$ term in Eq. (\ref{Hubbard}) is to renormalize it by means of a canonical transformation $e^{iS}$, which eliminates large-$U$ term from the block which contains no doubly occupied states, and which presumably contains all the low-energy eigenstates and thus the ground state, and then the transformed Hamiltonian can be obtained as,
\begin{eqnarray}\label{tJmodel}
H=-t\sum_{l\hat{\eta}\sigma}C^{\dagger}_{l\sigma}C_{l+\hat{\eta}\sigma}+t'\sum_{l\hat{\tau}\sigma}C^{\dagger}_{l\sigma}C_{l+\hat{\tau}\sigma}+\mu\sum_{l\sigma}
C^{\dagger}_{l\sigma}C_{l\sigma}+J\sum_{l\hat{\eta}}{\bf S}_{l}\cdot {\bf S}_{l+\hat{\eta}},
\end{eqnarray}
with the nearest-neighbors $\hat{\eta}=\pm\hat{x},\pm\hat{y}$, the next nearest-neighbors $\hat{\tau}=\pm\hat{x}\pm\hat{y}$, the magnetic exchange coupling constant $J=4t^{2}/U$, the spin operators ${\bf S}_{l}=(S^{\rm x}_{l},S^{\rm y}_{l},S^{\rm z}_{l})$. The kinetic-energy term in Eq. (\ref{tJmodel}) describes mobile charge carriers in the AF background, while the Heisenberg term in Eq. (\ref{tJmodel}) describes AF coupling between localized spins. In particular, the nearest-neighbor hopping integral $t$ in the kinetic-energy term is much larger than the magnetic exchange coupling constant $J$ in the Heisenberg term, and therefore the spin configuration is strongly rearranged due to the effect of the charge-carrier hopping $t$ on the spins, which leads to strong coupling between the charge and spin degrees of freedom of the electron. This transformed Hamiltonian (\ref{tJmodel}) is so-called $t$-$J$ model acting on a restricted Hilbert space without double electron occupancy, where there are three physical states {\it only},
\begin{eqnarray}\label{RHspace}
|0>,~~~ |\uparrow>,~~~ |\downarrow>.
\end{eqnarray}
The essential physics is cuprate superconductors obeying the $t$-$J$ model (\ref{tJmodel}), in which the hopping integrals of $t$ and $t'$ to states having higher energy $U$ are removed in favour of the magnetic exchange interaction $J$ \cite{Anderson08}. At half-filling, this $t$-$J$ model (\ref{tJmodel}) is reduced to an AF Heisenberg model, where the degree of freedom is local spin only. In particular, it has been demonstrated explicitly the local SU(2) gauge invariance of the Heisenberg model written in terms of electron operators with a constraint of one particle per site \cite{Affleck88}. In spite of its simple form, the $t$-$J$ model (\ref{tJmodel}) has been proved to be very difficult to analyze, analytically as well as numerically, because of the restriction of the motion of electrons in the restricted Hilbert space without double electron occupancy.

\subsection{Constraints and sum rules}\label{tJconstraint}

The strong electron correlation originates from a large on-site repulsion between two electrons occupying the same site in the Hubbard model (\ref{Hubbard}), which effectively translates into an elimination of double occupancy in the $t$-$J$ model (\ref{tJmodel}). There are two ways to implement the crucial requirement of no double occupancy \cite{Feng94}: either to solve the $t$-$J$ model (\ref{tJmodel}) combined with a single occupancy local constraint \cite{Zou88,Kotliar88,Yu92},
\begin{eqnarray}\label{constraint}
\sum_{\sigma}C^{\dagger}_{l\sigma}C_{l\sigma}\leq 1,
\end{eqnarray}
or to introduce the constrained electron operator \cite{Gros87}, replacing $C_{l\sigma}$ by,
\begin{eqnarray}\label{CEO}
\tilde{C}_{l\sigma}=C_{l\sigma}(1-n_{l-\sigma}).
\end{eqnarray}
In this section, we will use both representations to clarify this matter. As a consequence of the electron motion in the restricted Hilbert space (\ref{RHspace}) without double occupancy, the constrained electron operator $\tilde{C}_{l\sigma}$ satisfies following relation \cite{Feng94},
\begin{eqnarray}
\sum_{\sigma}\tilde{C}^{\dagger}_{l\sigma}\tilde{C}_{l\sigma}=\sum_{\sigma}C^{\dagger}_{l\sigma}C_{l\sigma}(1-n_{l-\sigma}) ,
\end{eqnarray}
this leads to a sum rule for the constrained electron,
\begin{eqnarray}\label{SR1}
\langle \sum_{\sigma}\tilde{C}^{\dagger}_{l\sigma}\tilde{C}_{l\sigma}\rangle = 1 - \delta  ,
\end{eqnarray}
where $\delta$ is the charge carrier doping concentration, and $\langle\cdot\cdot\cdot\rangle$ means thermodynamical average. On the other hand, the constrained electron operator $\tilde{C}_{l\sigma}$ obeys a special on-site anticommutation relation,
\begin{eqnarray}
\sum_{\sigma}\{\tilde{C}_{l\sigma},\tilde{C}^{\dagger}_{l\sigma}\}=2-\sum_{\sigma}C^{\dagger}_{l\sigma}C_{l\sigma},
\end{eqnarray}
with its expectation value,
\begin{eqnarray}
\langle\sum_{\sigma}\{\tilde{C}_{l\sigma},\tilde{C}^{\dagger}_{l\sigma}\}\rangle = 1 + \delta ,
\end{eqnarray}
which gives rise to a sum rule for the electron spectral function $A_{\sigma}({\bf k},\omega)$,
\begin{eqnarray}\label{SR2}
\sum_{\sigma}\int_{-\infty}^{\infty}{d\omega\over 2\pi} A_{\sigma}({\bf k},\omega)= 1 + \delta .
\end{eqnarray}
In the Hubbard model (\ref{Hubbard}), the large-$U$ forces a finite density of the many-electron states to split out of the band continuum of states to the high-energy side, forming the {\it upper Hubbard band} \cite{Anderson08}, then the remaining continuum (lower Hubbard band) is completely described by the $t$-$J$ model (\ref{tJmodel}). In this case, a doped charge carrier leaves behind an empty site. However, each such empty site can be occupied only by either a spin-up or spin-down electron in the restricted Hilbert space (\ref{RHspace}). In particular, just from the charge-carrier doping, the empty part of the electron spectrum at low energies has a weight of $2\delta$. Furthermore, as the charge-carrier doping annihilates one state in the filled part of the electron spectrum, there are $1-\delta$ electron state (per site) remaining below the Fermi surface. This gives rise to a total weight of the lower Hubbard band of $1+\delta$ \cite{Phillips10,Phillips101}. Eqs. (\ref{SR1}) and (\ref{SR2}) are the exact sum rules for the $t$-$J$ model (\ref{tJmodel}), and they should be preserved in an adequate treatment.

\subsection{Slave-particle theory}\label{Slave-particle-theory}

The high complexity in the $t$-$J$ model (\ref{tJmodel}) comes mainly from the electron single occupancy local constraint (\ref{constraint}). This electron single occupancy local constraint (\ref{constraint}) can be exactly taken into account only by numerical methods, such as the variational Monte Carlo technique \cite{Gros88,Lee88,Paramekanti01,Edegger07}, exact cluster diagonalization \cite{Stephan91}, and various realizations of the quantum Monte Carlo method \cite{Dagotto94}. However, the exact diagonalization is limited by system sizes, while the quantum Monte Carlo technique faces the negative sign problem for lower temperatures. Apart from these numerical techniques, an intuitively appealing approach to implement this electron single occupancy local constraint (\ref{constraint}) and the CSS scheme is the slave-particle approach \cite{Zou88,Kotliar88,Yu92}, where the physics of no double occupancy is taken into account by representing the constrained electron as a composite object created by,
\begin{eqnarray}\label{slave-particle}
C_{l\sigma}=a^{\dagger}_{l}f_{l\sigma},
\end{eqnarray}
with $a^{\dagger}_{l}$ as the slave boson and $f_{l\sigma}$ as the fermion or {\it vice versa}, i.e., $a^{\dagger}_{l}$ as the fermion and $f_{l\sigma}$ as the boson. This way the nonholonomic constraint (\ref{constraint}) is converted into a holonomic one  \cite{Yu92},
\begin{eqnarray}\label{slave-boson-constraint}
a^{\dagger}_{l}a_{l}+\sum_{\sigma}f^{\dagger}_{l\sigma}f_{l\sigma}= 1,
\end{eqnarray}
which means a given site cannot be occupied by more than one particle. In this slave-particle representation (\ref{slave-particle}), the charge degree of freedom of the constrained electron is described by the operator $a_{l}$, while the spin degree of freedom of the constrained electron is described by the operator $f_{l\sigma}$, and then the elementary charge and spin excitations are so-called holon and spinon, respectively. However, a new $U(1)$ gauge degree of freedom must be introduced to incorporate the single occupancy local constraint (\ref{slave-boson-constraint}), which means that the slave-particle representation should be invariant under a local $U(1)$ gauge transformation,
\begin{eqnarray}\label{slave-boson-gauge}
a_{l}\rightarrow a_{l}e^{i\theta_{l}}, ~~~ f_{l\sigma}\rightarrow f_{l\sigma}e^{i\theta_{l}},
\end{eqnarray}
and then all physical quantities should be invariant with respect to this transformation. This reflects that the holon $a_{l}$ or spinon $f_{l\sigma}$ {\it itself} is not gauge invariant, and they are strongly coupled by the $U(1)$ gauge field fluctuation \cite{Yu92,Baskaran88,Ioffe89,Lee06}. In this sense, the collective modes for the holon and spinon are not real and therefore they can not be interpreted as the physical excitations of the system \cite{Laughlin97}. Moreover, there are a number of difficulties in this slave-particle approach. First of all, in the slave-boson version, the AFLRO correlation is absent for zero doping \cite{Arovas88}, so that the ground-state energy is high compared with the numerical estimate  \cite{Lee88,Liang88}, and the Marshall sign rule \cite{Marshall55} is not obeyed. Alternatively, in the slave-fermion approach, the ground-state is the state with AFLRO for the undoped case and persists until very high doping ($\sim 60\%$) \cite{Yoshioka89}. In particular, it should be noted that in the actual calculations \cite{Kotliar88,Yu92}, the electron single occupancy local constraint (\ref{slave-boson-constraint}) is explicitly replaced by a global constraint, and therefore the representation space is much larger than the restricted Hilbert space (\ref{RHspace}) for the physical electron. The local nature of the constraint is of prime importance, and its violation may lead to some unphysical results \cite{Zhang93,Feng93,Guillou95}. This is why the crucial requirement for the $t$-$J$ model (\ref{tJmodel}) is to impose the electron single occupancy local constraint (\ref{constraint}).

\subsection{CP$^{1}$ representation}\label{CP1R}

For convenience in the following discussions at this section, another useful approach to implement the electron single occupancy local constraint (\ref{constraint}) and the CSS scheme has been developed \cite{Ioffe90}, where the physical electron in the $t$-$J$ model (\ref{tJmodel}) is decoupled as,
\begin{eqnarray}\label{CP1}
C_{l\sigma}=h^{\dagger}_{l}b_{l\sigma},
\end{eqnarray}
supplemented by the local constraint,
\begin{eqnarray}\label{CP1constraint}
\sum_{\sigma}b^{\dagger}_{l\sigma}b_{l\sigma}=1,
\end{eqnarray}
with the spinless fermion $h_{l}$ keeping track of the charge degree of freedom of the constrained electron (holon), while the spinful boson $b_{l\sigma}$ keeping track of the spin degree of freedom of the constrained electron (spinon), and then the electron single occupancy local constraint in Eq. (\ref{constraint}),
\begin{eqnarray}\label{constraint1}
\sum_{\sigma}C^{\dagger}_{l\sigma}C_{l\sigma}=1-h^{\dagger}_{l}h_{l}\leq 1,
\end{eqnarray}
is satisfied, where $n^{(h)}_{l}=h^{\dagger}_{l}h_{l}$ is the holon number at site $l$, equal to 1 or 0. This {\it decoupling scheme}, the so-called slave-fermion CP$^{1}$ representation, was proposed in Ref. \cite{Ioffe90}. However, it is similar to the slave-particle representation (\ref{slave-particle}), the spinons and holons in the slave-fermion CP$^{1}$ representation are not physical objects, and there is an arbitrary phase relation between the two related to the single occupancy local constraint (\ref{CP1constraint}) as \cite{Ioffe90},
\begin{eqnarray}\label{gauge}
h_{l}\rightarrow h_{l}e^{i\theta_{l}}, ~~~ b_{l\sigma}\rightarrow b_{l\sigma}e^{i\theta_{l}}.
\end{eqnarray}
This local $U(1)$ gauge degree of freedom, of course, is canceled for the physical electron. However, it should be noted \cite{Feng94,Feng04} from Eq. (\ref{CP1constraint}) that so long as $h^{\dagger}_{l}h_{l}=h^{\dagger}_{l\sigma}h_{l\sigma}=1$, $\sum_{\sigma}C^{\dagger}_{l\sigma}C_{l\sigma}=0$, no matter what is the value $\sum_{\sigma}b^{\dagger}_{l\sigma}b_{l\sigma}$. In the slave-fermion CP$^{1}$ representation, the choice $\sum_{\sigma}b^{\dagger}_{l\sigma}b_{l\sigma}=1$ is convenient, because it also guarantees the local condition $\sum_{\sigma}C^{\dagger}_{l\sigma}C_{l\sigma}=1$, when $h^{\dagger}_{l}h_{l}=0$. However, the local constraint $\sum_{\sigma}b^{\dagger}_{l\sigma}b_{l\sigma}=1$ means the presence of one boson (spin-up or down) on each site, i.e., a {\it spin} even to an empty site has been assigned. This will not affect the physical expectation values, because the charge carrier number expectation $\langle h^{\dagger}_{l}h_{l} \rangle=\langle h^{\dagger}_{l\sigma}h_{l\sigma}\rangle$ will remove all spurious effects. Nevertheless, the extra degrees of freedom will affect the partition function. In subsection \ref{Projection}, we \cite{Feng94} will define a projection operator to cure this defect. As a result, the sum rules (\ref{SR1}) and (\ref{SR2}) will be satisfied exactly.

\subsection{Fermion-spin transformation implement gauge invariant charge carrier}\label{FST}

The decoupling of the charge and spin degrees of freedom of electron is undoubtedly correct in one-dimensional interacting electron systems \cite{Haldane80}, where the charge and spin degrees of freedom of electron are represented by boson operators that describe the excitations of charge-density wave and spin-density wave, respectively. In particular, the typical behavior of the non-Fermi-liquid, showing up as CSS and vanishing of the quasiparticle residue, has been demonstrated theoretically within the one-dimensional $t$-$J$ model \cite{Ogata90}. Moreover, the excitations of the charge and spin degrees of freedom of electron as the real elementary excitations in the one-dimensional cuprates has been observed directly by the ARPES experiment \cite{Kim96}. Therefore both theoretical and experimental studies indicate that the existence of the real excitations of the charge and spin degrees of freedom of electron is common in one-dimensional interacting electron systems \cite{Haldane80,Ogata90,Kim96,Maekawa01}. However, the case in cuprate superconductors (two-dimensional strongly correlated electron systems) is very complex. Among the anomalous properties of cuprate superconductors in the normal-state in the underdoped and optimally doped regimes, a hallmark is the charge transport \cite{Phillips10,Kastner98,Timusk99}, where the low-energy conductivity deviates strongly from the Drude behavior, and is carried by $\delta$ charge carriers. This is very natural in that the low-energy spectral weight of the conductivity must vanish when $\delta\rightarrow 0$. It follows that a superconductor that forms out of the underdoped cuprates must have a superfluid density $\rho_{\rm s}$ given by this spectral weight \cite{Tallon95,Uemura89,Lee99}, so that the superfluid density $\rho_{\rm s}$ in the underdoped regime vanishes more or less linearly with the decrease of the charge-carrier doping concentration $\delta$. This in turn gives rise to the linear relation between $T_{\rm c}$ and $\rho_{\rm s}$ observed in cuprate superconductors in the underdoped regime \cite{Uemura89}. In corresponding to the non-Drude behavior of the low-energy conductivity in the underdoped and optimally doped regimes, the resistivity exhibits a linear temperature behavior over a wide range of temperatures. It has been argued that these experimental facts are a strong experimental evidence supporting the notion of CSS, since not even conventional electron-electron scattering would show the striking linear rise of scattering rate above the Debye frequency, and if there is no CSS, the phonons should affect these properties \cite{Anderson00}. In this case, a formal theory with the gauge invariant excitations of the charge and spin degrees of freedom of electron in the two-dimensional strongly correlated electron systems, i.e., the issue of whether the excitations of the charge and spin degrees of freedom of electron are real, is centrally important \cite{Laughlin97}. In this subsection, we \cite{Feng94,Feng04} start from the above slave-fermion CP$^{1}$ approach (\ref{CP1}) and show that if the electron single occupancy local constraint is treated properly, the constrained electron in the $t$-$J$ model (\ref{tJmodel}) can be decoupled by introducing the charge carrier and spin, where the collective mode for the charge carrier or spin is real and therefore can be interpreted as the physical excitation of the system.

First of all, we \cite{Feng94} examine the properties of the spinless bosons $b_{l}$. The spinless boson creation and annihilation operators are expressed in the infinite-dimensional Fock space as \cite{Schiff68},
\begin{eqnarray}\label{boson}
b^{\dagger}_{l}=\left(
\begin{array}{ccccc}
0 & 0 & 0 & 0 & ... \\
\sqrt{1} & 0 & 0 & 0 & ... \\
0 & \sqrt{2} & 0 & 0 & ... \\
0 & 0 & \sqrt{3} & 0 & ... \\
... & ... & ... & ... & ...
\end{array} \right) \,,~~~~~~
b_{l}=\left(
\begin{array}{ccccc}
0 & \sqrt{1} & 0 & 0 & ... \\
0 & 0 & \sqrt{2} & 0 & ... \\
0 & 0 & 0 & \sqrt{3} & ... \\
0 & 0 & 0 & 0 & ... \\
... & ... & ... & ... & ...
\end{array} \right) \,.
\end{eqnarray}
However, if the boson occupation number is restricted to $0$ or $1$, the spinless boson creation and annihilation operators in Eq. (\ref{boson}) are reduced immediately to the two-dimensional space as \cite{Feng94},
\begin{eqnarray}\label{HCboson}
b^{\dagger}_{l}=\left(
\begin{array}{cc}
0 & 0 \\
1 & 0
\end{array} \right) \,,~~~~~~
b_{l}=\left(
\begin{array}{cc}
0 & 1 \\
0 & 0
\end{array} \right) \,,
\end{eqnarray}
which are nothing but the spin-lowering $S^{-}_{l}$ and spin-raising $S^{+}_{l}$ operators for spin $S=1/2$. These spin operators $S^{-}_{l}$ and $S^{+}_{l}$ behave as fermions on the same site, and as bosons on different sites, and therefore satisfy the {\it hard-core} constraints $b_{l}b_{l}= b^{\dagger}_{l} b^{\dagger}_{l}=0$.

Now we turn to explore further the properties of the CP$^{1}$ bosons $b_{l\sigma}$. Since the CP$^{1}$ bosons $b_{l\uparrow}$ and $b_{l\downarrow}$ satisfy the CP$^{1}$ local constraint $\sum_{\sigma}b^{\dagger}_{l\sigma}b_{l\sigma}=1$ in Eq. (\ref{CP1constraint}), the empty and doubly occupied states have been ruled out, and only the spin-up and spin-down singly occupied spin states are allowed. In particular, due to the symmetry of the spin-up and spin-down states,
$\mid{\rm occupied}\rangle_{\uparrow}=\left (\begin{array}{cc} {1}\\{0}\end{array}\right)_{\uparrow}$ and
$\mid{\rm empty}\rangle_{\uparrow}=\left (\begin{array}{cc}{0}\\{1}\end{array}\right)_{\uparrow}$ are singly-occupied and empty spin-up, while
$\mid{\rm occupied}\rangle_{\downarrow}=\left(\begin{array}{cc}{0}\\ {1}\end{array}\right)_{\downarrow}$ and
$\mid{\rm empty}\rangle_{\downarrow}=\left (\begin{array}{cc} {1}\\ {0}\end{array}\right)_{\downarrow}$ are
singly-occupied and empty spin-down states, respectively. In this case, the CP$^{1}$ boson operators $b_{l\uparrow}$ and $b_{l\downarrow}$ together with the local constraint (\ref{CP1constraint}) can be represented in the basis $\left (\begin{array}{cc} {1}\\{0}\end{array}\right)$ and $\left (\begin{array}{cc}{0}\\{1}\end{array}\right)$ as \cite{Feng94,Feng04},
\begin{subequations}\label{CSSB}
\begin{eqnarray}
b_{\uparrow}&=&e^{i\Phi_{\uparrow}}\mid {\rm occupied}\rangle_{\downarrow}~_{\uparrow}\langle {\rm occupied}\mid=
e^{i\Phi_{\uparrow}}\left (
\begin{array}{cc}
{0} & {0}\\{1} & {0}
\end{array}
\right)=e^{i\Phi_{\uparrow}}S^{-}, ~~~\\
b_{\downarrow}&=&e^{i\Phi_{\downarrow}}\mid {\rm occupied}\rangle_{\uparrow}~_{\downarrow}\langle {\rm occupied}\mid=
e^{i\Phi_{\downarrow}}\left (
\begin{array}{cc}
{0} & {1}\\{0} & {0}
\end{array}
\right)= e^{i\Phi_{\downarrow}}S^{+},~~~
\end{eqnarray}
\end{subequations}
and then all the {\it hard-core boson} conditions, {\it i.e.}, $b_{l\sigma}b^{\dagger}_{l\sigma}+b^{\dagger}_{l\sigma}b_{l\sigma}=1$, $b^{\dagger}_{l\sigma}b^{\dagger}_{l\sigma}=b_{l\sigma}b_{l\sigma}=0$, (without summation over $\sigma$), are satisfied. As a result, the CP$^{1}$ boson operators $b_{l\uparrow}$ and $b_{l\downarrow}$ together with the local constraint (\ref{CP1constraint}) are identified with the spin lowering operator $S^{-}_{l}$ with an additional phase factor $e^{i\Phi_{l\uparrow}}$ and raising $S^{+}_{l}$ operator with an additional phase factor $e^{i\Phi_{l\downarrow}}$, respectively. Consequently, the corresponding CP$^{1}$ ordinary boson occupation space together with the local constraint of one boson per site is identified with the natural spin 1/2 representation space, while the phase factor $e^{i\Phi_{l\sigma}}$ in Eq. (\ref{CSSB}) is closely related to this transformation of the representation spaces, and therefore carries some messages of the spin degree of freedom of the constrained electron, especially, some messages of the spin configuration rearrangements in the doped case \cite{Feng04}. In this case, the electron decoupling form in Eq. (\ref{CP1}) can be expressed as,
\begin{eqnarray}\label{CSS1}
\tilde{C}_{l\uparrow}=h^{\dagger}_{l}e^{i\Phi_{l\uparrow}}S^{-}_{l},~~~~
\tilde{C}_{l\downarrow}=h^{\dagger}_{l}e^{i\Phi_{l\downarrow}}S^{+}_{l},
\end{eqnarray}
while the local $U(1)$ gauge transformation (\ref{gauge}) therefore is rewritten as,
\begin{eqnarray}\label{CSSgauge}
h_{l}\rightarrow h_{l}e^{i\theta_{l}}, ~~~ \Phi_{l\sigma}\rightarrow\Phi_{l\sigma}+\theta_{l}.
\end{eqnarray}
In particular, the phase factor $e^{i\Phi_{l\sigma}}$ in Eq. (\ref{CSS1}) can be incorporated into the spinless fermion operator $h^{\dagger}_{l}$, and then as the solution of the electron single occupancy constraint (\ref{constraint}) $\sum_{\sigma}C^{\dagger}_{l\sigma}C_{l\sigma}\leq 1$ under CP$^{1}$ slave-fermion convention  (\ref{CP1constraint}) $\sum_{\sigma}b^{\dagger}_{l\sigma}b_{l\sigma}=1$, we find the following transformation \cite{Feng94,Feng04},
\begin{eqnarray}\label{CSS}
\tilde{C}_{l\uparrow}=h^{\dagger}_{l\uparrow}S^{-}_{l},~~~~
\tilde{C}_{l\downarrow}=h^{\dagger}_{l\downarrow}S^{+}_{l},
\end{eqnarray}
where the {\it spinful fermion} operator $h_{l\sigma}=e^{-i\Phi_{l\sigma}}h_{l}$ represents the charge degree of freedom of the constrained electron together with some messages of the spin degree of freedom (charge carrier), while the spin operator $S_{l}$ represents the spin degree of freedom of the constrained electron, and then the electron single occupancy local constraint (\ref{constraint}),
\begin{eqnarray}\label{CSSconstraint}
\sum_{\sigma}C^{\dagger}_{l\sigma}C_{l\sigma}&=&S^{+}_{l}h_{l\uparrow}h^{\dagger}_{l\uparrow}S^{-}_{l}+S^{-}_{l}h_{l\downarrow}h^{\dagger}_{l\downarrow}S^{+}_{l}=h_{l} h^{\dagger}_{l}(S^{+}_{l}S^{-}_{l}+S^{-}_{l}S^{+}_{l})\nonumber\\
&=&1-h^{\dagger}_{l}h_{l}\leq 1,
\end{eqnarray}
is always satisfied in actual calculations. In other words, the electron single occupancy local constraint (\ref{constraint}) is implemented exactly using the transformation (\ref{CSS}). This electron decoupling form (\ref{CSS}) is called the {\it fermion-spin transformation} \cite{Feng94,Feng04}. Since the spinless fermion $h_{l}$ and spin operators $S^{+}_{l}$ and $S^{-}_{l}$ obey the anticommutation relation and Pauli spin algebra, respectively, it is then easy to show that the spinful fermion $h_{l\sigma}$ also obeys the same anticommutation relation as the spinless fermion $h_{l}$. However, in contrast to the holon and spinon in the slave-particle approach (\ref{slave-particle}) and the slave-fermion CP$^{1}$ formalism (\ref{CP1}), the charge carrier $h_{l\sigma}$ {\it or} spin $S_{l}$ {\it itself} is invariant under the local $U(1)$ gauge transformation (\ref{CSSgauge}), i.e.,
\begin{eqnarray}\label{fermion-spin-gauge}
h_{l\sigma}=h_{l}e^{-i\Phi_{l\sigma}}\rightarrow h_{l}e^{i\theta_{l}}e^{-i(\Phi_{l\sigma}+\theta_{l})}=h_{l}e^{-i\Phi_{l\sigma}}=h_{l\sigma}, ~~~ S_{l}\rightarrow S_{l},
\end{eqnarray}
which leads to that all physical quantities from charge carriers {\it or} spins are invariant with respect to the gauge transformation (\ref{CSSgauge}). In this sense, the collective mode for the charge carrier {\it or} spin is real and therefore can be interpreted as the physical excitation of the system \cite{Laughlin97}.

The essential physics of the fermion-spin transformation (\ref{CSS}) is simple: at half-filling, the $t$-$J$ model (\ref{tJmodel}) is reduced as an AF Heisenberg model, where each lattice site is singly occupied by a spin-up or spin-down electron, then the electron spins (the spin degree of freedom) are coupled antiferromagnetically with AFLRO. With the charge-carrier doping, the spin configuration must be rearranged to provide the lowest magnetic energy. However, at the same time, the holon (the charge degree of freedom) can feel self-consistently the effect of the spin configuration rearrangements through the strong coupling between the two degrees of freedom, and then the charge carrier arises from the binding of the holon and the phase factor $e^{i\Phi_{l\sigma}}$ carrying some messages of the spin configuration rearrangements due to the presence of the doped charge carrier itself. In particular, when a charge carrier at site $l$ hops to site $l'$ in the $t$-$J$ model (\ref{tJmodel}), the spin configurations are optimized by the fermion-spin transformation (\ref{CSS}), indicating that the frustration effect of charge carrier hopping in the spin background is well taken care of by the fermion-spin transformation (\ref{CSS}). In fact, the representation of the {\it hard-core boson} in terms of spin raising and lowering operators is essential, because whenever a charge-carrier hops it gives rise immediately to a change of the spin background as a result of careful treatment of the electron single occupancy local constraint. This is why the $t$-term is so efficient in destroying the AFLRO \cite{Feng94,Feng96}.

The fermion-spin theory (\ref{CSS}) also indicates that the constrained electron is a composite object, with charge carriers and spins being the physical excitations. However, the charge-carrier quasiparticle and spin excitation are strongly renormalized each other because of the coupling between the two degrees of freedom. In this case, three basic low-energy excitations for the charge-carrier quasiparticles, the spin excitations, and the electron quasiparticles, respectively, emerge as the propagating modes in a doped Mott insulator \cite{Phillips10}, with the charge-carrier quasiparticles that are responsible for the charge transport, and the spin excitations dominate the spin response, while as a result of the charge-spin recombination, the electron quasiparticles govern the electronic properties \cite{Feng15,Feng97a,Guo06,Feng08}.

\subsection{Fermion-spin representation - a natural representation for constrained electron}\label{natural-representation}

Now we show that in the {\it decoupling scheme}, the fermion-spin transformation (\ref{CSS}) is a natural representation for the constrained electron defined in a restricted Hilbert space without double electron occupancy \cite{Feng08}. In Eq. (\ref{CEO}), the constrained electron operators $\tilde{C}^{\dagger}_{l\sigma}$ and $\tilde{C}_{l\sigma}$ are expressed in terms of the unconstrained electron operators
$C^{\dagger}_{l\sigma}$ and $C_{l\sigma}$ as $\tilde{C}^{\dagger}_{l\sigma}=C^{\dagger}_{l\sigma}(1-n_{l-\sigma})$ and $\tilde{C}_{l\sigma}=C_{l\sigma}(1-n_{l-\sigma})$, respectively. Although the constrained electron operator $\tilde{C}^{\dagger}_{l\sigma}$ ($\tilde{C}_{l\sigma}$) does not create (destroy) any doubly occupied sites \cite{Anderson08}, the unconstrained electron operators $C^{\dagger}_{l\sigma}$ and $C_{l\sigma}$ are thought to be operators to operating within the full Hilbert space. In particular, the constrained electron operators $\tilde{C}_{l\uparrow}$ and $\tilde{C}_{l\downarrow}$ can be rewritten as \cite{Anderson08},
\begin{subequations}\label{CEO1}
\begin{eqnarray}
\tilde{C}_{l\uparrow}&=&C_{l\uparrow}(1-n_{l\downarrow})=C_{l\downarrow}C^{\dagger}_{l\downarrow}C_{l\uparrow}=C_{l\downarrow}S^{-}_{l},\\
\tilde{C}_{l\downarrow}&=&C_{l\downarrow}(1-n_{l\uparrow})=C_{l\uparrow}C^{\dagger}_{l\uparrow}C_{l\downarrow}=C_{l\uparrow}S^{+}_{l},
\end{eqnarray}
\end{subequations}
where the {\it spin index} $\downarrow$ ($\uparrow$) of the unconstrained electron operator $C_{l\downarrow}$ ($C_{l\uparrow}$) in the right-hand side in Eq. (\ref{CEO1}) is not an independent degree of freedom, since the spin fluctuation of the system is mainly described by the spin operator $S_{l}$. In other words, in the constrained electron operators (\ref{CEO1}), the unconstrained electron operator $C_{i\sigma}$ in the right-hand side in Eq. (\ref{CEO1}) mainly describes charge degree of freedom of the constrained electron together with some messages of the spin degree of freedom, while the spin operator $S_{l}$ represents the spin degree of freedom of the constrained electron. Furthermore, the constrained electron operators in Eq. (\ref{CEO1}) are exactly same as quoted in Eq. (\ref{CSS}) in the fermion-spin representation if the constrained electron (\ref{CEO1}) is {\it decoupled} according to its charge and spin degrees of freedom, i.e., in the {\it decoupling scheme}, the unconstrained operator $C_{l\sigma}$ and spin operator $S_{l}$ in the right-hand side in Eq. (\ref{CEO1}) are {\it decoupled} as two independent operators $C_{l\sigma}$ and $S_{l}$, and then $C_{l\sigma}$ and $S_{l}$ in Eq. (\ref{CEO1}) commute each other in the decoupling scheme. To see this point clearly, the constrained electron operators $\tilde{C}_{l\uparrow}$ and $\tilde{C}_{l\downarrow}$ in Eq. (\ref{CEO1}) in the {\it decoupling scheme} can be rewritten in terms of a particle-hole transformation for the unconstrained operator $C_{l\sigma}\rightarrow h^{\dagger}_{l-\sigma}$ as,
\begin{eqnarray}
\tilde{C}_{l\uparrow}=h^{\dagger}_{l\uparrow}S^{-}_{l},~~~~
\tilde{C}_{l\downarrow}=h^{\dagger}_{l\downarrow}S^{+}_{l},
\end{eqnarray}
which are exactly same as quoted in Eq. (\ref{CSS}) in the CSS fermion-spin representation.

\subsection{Projection operator}\label{Projection}

In the local representation, the restricted Hilbert space without double electron occupancy in the $t$-$J$ model (\ref{tJmodel}) consists of three states, $|0\rangle$, $|\uparrow \rangle$, $|\downarrow \rangle$ as we have mentioned in Eq. (\ref{RHspace}). However, in the fermion-spin transformation (\ref{CSS}), there are four states
$|{\rm charge}\rangle\otimes |{\rm spin}\rangle$, namely $|1,\uparrow\rangle$, $|1,\downarrow \rangle$, $|0,\uparrow\rangle$, and $|0,\downarrow\rangle$, where 1 or 0 means charge-carrier occupation or empty. In this case, a projection operator $P$ can be introduced to remove the extra degrees of freedom \cite{Feng94}. The matrix elements of this projection operator can be defined as,
\begin{eqnarray}
P_{\kappa\alpha}\equiv |\kappa\rangle\langle\alpha |,
\end{eqnarray}
where $|\kappa\rangle$ is one of the bases of the physical states, while $|\alpha\rangle$ is one of the bases in the fermion-spin representation space $|{\rm charge}\rangle\otimes |{\rm spin}\rangle$. Since the space dimensions of $|\kappa\rangle$ and $|\alpha\rangle$ are different, the usual relations for the projection operator $P^{2}=P=P^{\dagger}$ are not satisfied. Using this projection operator, the constrained electron operators in the restricted Hilbert space of no double occupancy can be defined as,
\begin{subequations}\label{CEP1}
\begin{eqnarray}
\tilde{C}_{l\uparrow}&=&P_{l}h^{\dagger}_{l\uparrow}S^{-}_{l}P^{\dagger}_{l},~~~~~
\tilde{C}^{\dagger}_{l\uparrow}=P_{l}h_{l\uparrow}S^{+}_{l}P^{\dagger}_{l},\\
\tilde{C}_{l\downarrow}&=&P_{l}h^{\dagger}_{l\downarrow}S^{+}_{l}P^{\dagger}_{l},~~~~~
\tilde{C}^{\dagger}_{l\downarrow}=P_{l}h_{l\downarrow}S^{-}_{l}P^{\dagger}_{l},
\end{eqnarray}
\end{subequations}
where $P_{l}$ is the projection operator for the site $l$, and $P^{\dagger}_{l}$ is the Hermitian conjugate of $P_{l}$. Making use of the matrix representation of the charge-carrier operators,
\begin{eqnarray}
h^{\dagger}_{l\sigma}=\left(
\begin{array}{cc}
0 & 1 \\
0 & 0
\end{array} \right) \,,~~~
h_{l\sigma}=\left(
\begin{array}{cc}
0 & 0 \\
1 & 0
\end{array} \right) \,,
\end{eqnarray}
all these operators (\ref{CEP1}) in matrix form can be obtained explicitly (see the Appendix \ref{matrix}). In particular, the constrained electron operators in the basis of the physical states $|0\rangle$, $|\uparrow\rangle$, $|\downarrow\rangle$ can be written as \cite{Feng94},
\begin{subequations}\label{CEP2}
\begin{eqnarray}
\tilde{C}_{l\uparrow}&=&\left(
\begin{array}{ccc}
0 & 1 & 0 \\
0 & 0 & 0  \\
0 & 0 & 0
\end{array} \right) \,,~~~~~
\tilde{C}^{\dagger}_{l\uparrow}=\left(
\begin{array}{ccc}
0 & 0 & 0 \\
1 & 0 & 0  \\
0 & 0 & 0
\end{array} \right) \,,~~~\\
\tilde{C}_{l\downarrow}&=&\left(
\begin{array}{ccc}
0 & 0 & 1 \\
0 & 0 & 0  \\
0 & 0 & 0
\end{array} \right) \,,~~~~~
\tilde{C}^{\dagger}_{l\downarrow}=\left(
\begin{array}{ccc}
0 & 0 & 0 \\
0 & 0 & 0  \\
1 & 0 & 0
\end{array} \right) \,,~~~
\end{eqnarray}
\end{subequations}
which are nothing but the Hubbard $X$ operators $X_{0\uparrow}$, $X_{0\downarrow}$, etc. \cite{Hubbard63}. It is then straightforward to check that
\begin{subequations}\label{SR3}
\begin{eqnarray}
\sum_{\sigma}\tilde{C}^{\dagger}_{l\sigma}\tilde{C}_{l\sigma}= 1-n^{(h)}_{l},\\
\sum_{\sigma}\{\tilde{C}_{l\sigma},\tilde{C}^{\dagger}_{l\sigma}\}=1+n^{(h)}_{l},
\end{eqnarray}
\end{subequations}
with the charge-carrier number operator,
\begin{eqnarray}
n^{(h)}_{l}=\left(
\begin{array}{ccc}
1 & 0 & 0 \\
0 & 0 & 0 \\
0 & 0 & 0
\end{array} \right) \,.
\end{eqnarray}
Taking expectation values of Eq. (\ref{SR3}), one sees immediately that the sum rules (\ref{SR1}) and (\ref{SR2}) are exactly satisfied. It is therefore shown that the fermion-spin transformation (\ref{CSS}) defined with an additional projection operator $P$ satisfies exactly the single occupancy local constraint and all sum rules, i.e., they are an exact mapping \cite{Feng94}.

However, the projection operator $P$ is cumbersome to handle in many cases, and it has been dropped in the actual calculations \cite{Feng94,Feng04,Feng03,Feng06,Feng06a,Feng12}. Now let us see which of these properties are still preserved and what kind of errors we are committing in such approximate treatments. First of all, the electron single occupancy local constraint (\ref{CSSconstraint}) is exactly obeyed even in the mean-field (MF) level \cite{Feng94,Feng04}. For comparison, it should be noted that in the usual slave-particle approach \cite{Kotliar88,Yu92,Arovas88,Yoshioka89}, the electron single occupancy local constraint is explicitly replaced by a global constraint in the MF approximation, and therefore the representation space is much larger than the representation space for the physical electron. From this point of view, the treatment of the electron single occupancy local constraint for the physical electron in the fermion-spin theory is much better than the slave-particle approach, and therefore the MF results \cite{Feng94,Feng04,Feng97a} based on the fermion-spin theory even without projection operator are better than those obtained within the framework of the slave-particle MF theory. Secondly, those expectation values of electron operators, including spin-spin correlation functions, which should vanish, actually do not appear due to the presence of the charge-carrier number operator $n^{(h)}_{l}=h^{\dagger}_{l\sigma}h_{l\sigma}=h^{\dagger}_{l}h_{l}$. By adding the {\it extra spin} degree of freedom to an empty site, we are making errors of the order $\delta$ in counting the number of the spin states \cite{Feng94,Feng04,Plakida02}, which is negligible for small doping. To show this point clearly, we can map electron operator $C_{l\sigma}$ with the electron single occupancy local constraint (\ref{constraint}) onto the slave-fermion formulism \cite{Yu92} as $C_{l\sigma}=h^{\dagger}_{l}a_{l\sigma}$ supplemented by the local constraint $h^{\dagger}_{l}h_{l}+\sum_{\sigma}a^{\dagger}_{l\sigma}a_{l\sigma}=1$. We can solve the local constraint in the slave-fermion formulism by rewriting the boson operator $a_{l\sigma}$ in terms of the CP$^{1}$ boson operator $b_{l\sigma}$ as $a_{l\sigma}=b_{l\sigma}\sqrt{1-h^{\dagger}_{l} h_{l}}$ supplemented by the local constraint
$\sum_{\sigma}b^{\dagger}_{l\sigma}b_{l\sigma}=1$. As we have mentioned in subsection \ref{FST}, the CP$^{1}$ boson operators $b_{l\uparrow}$ and $b_{l\downarrow}$ with the local constraint $\sum_{\sigma}b^{\dagger}_{l\sigma}b_{l\sigma}=1$ can be identified with the spin lowering and raising operators, respectively, defined with an additional phase factor, therefore the projection operator is approximately related to the charge-carrier number operator by $P_{l}\sim\sqrt{1-h^{\dagger}_{l\sigma}h_{l\sigma}} =\sqrt{1-h^{\dagger}_{l}h_{l}}$, and its main role is to remove the spurious spin when there is a charge carrier at a given site $l$ \cite{Feng04}. Thirdly, the essential physics of the local $U(1)$ gauge invariant charge carrier and spin is still kept \cite{Feng04}, which guarantees that the collective mode for the charge carrier {\it or} spin is real and therefore can be interpreted as the physical excitation of the system \cite{Laughlin97}.

\subsection{Summary}

In this section, a brief review is given for the fermion-spin theory. In the framework of the fermion-spin theory, the constrained electron is decoupled as a product of a gauge invariant charge carrier and a localized spin, with the charge carrier keeping track of the charge degree of freedom together with some effects of the spin configuration rearrangements due to the presence of the doped charge carrier itself, while the spin operator keeping track of the spin degree of freedom. The main advantage of this fermion-spin theory is that the electron on-site local constraint for single occupancy is exactly satisfied even in the MF approximation. In particular, in the {\it decoupling scheme}, the fermion-spin representation is a natural representation of the constrained electron defined in a restricted Hilbert space without double electron occupancy.

\section{Kinetic-energy driven superconducting mechanism}\label{KEDM}

The theory of superconductivity in cuprate superconductors remains one of the most important outstanding problem in the field of condensed matter physics. From the experimental front, the measured result from optical spectroscopy on cuprate superconductors with simultaneous time and frequency resolution demonstrates that bosonic excitations of electronic origin are the most important factor in the formation of the SC-state \cite{Conte12}. In particular, the combined RIXS-INS experimental data have identified spin excitations with high intensity over a large part of momentum space, and shown that spin excitations exist across the entire range of the SC dome \cite{Fujita12,Dean14}, which provides a clear link between the pairing mechanism and spin excitations. In this case, a question is raised whether the spin excitation, which is a generic consequence of the strong Coulomb interaction in the large-$U$ Hubbard model (\ref{Hubbard}), can mediate charge-carrier pairing in cuprate superconductors in analogy to the phonon-mediate pairing mechanism in conventional superconductors \cite{Monthoux07,Miller11}? On the theoretical side, it has been shown that the BCS theory is not specific to a phonon-mediated interaction, and other bosonic excitation can also serve as the pairing glue \cite{Monthoux07,Miller11}. In this section, we review briefly the kinetic-energy driven superconductivity \cite{Feng03,Feng06,Feng06a}, where the effective attractive interaction between charge carriers originates in their coupling to spin excitations, and therefore the spin excitation has been served as the pairing glue.

Since the $t$-$J$ model (\ref{tJmodel}) is obtained from the Hubbard model (\ref{Hubbard}) as mentioned in section \ref{CSSFST}, taking the large-$U$ limit and making certain approximations \cite{Gros87}, there is a mixing of kinetic energy and potential energy (Coulombic contribution) in going from the Hubbard model (\ref{Hubbard}) to the $t$-$J$ model (\ref{tJmodel}), i.e., the original kinetic energy in the Hubbard model (\ref{Hubbard}) has been reorganized as the kinetic energy of the lower Hubbard band in the $t$-$J$ model (\ref{tJmodel}), which therefore contains a strong Coulombic contribution due to the restriction of no doubly occupancy of a given site. On the other hand, the magnetic exchange energy in the $t$-$J$ model (\ref{tJmodel}) also has mixed electronic kinetic and Coulombic origins, since the intersite magnetic exchange interaction in the $t$-$J$ model (\ref{tJmodel}) is due to the mixing of states in the lower and upper Hubbard bands on adjacent sites, i.e., the virtual hopping, which is enabled by AF alignment, and involves kinetic energy \cite{Singh07}. In the fermion-spin representation (\ref{CSS}), the $t$-$J$ model (\ref{tJmodel}) can be rewritten as \cite{Feng04},
\begin{eqnarray}\label{CSStJmodel}
H&=&t\sum_{l\hat{\eta}}(h^{\dagger}_{l+\hat{\eta}\uparrow}h_{l\uparrow}S^{+}_{l}S^{-}_{l+\hat{\eta}}+h^{\dagger}_{l+\hat{\eta}\downarrow}h_{l\downarrow}S^{-}_{l}S^{+}_{l+\hat{\eta}}) -t'\sum_{l\hat{\tau}}(h^{\dagger}_{l+\hat{\tau}\uparrow}h_{l\uparrow}S^{+}_{l}S^{-}_{l+\hat{\tau}}\nonumber\\
&+&h^{\dagger}_{l+\hat{\tau}\downarrow}h_{l\downarrow}S^{-}_{l}S^{+}_{l+\hat{\tau}}) -\mu\sum_{l\sigma} h^{\dagger}_{l\sigma}h_{l\sigma}+J_{{\rm eff}}\sum_{l\hat{\eta}}{\bf S}_{l} \cdot{\bf S}_{l+\hat{\eta}},~~~~
\end{eqnarray}
where $J_{{\rm eff}}=(1-\delta)^{2}J$, $\delta=\langle h^{\dagger}_{l\sigma}h_{l\sigma}\rangle=\langle h^{\dagger}_{l}h_{l}\rangle$ is the charge-carrier doping concentration, while the electron degrees of freedom in the original $t$-$J$ model (\ref{tJmodel}) has been replaced by two degrees of freedom associated with the doped charge carrier within a spin correlation background related to the electronic structure of the AF correlation phase. As an important consequence, the mixing of kinetic energy and potential energy in the kinetic-energy term of the original $t$-$J$ model (\ref{tJmodel}) has been released as the interaction between charge carriers and spins in the $t$-$J$ model (\ref{CSStJmodel}), which therefore dominates the essential physics of cuprate superconductors, while the magnetic exchange energy term is to form an adequate spin configuration only \cite{Anderson00,Anderson02}, which is strongly rearranged due to the effect of the charge-carrier hopping $t$ on the spins. In particular, since the electron single occupancy local constraint (then the strong electron correlation) in the $t$-$J$ model (\ref{CSStJmodel}) has been taken into account properly within the fermion-spin representation (\ref{CSS}), the remaining analysis, including the treatment of the interaction between charge carriers and spins in the $t$-$J$ model (\ref{CSStJmodel}), can be fully carried out by a perturbation theory based on the conventional many-body technique as we do for the electron-phonon interaction underlying conventional superconductors.

In a superconductor, electrons form Cooper pairs with an energy gap in the single-particle excitation spectrum when they become superconductors. This gap corresponds to the energy for breaking a Cooper pair of electrons and creating two quasiparticles, therefore measures the strength of the binding of two electrons into a Cooper pair. In particular, electron Cooper pairs in conventional superconductors are formed in momentum space, not in real space, because the onset of the long-range phase coherence occurs due to the overlap of the electron Cooper pair wavefunctions \cite{Bardeen57,Schrieffer64,Mourachkine04}. As a consequence, the order parameter in the characterization of the SC transition and the Cooper pair wavefunctions are the same: the order parameter is a {\it magnified} version of the Cooper pair wavefunctions \cite{Mourachkine04}. On the other hand, within the kinetic-energy driven SC mechanism \cite{Feng03,Feng06,Feng06a}, the charge-carrier pairing also occurs in momentum space, then the order parameter in the characterization of the SC transition in cuprate superconductors and the charge-carrier pair wavefunctions are the same, i.e., the charge-carrier pairing and the onset of the phase coherence take place simultaneously at the charge-carrier pair transition temperature, then the SC transition temperature $T_{\rm c}$ is identical to the charge-carrier pair transition temperature. However, cuprate superconductors are doped Mott insulators, where charge carriers (then electrons) couple to spin excitations \cite{Fujita12,Eschrig06}. In this case, we first consider electron Cooper pair in real space, since it can give some insights into the nature of the SC-state. For example, the amplitude of the electron Cooper pair between the electrons at sites $l$ and $l'$ can be expressed in the fermion-spin representation (\ref{CSS}) as \cite{Feng03,Feng06,Feng06a},
\begin{eqnarray}\label{ECpair}
\Delta(l-l')=\langle C^{\dagger}_{l\uparrow}C^{\dagger}_{l'\downarrow}-C^{\dagger}_{l\downarrow}C^{\dagger}_{l'\uparrow}\rangle=\langle h_{l\uparrow}h_{l'\downarrow}S^{+}_{l} S^{-}_{l'}-h_{l\downarrow}h_{l'\uparrow}S^{-}_{l}S^{+}_{l'}\rangle.
\end{eqnarray}
In the doped regime without an AFLRO, charge carriers move in the spin liquid background, where the spin correlation functions $\langle S^{+}_{l}S^{-}_{l'}\rangle= \langle S^{-}_{l}S^{+}_{l'}\rangle=\chi_{l-l'}$, and then the amplitude of the electron Cooper pair in Eq. (\ref{ECpair}) can be rewritten as,
\begin{eqnarray}\label{ECpair1}
\Delta(l-l')=-\chi_{l-l'}\Delta_{\rm h}(l-l'),
\end{eqnarray}
with the amplitude of the charge-carrier pair,
\begin{eqnarray}\label{CCpair}
\Delta_{\rm h}(l-l')=\langle h_{l'\downarrow}h_{l\uparrow}-h_{l'\uparrow} h_{l\downarrow}\rangle,
\end{eqnarray}
which shows clearly that the amplitude of the electron Cooper pair is closely related to the corresponding one of the charge-carrier pair, and is proportional to the number of the doped charge carriers, but not to the number of electrons. In this case, the {\it essential physics} in the SC-state of cuprate superconductors is dominated by the corresponding
one in the charge-carrier pairing state. The charge-carrier pairing in momentum space can be considered as a {\it collective} phenomenon, while that in real space as {\it individual} \cite{Mourachkine04}. Of course, the transition into the SC-state always occurs in momentum space. However, in the extremely low-doped regime with an AFLRO, where the spin correlation functions $\langle S^{+}_{l}S^{-}_{l'}\rangle\neq\langle S^{-}_{l}S^{+}_{l'}\rangle$, and then the conduct is disrupted by this AFLRO. In the following discussions, we \cite{Feng03,Feng06,Feng06a} only focus on the case without AFLRO, and leave the case with AFLRO for the further discussions.

\subsection{Charge-carrier and spin Green's functions}\label{CCSGF}

For convenience in the following discussions, we first define the charge-carrier diagonal and off-diagonal Green's functions as \cite{Feng04},
\begin{subequations}\label{DTHGF}
\begin{eqnarray}
g(l-l',t-t')&=&\langle\langle h_{l\sigma}(t);h^{\dagger}_{l'\sigma}(t')\rangle\rangle ,\\
\Gamma(l-l',t-t')&=&\langle\langle h_{l\downarrow}(t);h_{l'\uparrow}(t')\rangle\rangle ,\\
\Gamma^{\dagger}(l-l',t-t')&=&\langle\langle h^{\dagger}_{l\uparrow}(t);h^{\dagger}_{l'\downarrow}(t')\rangle\rangle ,
\end{eqnarray}
\end{subequations}
respectively, and the spin Green's functions as,
\begin{subequations}\label{DTSGF}
\begin{eqnarray}
D(l-l',t-t')&=&\langle\langle S^{+}_{l}(t);S^{-}_{l'}(t')\rangle\rangle,\\
D_{\rm z}(l-l',t-t')&=&\langle\langle S^{\rm z}_{l}(t);S^{\rm z}_{l'}(t')\rangle\rangle ,
\end{eqnarray}
\end{subequations}
where $\langle\langle \ldots \rangle\rangle$ is an average over the ensemble. Since the spin system in the $t$-$J$ model (\ref{CSStJmodel}) is anisotropic away from half-filling, therefore two spin Green's functions $D(l-l',t-t')$ and $D_{\rm z}(l-l',t-t')$ have been defined to describe properly spin propagations \cite{Feng04}.

\subsection{Equation of motion method}\label{EOOM}

Since the spin operators obey the Pauli algebra, then our goal is to evaluate the charge-carrier and spin Green's functions directly for the fermion and spin operators in terms of the equation of motion method. In the framework of the equation of motion, the time-Fourier transform of the two-time Green's function,
\begin{eqnarray}
G(\omega)=\langle\langle A;A^{\dagger}\rangle\rangle_{\omega}
\end{eqnarray}
satisfies the equation \cite{Zubarev60},
\begin{eqnarray}
\omega\langle\langle A;A^{\dagger}\rangle\rangle_{\omega}=\langle [A,A^{\dagger}]\rangle+\langle\langle [A,H];A^{\dagger}\rangle\rangle_{\omega}.
\end{eqnarray}
If the orthogonal operator $L$ is defined as, $[A,H]=\zeta A-iL$ with $\langle [L,A^{\dagger}]\rangle =0$, the full Green's function can be expressed as,
\begin{eqnarray}\label{EOOMGM}
G(\omega)=G^{(0)}(\omega)+{1\over \varsigma^{2}}G^{(0)}(\omega)\langle\langle L;L^{\dagger}\rangle\rangle_{\omega}G(\omega),
\end{eqnarray}
where $\varsigma=\langle [A,A^{\dagger}]\rangle$, and the MF Green's function,
\begin{eqnarray}
G^{(0)}(\omega)={\varsigma\over\omega-\zeta}.
\end{eqnarray}
It has been shown \cite{Zubarev60} that if the self-energy $\Sigma(\omega)$ is identified as the irreducible part of $\langle\langle L;L^{\dagger}\rangle\rangle_{\omega}$, the full Green's function (\ref{EOOMGM}) can be evaluated as,
\begin{eqnarray}
G(\omega)={\varsigma\over \omega-\zeta-\Sigma(\omega)},
\end{eqnarray}
with the self-energy,
\begin{eqnarray}
\Sigma(\omega)={1\over\varsigma} \langle\langle L;L^{\dagger}\rangle\rangle ^{irr}_{\omega}.
\end{eqnarray}
In the framework of the diagrammatic technique, this self-energy $\Sigma(\omega)$ corresponds to the contribution of irreducible diagrams.

\subsection{Mean-field approximation}\label{MFA}

In the MF approximation, the $t$-$J$ model (\ref{CSStJmodel}) can be decoupled as \cite{Feng04},
\begin{subequations}\label{MFtJmodel}
\begin{eqnarray}
H_{\rm MFA}&=&H_{\rm t}+H_{\rm J}-2NZt\phi_{1}\chi_{1}+2NZt'\phi_{2}\chi_{2}, \\
H_{\rm t}&=&\chi_{1}t\sum_{l\hat{\eta}\sigma}h^{\dagger}_{l+\hat{\eta}\sigma}h_{l\sigma}-\chi_{2}t'\sum_{l\hat{\tau}\sigma}h^{\dagger}_{l+\hat{\tau}\sigma}h_{l\sigma}
-\mu\sum_{l\sigma}h^{\dagger}_{l\sigma}h_{l\sigma}, ~~~~~\\
H_{\rm J}&=& {1\over 2}J_{\rm eff}\sum_{l\hat{\eta}}[\epsilon(S^{+}_{l}S^{-}_{l+\hat{\eta}}+S^{-}_{l}S^{+}_{l+\hat{\eta}})+2S^{\rm z}_{l}S^{\rm z}_{l+\hat{\eta}}]\nonumber\\
&-&t'\phi_{2}\sum_{l\hat{\tau}}(S^{+}_{l}S^{-}_{l+\hat{\tau}}+S^{-}_{l}S^{+}_{l+\hat{\tau}}),~~~~~~~~~
\end{eqnarray}
\end{subequations}
where the anisotropic parameter $\epsilon=1+2t\phi_{1}/J_{\rm eff}$, the charge-carrier's particle-hole parameters $\phi_{1}=\langle h^{\dagger}_{l\sigma}h_{l+\hat{\eta}\sigma} \rangle$ and $\phi_{2}=\langle h^{\dagger}_{l\sigma}h_{l+\hat{\tau}\sigma}\rangle$, the spin correlation functions $\chi_{1}=\langle S_{l}^{+}S_{l+\hat{\eta}}^{-}\rangle$ and $\chi_{2}=\langle S_{l}^{+}S_{l+\hat{\tau}}^{-}\rangle$, $Z$ is the number of the nearest-neighbor or next nearest-neighbor sites on a square lattice, and $N$ is the number of lattice sites.

Within the framework of the equation of motion method, it is easy to find the MF charge-carrier Green's function as \cite{Feng04},
\begin{eqnarray}\label{MFHGF}
g^{(0)}({\bf k},\omega)={1\over \omega-\xi_{\bf k}},
\end{eqnarray}
where the MF charge-carrier excitation spectrum,
\begin{eqnarray}\label{MFCCS}
\xi_{\bf k}=Zt\chi_{1}\gamma_{\bf k}-Zt'\chi_{2}\gamma_{\bf k}'-\mu,
\end{eqnarray}
with $\gamma_{\bf k}=(1/Z)\sum_{\hat{\eta}}e^{i{\bf k}\cdot\hat{\eta}}=({\rm cos} k_{x}+{\rm cos}k_{y})/2$, and $\gamma_{\bf k}'=(1/Z)\sum_{\hat{\tau}}e^{i{\bf k}\cdot\hat{\tau}}= {\rm cos} k_{x}{\rm cos}k_{y}$. However, in the doped regime without an AFLRO, i.e., $\langle S^{\rm z}_{l}\rangle =0$, the MF spin Green's functions can be discussed in terms of the Kondo-Yamaji decoupling scheme \cite{Kondo72}, which is a stage one-step further than the Tyablikov's decoupling scheme \cite{Tyablikov67}. After a straightforward calculation, the MF spin Green's functions have been obtained explicitly as \cite{Feng04},
\begin{subequations}\label{Two-MFSGF}
\begin{eqnarray}
D^{(0)}({\bf k},\omega)&=&{B_{\bf k}\over 2\omega_{\bf k}}\left ({1\over \omega-\omega_{\bf k}}-{1\over\omega+\omega_{\bf k}}\right ),\label{MFSGF}\\
D^{(0)}_{\rm z}({\bf k},\omega)&=& {B_{\rm z}({\bf k})\over 2\omega_{\rm z}({\bf k})}\left ({1\over\omega-\omega_{\rm z}({\bf k})}-{1\over \omega+\omega_{\rm z} ({\bf k})}
\right ),\label{MFSGFZ}
\end{eqnarray}
\end{subequations}
with the MF spin excitation spectra,
\begin{subequations}
\begin{eqnarray}
\omega^{2}_{\bf k}&=&\lambda_{1}^{2}\left [{1\over 2}\epsilon\left (A_{1}-{1\over 2}\alpha\chi^{\rm z}_{1}-\alpha\chi_{1}\gamma_{\bf k}\right)(\epsilon-\gamma_{\bf k})+\left (A_{2} -{1\over 2Z}\alpha\epsilon\chi_{1}-\alpha \epsilon\chi^{\rm z}_{1}\gamma_{\bf k}\right )\right .\nonumber\\
&\times&\left . (1-\epsilon\gamma_{\bf k})\right ]+\lambda_{2}^{2}\left [\alpha\left (\chi^{\rm z}_{2}\gamma_{\bf k}'-{3\over 2Z}\chi_{2}\right )\gamma_{\bf k}'+{1\over 2}\left (A_{3}-{1\over 2}\alpha\chi^{\rm z}_{2}\right )\right ] \nonumber\\
&+&\lambda_{1}\lambda_{2} \left [\alpha\chi^{\rm z}_{1}(1-\epsilon\gamma_{\bf k})\gamma_{\bf k}'+{1\over 2}\alpha(\chi_{1}\gamma_{\bf k}'-C_{3})(\epsilon-\gamma_{\bf k})+\alpha \gamma_{\bf k}'(C^{\rm z}_{3}-\epsilon\chi^{\rm z}_{2}\gamma_{\bf k})\right . \nonumber\\
&-&\left. {1\over 2}\alpha\epsilon (C_{3}-\chi_{2}\gamma_{\bf k})\right ], \label{MFSES}\\
\omega^{2}_{\rm z}({\bf k})&=&\epsilon\lambda^{2}_{1}\left (\epsilon A_{1}-{1\over Z}\alpha\chi_{1}-\alpha\chi_{1}\gamma_{\bf k}\right )(1-\gamma_{\bf k})+\lambda^{2}_{2}A_{3}
(1-\gamma_{\bf k}')\nonumber\\
&+&\lambda_{1}\lambda_{2}[\alpha\epsilon C_{3}(\gamma_{\bf k}-1)+\alpha(\chi_{2}\gamma_{\bf k}-\epsilon C_{3})(1 -\gamma_{\bf k}')],\label{MFSESZ}
\end{eqnarray}
\end{subequations}
and the functions,
\begin{subequations}
\begin{eqnarray}
B_{\bf k}&=&\lambda_{1}[2\chi^{\rm z}_{1}(\epsilon\gamma_{\bf k}-1)+\chi_{1}(\gamma_{\bf k}-\epsilon)]-\lambda_{2}(2\chi^{\rm z}_{2}\gamma_{\bf k}'-\chi_{2}), \\
B_{\rm z}({\bf k}) &=&\epsilon\chi_{1}\lambda_{1}(\gamma_{\bf k}-1)-\chi_{2}\lambda_{2}(\gamma_{\bf k}'-1),
\end{eqnarray}
\end{subequations}
where $\lambda_{1}=2ZJ_{\rm eff}$, $\lambda_{2}=4Z\phi_{2}t'$, $A_{1}=\alpha C_{1}+(1-\alpha)/(2Z)$, $A_{2}=\alpha C^{\rm z}_{1}+(1-\alpha)/(4Z)$, $A_{3}=\alpha C_{2}+(1-\alpha) /(2Z)$, the spin correlation functions $\chi^{\rm z}_{1}=\langle S_{l}^{\rm z}S_{l+\hat{\eta}}^{\rm z} \rangle$, $\chi^{\rm z}_{2}=\langle S_{l}^{\rm z}S_{l+\hat{\tau}}^{\rm z} \rangle$, $C_{1}=(1/Z^{2})\sum_{\hat{\eta},\hat{\eta'}}\langle S_{l+\hat{\eta}}^{+}S_{l+\hat{\eta'}}^{-} \rangle$, $C^{\rm z}_{1}=(1/Z^{2})\sum_{\hat{\eta},\hat{\eta'}}\langle S_{l+\hat{\eta}}^{z}S_{l+\hat{\eta'}}^{z}\rangle$, $C_{2}=(1/Z^{2})\sum_{\hat{\tau},\hat{\tau'}}\langle S_{l+\hat{\tau}}^{+}S_{l+\hat{\tau'}}^{-}\rangle$, $C_{3}=(1/Z) \sum_{\hat{\tau}}\langle S_{l+\hat{\eta}}^{+}S_{l+\hat{\tau}}^{-}\rangle$, and $C^{\rm z}_{3}=(1/Z)\sum_{\hat{\tau}}\langle S_{l+\hat{\eta}}^{\rm z}S_{l+\hat{\tau}}^{\rm z}\rangle$. In order to satisfy the sum rule of the correlation function $\langle S^{+}_{l}S^{-}_{l}\rangle=1/2$ in the case without AFLRO, the important decoupling parameter $\alpha$ has been introduced in the above calculation \cite{Kondo72,Feng04}, which can be regarded as the vertex correction.

\subsection{Kinetic-energy driven superconductivity}\label{KEDS}

In the following discussions, we \cite{Feng03,Feng06,Feng06a} show that given the MF solution of the $t$-$J$ model (\ref{CSStJmodel}) in subsection \ref{MFA} and by including the fluctuation around it due to the interaction between charge carriers and spins directly from the kinetic energy, we can obtain a formalism for the charge-carrier pairing which can be used to compute $T_{\rm c}$ on the first-principles as can be done for conventional superconductors. In particular, this formalism also can give a description of physical quantities which are consistent with the rather severe set by experiments. The interaction between charge carriers and spins directly from the kinetic energy in the $t$-$J$ model (\ref{CSStJmodel}) is quite strong, which can induce the SC-state in the particle-particle channel by the exchange of spin excitations in the higher power of the doping concentration \cite{Feng03,Feng06,Feng06a}. To see this point clearly, the self-consistent equations that are satisfied by the full charge-carrier diagonal and off-diagonal Green's functions are obtained in terms of the Eliashberg's strong coupling formalism \cite{Eliashberg60,Mahan81} as,
\begin{subequations}\label{HGF}
\begin{eqnarray}
g({\bf k},\omega)&=&g^{(0)}({\bf k},\omega)+g^{(0)}({\bf k},\omega)[\Sigma^{({\rm h})}_{1}({\bf k},\omega)g({\bf k},\omega)-\Sigma^{({\rm h})}_{2}(-{\bf k},-\omega) \Gamma^{\dagger}({\bf k},\omega)],~~~~~~~ \\
\Gamma^{\dagger}({\bf k},\omega)&=&g^{(0)}(-{\bf k},-\omega)[\Sigma^{({\rm h})}_{1}(-{\bf k},-\omega)\Gamma^{\dagger}(-{\bf k},-\omega)+\Sigma^{({\rm h})}_{2}(-{\bf k},-\omega) g({\bf k},\omega)],~~~~~~~
\end{eqnarray}
\end{subequations}
where the self-energies $\Sigma^{({\rm h})}_{1}({\bf k},\omega)$ in the particle-hole channel and $\Sigma^{({\rm h})}_{2}({\bf k},\omega)$ in the particle-particle channel are evaluated from the spin bubble as \cite{Feng03,Feng06,Feng06a},
\begin{subequations}\label{SE}
\begin{eqnarray}
\Sigma^{({\rm h})}_{1}({\bf k},i\omega_{n})&=&{1\over N^{2}}\sum_{{\bf p,p'}}\Lambda^{2}_{{\bf p}+{\bf p}'+{\bf k}}{1\over \beta}\sum_{ip_{m}}g({\bf p}+{\bf k},ip_{m}+i\omega_{n}) \Pi({\bf p},{\bf p}',ip_{m})\nonumber\\
&=&{1\over N}\sum_{\bf p}{1\over \beta}\sum_{ip_{m}}V_{\rm eff}({\bf k},{\bf p},ip_{m})g({\bf p}+{\bf k},ip_{m}+i\omega_{n}), \label{self-energy-1}\\
\Sigma^{({\rm h})}_{2}({\bf k},i\omega_{n})&=&{1\over N^{2}}\sum_{{\bf p,p'}}\Lambda^{2}_{{\bf p}+{\bf p}'+{\bf k}}{1\over \beta}\sum_{ip_{m}}\Gamma^{\dagger}(-{\bf p}-{\bf k}, -ip_{m}-i\omega_{n})\Pi({\bf p},{\bf p}',ip_{m})\nonumber\\
&=&{1\over N}\sum_{\bf p}{1\over\beta}\sum_{ip_{m}}V_{\rm eff}({\bf k},{\bf p},ip_{m})\Gamma^{\dagger}(-{\bf p}-{\bf k},-ip_{m}-i\omega_{n}), \label{self-energy-2}
\end{eqnarray}
\end{subequations}
respectively, with $\Lambda_{{\bf k}}=Zt\gamma_{\bf k}-Zt'\gamma_{\bf k}'$, the effective charge carrier interaction,
\begin{eqnarray}\label{Interaction}
V_{\rm eff}({\bf k},{\bf p},\omega)={1\over N}\sum_{\bf p'}\Lambda^{2}_{{\bf p}+{\bf p}'+{\bf k}}\Pi({\bf p},{\bf p}',\omega),
\end{eqnarray}
and the spin bubble,
\begin{eqnarray}\label{SB}
\Pi({\bf p},{\bf p}',ip_{m})={1\over\beta}\sum_{ip'_{m}}D^{(0)}({\bf p'},ip_{m}')D^{(0)}({\bf p}'+{\bf p},ip_{m}'+ip_{m}).
\end{eqnarray}
This spin-excitation-mediated interaction (\ref{Interaction}) is a key to the SC transition in cuprate superconductors. Since both the pairing force and charge-carrier pair order parameter have been incorporated into the self-energy $\Sigma^{({\rm h})}_{2}({\bf k}, \omega)$, it is called the charge-carrier pair gap in the charge-carrier excitation spectrum,
\begin{eqnarray}\label{CCPGF1}
\bar{\Delta}_{\rm h}({\bf k},\omega)=\Sigma^{({\rm h})}_{2}({\bf k},\omega),
\end{eqnarray}
which corresponds to the energy for breaking a charge-carrier pair and creating two charge-carrier quasiparticles. On the other hand, the self-energy
$\Sigma^{({\rm h})}_{1}({\bf k},\omega)$ renormalizes the MF charge-carrier spectrum, and therefore it describes the charge-carrier quasiparticle coherence. In particular, the self-energy $\Sigma^{({\rm h})}_{2}({\bf k},\omega)$ is an even function of $\omega$, while the self-energy $\Sigma^{({\rm h})}_{1}({\bf k},\omega)$ is not. In this case, the self-energy $\Sigma^{({\rm h})}_{1}({\bf k},\omega)$ can be broken up into its symmetric and antisymmetric parts as,
$\Sigma^{({\rm h})}_{1}({\bf k},\omega)=\Sigma^{({\rm h})}_{\rm 1e}({\bf k},\omega)+\omega\Sigma^{({\rm h})}_{\rm 1o}({\bf k},\omega)$, then both
$\Sigma^{({\rm h})}_{\rm 1e}({\bf k},\omega)$ and $\Sigma^{({\rm h})}_{\rm 1o}({\bf k},\omega)$ are an even function of $\omega$. Moreover, the antisymmetric part
$\Sigma^{({\rm h})}_{\rm 1o}({\bf k},\omega)$ of the self-energy $\Sigma^{({\rm h})}_{1}({\bf k},\omega)$ is directly related to the charge-carrier quasiparticle coherent weight as,
\begin{eqnarray}\label{CCQCW1}
{1\over Z_{\rm hF}({\bf k},\omega)}=1-{\rm Re}\Sigma^{({\rm h})}_{\rm 1o}({\bf k},\omega).
\end{eqnarray}
As a first step of discussions, we \cite{Feng03,Feng06,Feng06a} study the kinetic-energy driven superconductivity, and therefore only focus on the low-energy behavior. In this case, the charge-carrier pair gap and quasiparticle coherent weight can be generally discussed in the static limit, i.e., $\bar{\Delta}_{\rm h}({\bf k})=\Sigma^{({\rm h})}_{2}({\bf k}, \omega)\mid_{\omega=0}$, and
$Z^{-1}_{\rm hF}({\bf k})=1-{\rm Re}\Sigma^{({\rm h})}_{\rm 1o}({\bf k},\omega)\mid_{\omega=0}$. As in conventional superconductor \cite{Eliashberg60,Mahan81}, the retarded function ${\rm Re}\Sigma^{(\rm h)}_{\rm 1e}({\bf k},\omega)\mid_{\omega=0}$ just renormalizes the chemical potential. Although $Z_{\rm hF}({\bf k})$ still is a function of momentum, the momentum dependence may be unimportant in a qualitative discussion, and therefore the wave vector ${\bf k}$ in $Z_{\rm hF}({\bf k})$ can be chosen as,
\begin{eqnarray}\label{CCQCW}
{1\over Z_{\rm hF}}=1-{\rm Re}\Sigma^{(\rm h)}_{\rm 1o}({\bf k},\omega=0)\mid_{{\bf k}=[\pi,0]},
\end{eqnarray}
just as it has been done in the experiments \cite{DLFeng00,Ding01}. Moreover, this charge-carrier quasiparticle coherent weight $Z_{\rm hF}$ reduces the charge-carrier (then electron) quasiparticle bandwidth, and suppresses the spectral weight of the single-particle excitation spectrum, then the energy scale \cite{Guo06} of the quasiparticle band is controlled by the magnetic exchange coupling $J$. In particular, this charge-carrier quasiparticle coherence antagonizes superconductivity, and then $T_{\rm c}$ is depressed to low temperatures \cite{Feng03,Feng06,Feng06a}. On the other hand, the s-wave component of the charge-carrier pair gap is suppressed heavily by $Z_{\rm hF}$, and then the SC-state is dominated by the d-wave component,
\begin{eqnarray}\label{CCPGF}
\bar{\Delta}_{\rm h}({\bf k})=\bar{\Delta}_{\rm h}\gamma^{(\rm d)}_{{\bf k}},
\end{eqnarray}
with $\gamma^{(\rm d)}_{{\bf k}}=({\rm cos}k_{x}-{\rm cos}k_{y})/2$, which is consistent with the experimental fact \cite{Hardy93,Tsuei00} that the charge-carrier pair state in cuprate superconductors has a dominant d-wave symmetry over a wide range of the doping concentration, around the optimal doping. With the above static limit approximation, the full charge-carrier diagonal and off-diagonal Green's functions in Eq. (\ref{HGF}) are obtained explicitly as,
\begin{subequations}\label{BCSHGF}
\begin{eqnarray}
g({\bf k},\omega)&=&Z_{\rm hF}\left ({U^{2}_{{\rm h}{\bf k}}\over\omega-E_{{\rm h}{\bf k}}}+{V^{2}_{{\rm h}{\bf k}}\over\omega+E_{{\rm h}{\bf k}}}\right ),\label{BCSHDGF}\\
\Gamma^{\dagger}({\bf k},\omega)&=&-Z_{\rm hF}{\bar{\Delta}_{\rm hZ}({\bf k})\over 2E_{{\rm h}{\bf k}}}\left ({1\over \omega-E_{{\rm h}{\bf k}}}-{1\over\omega
+E_{{\rm h}{\bf k}}}\right ),\label{BCSHODGF}
\end{eqnarray}
\end{subequations}
where $E_{{\rm h}{\bf k}}=\sqrt{\bar{\xi}^{2}_{{\bf k}}+\mid\bar{\Delta}_{\rm hZ}({\bf k})\mid^{2}}$ is the charge-carrier quasiparticle energy spectrum,
$\bar{\xi}_{{\bf k}}=Z_{\rm hF}\xi_{{\bf k}}$ is the renormalized charge-carrier excitation spectrum, and $\bar{\Delta}_{\rm hZ}({\bf k})=Z_{\rm hF}\bar{\Delta}_{\rm h}({\bf k})$ is the renormalized charge-carrier pair gap, while the charge-carrier quasiparticle coherence factors,
\begin{subequations}\label{BCSCF}
\begin{eqnarray}
U^{2}_{{\rm h}{\bf k}}={1\over 2}\left (1+{\bar{\xi_{{\bf k}}}\over E_{{\rm h}{\bf k}}}\right ),\\
V^{2}_{{\rm h}{\bf k}}={1\over 2}\left (1-{\bar{\xi_{{\bf k}}}\over E_{{\rm h}{\bf k}}}\right ),
\end{eqnarray}
\end{subequations}
with the constraint $U^{2}_{{\rm h}{\bf k}}+V^{2}_{{\rm h}{\bf k}}=1$ for any wave vector ${\bf k}$ (normalization). In spite of the pairing mechanism driven by the kinetic energy by the exchange of spin excitations, the results in Eqs. (\ref{BCSHGF}) and (\ref{BCSCF}) are the standard BCS expressions for a d-wave charge-carrier pair state. However, as a natural consequence of the charge-spin recombination, this charge-carrier pair state also leads to form the electron pairing state, and then the obtained d-wave BCS-like formalism for the electron pairing \cite{Feng03,Feng06,Feng06a,Feng15,Guo07} indicates clearly the Bogoliubov quasiparticle nature of the SC quasiparticle peak. As in conventional superconductors, the Bogoliubov quasiparticle in cuprate superconductors does not carry definite charge, and is a coherent combination of the particle and its absence, then the SC coherence of low-energy excitations and the related quantities can be discussed on the first-principles basis much as can be done for conventional superconductors. Moreover, the AFSRO correlation has been incorporated into the SC-state through the spin's order parameters entering into the charge-carrier self-energies (\ref{SE}) in the particle-particle and particle-hole channels, therefore there is a coexistence of the SC-state and AFSRO correlation, and then AFSRO fluctuation persists into superconductivity.

\subsection{Self-consistent equations}\label{TSCE}

With the help of the full charge-carrier Green's functions in Eq. (\ref{BCSHGF}) and spin Green's function in Eq. (\ref{MFSGF}), the self-energies $\Sigma^{({\rm h})}_{1}({\bf k}, \omega)$ and $\Sigma^{({\rm h})}_{2}({\bf k},\omega)$ are evaluated explicitly as \cite{Feng03,Feng06,Feng06a},
\begin{subequations}\label{SE1}
\begin{eqnarray}
\Sigma^{({\rm h})}_{1}({\bf k},\omega)&=&{1\over N^{2}}\sum_{{\bf pp'}\nu}(-1)^{\nu+1}\Omega^{({\rm h})}_{\bf pp'k}\left [U^{2}_{{\rm h}{\bf p}+{\bf k}}\left ({F^{(\nu)}_{{\rm 1h} {\bf pp'k}}\over\omega+\omega_{\nu{\bf p}{\bf p}'}-E_{{\rm h}{\bf p}+{\bf k}}}+{F^{(\nu)}_{{\rm 2h}{\bf pp'k}}\over\omega-\omega_{\nu{\bf p}{\bf p}'} -E_{{\rm h}{\bf p}+{\bf k}}} \right )\right. \nonumber\\
&+&\left . V^{2}_{{\rm h}{\bf p}+{\bf k}}\left ({F^{(\nu)}_{{\rm 1h}{\bf pp'k}}\over\omega-\omega_{\nu{\bf p}{\bf p}'}+E_{{\rm h}{\bf p}+{\bf k}}}+{F^{(\nu)}_{{\rm 2h}{\bf pp'k}} \over\omega+\omega_{\nu{\bf p}{\bf p}'}+E_{{\rm h}{\bf p}+{\bf k}}}\right )\right ],\label{PHSE}\\
\Sigma^{({\rm h})}_{2}({\bf k},\omega)&=&{1\over N^{2}}\sum_{{\bf pp'}\nu}(-1)^{\nu}\Omega^{({\rm h})}_{\bf pp'k}{\bar{\Delta}_{\rm hZ}({\bf p}+{\bf k})\over 2E_{{\rm h}{\bf p}+  {\bf k}}}\left [\left ({F^{(\nu)}_{{\rm 1h}{\bf pp'k}}\over\omega+\omega_{\nu{\bf p}{\bf p}'}-E_{{\rm h}{\bf p}+{\bf k}}}+{F^{(\nu)}_{{\rm 2h}{\bf pp'k}}\over\omega-\omega_{\nu{\bf p} {\bf p}'}-E_{{\rm h}{\bf p}+{\bf k}}}\right )\right .\nonumber\\
&-&\left . \left ({F^{(\nu)}_{{\rm 1h}{\bf pp'k}}\over\omega-\omega_{\nu{\bf p}{\bf p}'}+E_{{\rm h}{\bf p}+{\bf k}}}+{F^{(\nu)}_{{\rm 2h}{\bf pp'k}}\over\omega+\omega_{\nu{\bf p} {\bf p}'}+E_{{\rm h}{\bf p}+{\bf k}}}\right )\right ], \label{PPSE}
\end{eqnarray}
\end{subequations}
respectively, with $\nu=1,2$, $\Omega^{({\rm h})}_{\bf pp'k}=Z_{\rm hF}\Lambda^{2}_{{\bf p}+{\bf p}'+{\bf k}}B_{{\bf p}'}B_{{\bf p}+{\bf p}'}/(4\omega_{{\bf p}'}\omega_{{\bf p}+{\bf p}'})$, $\omega_{\nu{\bf p}{\bf p}'}=\omega_{{\bf p}+{\bf p}'}-(-1)^{\nu}\omega_{\bf p'}$, and the functions,
\begin{subequations}
\begin{eqnarray}
F^{(\nu)}_{{\rm 1h}{\bf pp'k}}&=&n_{\rm F}(E_{{\rm h}{\bf p}+{\bf k}})\{1+n_{\rm B}(\omega_{{\bf p}'+{\bf p}})+n_{\rm B}[(-1)^{\nu+1}\omega_{\bf p'}]\}\nonumber\\
&+& n_{\rm B}(\omega_{{\bf p}'+{\bf p}}) n_{\rm B}[(-1)^{\nu+1}\omega_{\bf p'}], \\
F^{(\nu)}_{{\rm 2h}{\bf pp'k}}&=&[1-n_{\rm F}(E_{{\rm h}{\bf p}+{\bf k}})]\{1+n_{\rm B}(\omega_{{\bf p}'+{\bf p}})+n_{\rm B}[(-1)^{\nu+1}\omega_{\bf p'}]\}\nonumber\\
&+&n_{\rm B}(\omega_{{\bf p}'+{\bf p}}) n_{\rm B}[(-1)^{\nu+1} \omega_{\bf p'}],
\end{eqnarray}
\end{subequations}
where $n_{\rm B}(\omega)$ and $n_{\rm F}(\omega)$ are the boson and fermion distribution functions, respectively. In this case, the charge-carrier quasiparticle coherent weight $Z_{\rm hF}$ and charge-carrier pair gap parameter $\bar{\Delta}_{\rm h}$ satisfy following two self-consistent equations,
\begin{subequations}\label{SCE1}
\begin{eqnarray}
{1\over Z_{\rm hF}}&=&1+{1\over N^{2}}\sum_{{\bf pp'}\nu}(-1)^{\nu+1}\Omega^{({\rm h})}_{{\bf pp'}{\bf k}_{\rm A}}\left ({F^{(\nu)}_{{\rm 1h}{\bf pp}'{\bf k}_{\rm A}}\over (\omega_{\nu{\bf p}{\bf p}'}-E_{{\rm h}{\bf p}+{\bf k}_{\rm A}})^{2}}+{F^{(\nu)}_{{\rm 2h}{\bf pp}'{\bf k}_{\rm A}}\over (\omega_{\nu{\bf p}{\bf p}'}+E_{{\rm h}{\bf p}+{\bf k}_{\rm A}})^{2}}\right ), ~~~~~~~~\label{CCQCWSCE}\\
1&=&{4\over N^{3}}\sum_{{\bf pp'k}\nu}(-1)^{\nu}Z_{\rm hF}\Omega^{({\rm h})}_{\bf pp'k}{\gamma^{({\rm d})}_{\bf k}\gamma^{({\rm d})}_{{\bf p}+{\bf k}}\over E_{{\rm h}{\bf p}+{\bf k} }}\left ({F^{(\nu)}_{{\rm 1h}{\bf pp'k}}\over\omega_{\nu{\bf p}{\bf p}'}-E_{{\rm h}{\bf p}+{\bf k}}}-{F^{(\nu)}_{{\rm 2h}{\bf pp'k}}\over \omega_{\nu{\bf p}{\bf p}'}+E_{{\rm h}{\bf p}+{\bf k}}}\right ), ~~~~~~~~\label{CCPGPSCE}
\end{eqnarray}
\end{subequations}
respectively, where ${\bf k}_{\rm A}=[\pi,0]$. These two equations (\ref{CCQCWSCE}) and (\ref{CCPGPSCE}) must be solved simultaneously with following self-consistent equations \cite{Feng03,Feng06,Feng06a,Feng12,Guo07,Huang13,Zhao12},
\begin{subequations}\label{SCE2}
\begin{eqnarray}
\phi_{1}&=&{1\over 2N}\sum_{{\bf k}}\gamma_{{\bf k}}Z_{\rm hF}\left (1-{\bar{\xi_{{\bf k}}}\over E_{{\rm h}{\bf k}}}{\rm tanh} [{1\over 2}\beta E_{{\rm h}{\bf k}}]\right ),\\
\phi_{2}&=&{1\over 2N}\sum_{{\bf k}}\gamma_{{\bf k}}'Z_{\rm hF}\left (1-{\bar{\xi_{{\bf k}}}\over E_{{\rm h}{\bf k}}}{\rm tanh}[{1\over 2}\beta E_{{\rm h}{\bf k}}]\right ),\\
\delta &=& {1\over 2N}\sum_{{\bf k}}Z_{\rm hF}\left (1-{\bar{\xi_{{\bf k}}}\over E_{{\rm h}{\bf k}}}{\rm tanh}[{1\over 2}\beta E_{{\rm h}{\bf k}}] \right ),\\
\chi_{1}&=&{1\over N}\sum_{{\bf k}}\gamma_{{\bf k}} {B_{{\bf k}}\over 2\omega_{{\bf k}}}{\rm coth} [{1\over 2}\beta\omega_{{\bf k}}], \\
\chi_{2}&=&{1\over N}\sum_{{\bf k}}\gamma_{{\bf k}}'{B_{{\bf k}}\over 2\omega_{{\bf k}}}{\rm coth} [{1\over 2}\beta\omega_{{\bf k}}],\\
C_{1}&=&{1\over N}\sum_{{\bf k}}\gamma^{2}_{{\bf k}} {B_{{\bf k}}\over 2\omega_{{\bf k}}}{\rm coth}[{1\over 2}\beta\omega_{{\bf k}}],\\
C_{2}&=&{1\over N}\sum_{{\bf k}}\gamma'^{2}_{{\bf k}} {B_{{\bf k}}\over 2\omega_{{\bf k}}}{\rm coth}  [{1\over 2}\beta\omega_{{\bf k}}], \\
C_{3}&=&{1\over N}\sum_{{\bf k}}\gamma_{{\bf k}}\gamma_{{\bf k}}'{B_{{\bf k}}\over 2\omega_{{\bf k}}}{\rm coth}[{1\over 2}\beta\omega_{{\bf k}}],\\
{1\over 2} &=&{1\over N}\sum_{{\bf k}}{B_{{\bf k}} \over 2\omega_{{\bf k}}}{\rm coth} [{1\over 2}\beta\omega_{{\bf k}}],\label{SCE2i}\\
\chi^{z}_{1}&=&{1\over N}\sum_{{\bf k}}\gamma_{{\bf k}} {B_{z}({\bf k})\over 2\omega_{z}({\bf k})}{\rm coth}[{1\over 2}\beta\omega_{z}({\bf k})],\\
\chi^{z}_{2}&=& {1\over N}\sum_{{\bf k}}\gamma_{{\bf k}}'{B_{z}({\bf k})\over 2\omega_{z}({\bf k})}{\rm coth}[{1\over 2}\beta\omega_{z}({\bf k})], \\
C^{z}_{1}&= &{1\over N}\sum_{{\bf k}}\gamma^{2}_{{\bf k}}{B_{z}({\bf k})\over 2\omega_{z}({\bf k})}{\rm coth}[{1\over 2}\beta\omega_{z}({\bf k})], \\
C^{z}_{3}&=&{1\over N}\sum_{{\bf k}}\gamma_{{\bf k}}\gamma_{{\bf k}}'{B_{z}({\bf k})\over 2\omega_{z}({\bf k})}{\rm coth} [{1\over 2}\beta\omega_{z}({\bf k})],
\end{eqnarray}
\end{subequations}
then all the order parameters, the decoupling parameter $\alpha$, and the chemical potential $\mu$ are determined by the self-consistent calculation without using any adjustable parameters \cite{Feng03,Feng06,Feng06a,Feng12,Guo07,Huang13,Zhao12}.


\subsection{Doping dependence of charge-carrier pair gap and coupling strength}\label{KEDSQC}

\begin{figure}[h!]
\centering
\includegraphics[scale=0.75]{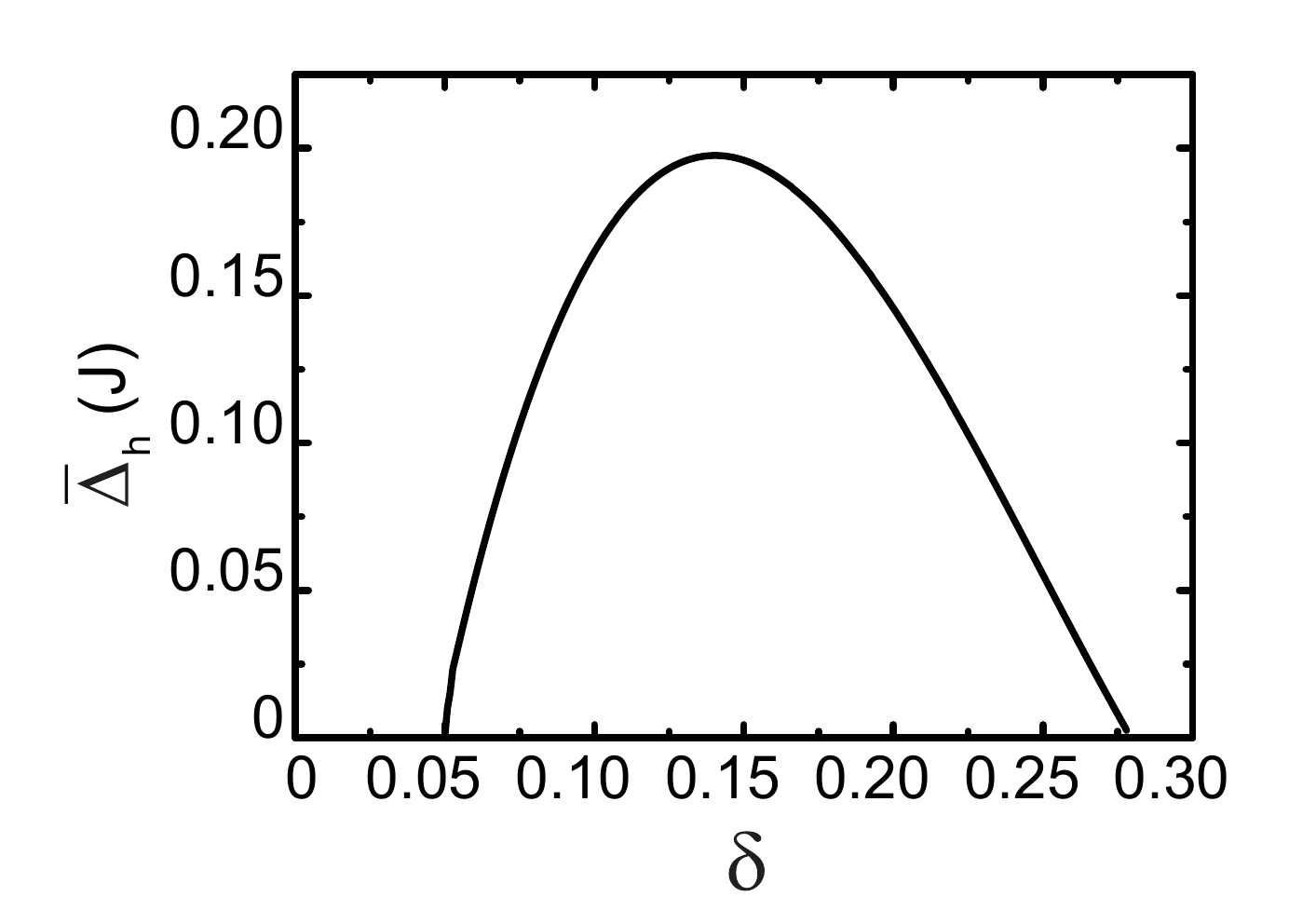}
\caption{The charge-carrier pair gap parameter as a function of doping with $T=0.002J$ for $t/J=2.5$ and $t'/t=0.3$. [From Ref. \cite{Huang13}.] \label{pair-gap-parameter-doping}}
\end{figure}

\begin{figure}[h!]
\centering
\includegraphics[scale=0.6]{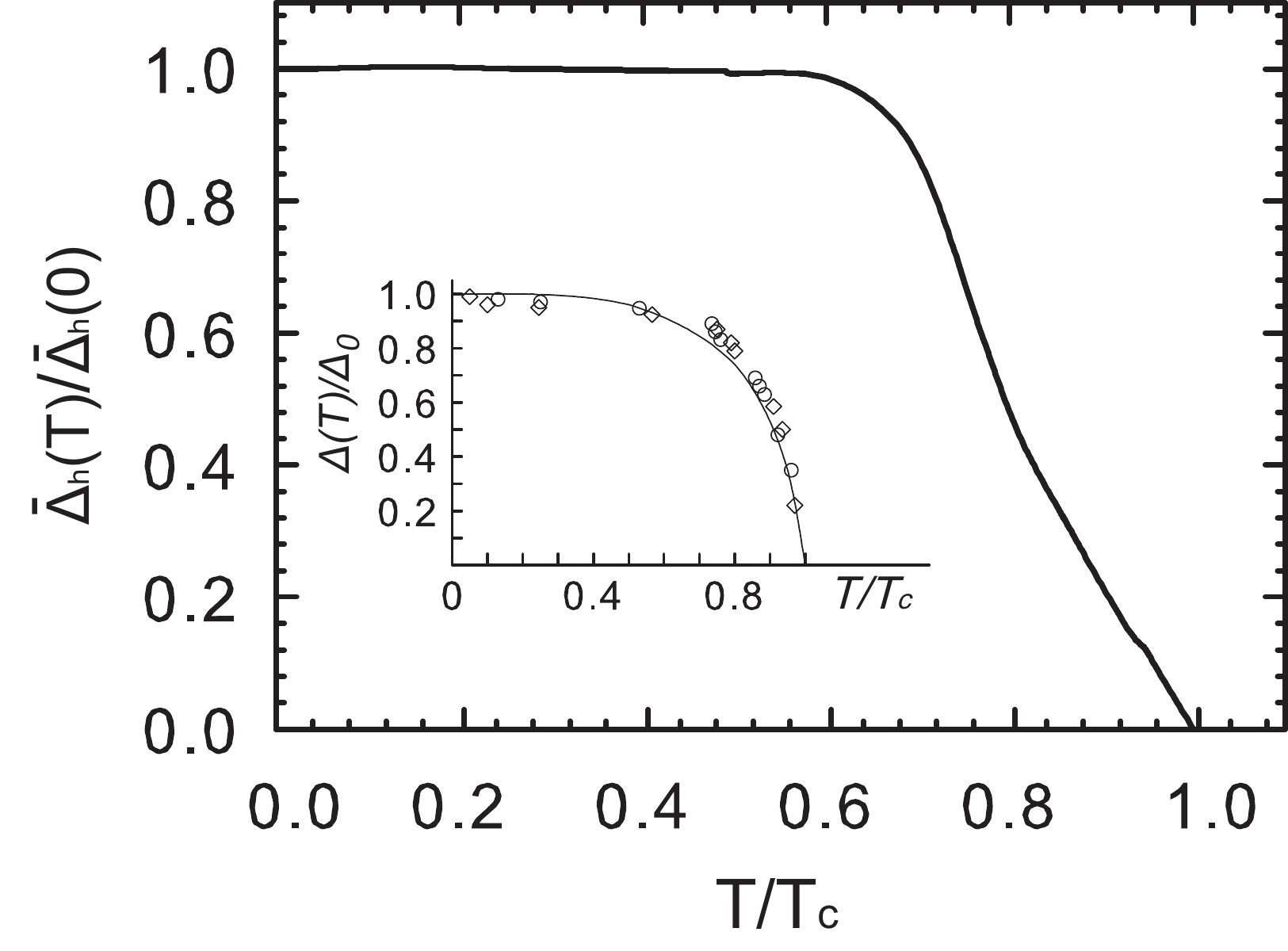}
\caption{The charge-carrier pair gap parameter as a function of temperature at $\delta=0.09$ for $t/J=2.5$ and $t'/t=0.3$. Inset: the corresponding experimental data of the pair gap parameter for the underdoped Bi$_{2}$Sr$_{2}$Ca$_{2}$Cu$_{3}$O$_{10+\delta}$ taken from Ref. \cite{Vishik10}. [From Ref. \cite{Zhao12}.] \label{pair-gap-parameter-temp}}
\end{figure}

The above equations (\ref{SCE1}) and (\ref{SCE2}) have been calculated self-consistently \cite{Feng03,Feng06,Feng12,Guo07,Huang13,Zhao12}, and the result \cite{Huang13} of the charge-carrier pair gap parameter $\bar{\Delta}_{\rm h}$ as a function of doping for parameters $t/J=2.5$ and $t'/t=0.3$ with temperature $T=0.002J$ is shown in Fig. \ref{pair-gap-parameter-doping}, where the charge-carrier pair gap parameter $\bar{\Delta}_{\rm h}$ takes a domelike shape with the underdoped and overdoped regimes on each side of the {\it optimal doping} $\delta_{\rm optimal}\approx 0.15$, where $\bar{\Delta}_{\rm h}$ reaches its maximum. The Andreev reflection experiments measure directly the binding energy of the charge-carrier pair \cite{Deutscher05}, while the Raman scattering and INS are the reverse experiments of the Andreev reflection in which they excite a charge-carrier pair out of the condensate energy \cite{Fujita12,Hufner08,Devereaux07}. All the experimental data from the Andreev reflection, Raman scattering and INS \cite{Fujita12,Hufner08,Deutscher05,Devereaux07} indicate that the charge-carrier pair gap parameter follows $T_{\rm c}$ as a function of doping. The domelike shape of the doping dependence of $\bar{\Delta}_{\rm h}$ in Fig. \ref{pair-gap-parameter-doping} obtained within the framework of the kinetic-energy driven SC mechanism is qualitatively consistent with these experimental data \cite{Fujita12,Hufner08,Deutscher05,Devereaux07}. In particular, since the charge-carrier pair order is established through an emergence of the charge-carrier quasiparticle, the charge-carrier pair state (then the SC-state) is controlled by both the charge-carrier pair gap $\bar{\Delta}_{\rm h}({\bf k})$ and charge-carrier quasiparticle coherence $Z_{\rm hF}$, which is reflected directly from the self-consistent equations (\ref{CCQCWSCE}) and (\ref{CCPGPSCE}). Moreover, this charge-carrier pair gap parameter $\bar{\Delta}_{\rm h}$ is strongly temperature dependent. To show this point clearly, the charge-carrier pair gap parameter $\bar{\Delta}_{\rm h}$ as a function of temperature \cite{Zhao12} for $t/J=2.5$ and $t'/t=0.3$ at doping $\delta=0.09$ is shown in Fig. \ref{pair-gap-parameter-temp}. For comparison, the corresponding experimental result of the pair gap parameter \cite{Vishik10} for the underdoped Bi$_{2}$Sr$_{2}$Ca$_{2}$Cu$_{3}$O$_{10+\delta}$ is also shown in Fig. \ref{pair-gap-parameter-temp} (inset). This calculated result in Fig. \ref{pair-gap-parameter-temp} indicates that the charge-carrier pair gap parameter $\bar{\Delta}_{\rm h}$ follows qualitatively a BCS-type temperature dependence, i.e., it decreases with increasing temperatures, and eventually vanishes at $T_{\rm c}$, which is also in qualitative agreement with the experimental data \cite{Vishik10} of cuprate superconductors.

\begin{figure}[h!]
\centering
\includegraphics[scale=1.0]{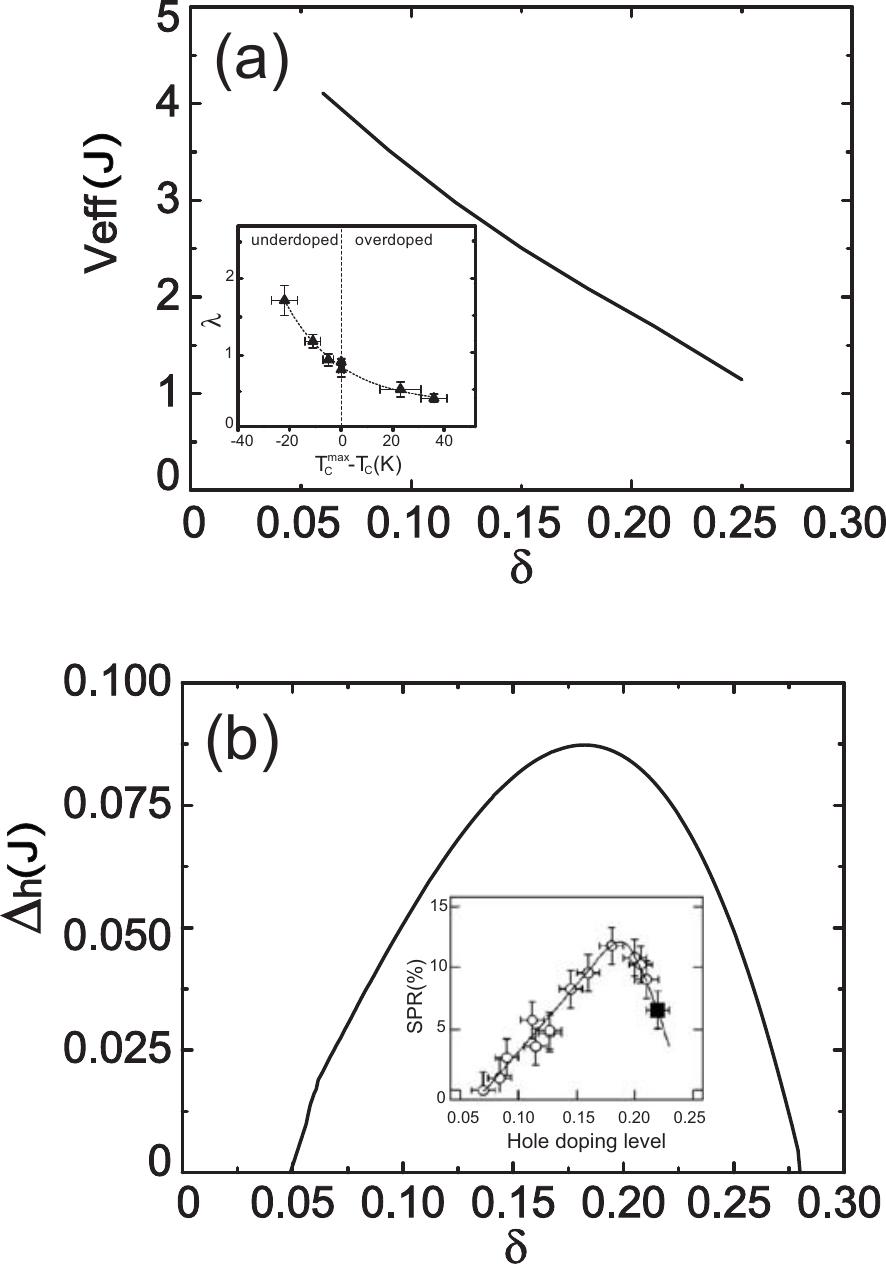}
\caption{(a) The coupling strength and (b) the charge-carrier pair order parameter as a function of doping with $T=0.002J$ for $t/J=2.5$ and $t'/t=0.3$. Inset in (a): the corresponding experimental data of the coupling strength for Bi$_{2}$Sr$_{2}$Ca$_{2}$Cu$_{3}$O$_{8+\delta}$ taken from Ref. \cite{Johnson01}, and inset in (b): the corresponding experimental data of the pair order parameter for Bi$_{2}$Sr$_{2}$Ca$_{2}$Cu$_{3}$O$_{8+\delta}$ taken from Ref. \cite{He04}. [From Ref. \cite{Huang13}.] \label{V-pair-order-parameter-doping}}
\end{figure}

In the framework of the kinetic-energy driven SC mechanism, spin excitations are directly coupled to charge-carrier pairs, and then the strength of this coupling with each charge-carrier pair is measured by the charge-carrier pair gap parameter 2$\bar{\Delta}_{\rm h}$. If the strength of the coupling exceeds the pairing energy 2$\bar{\Delta}_{\rm h}$, the charge-carrier pairs will immediately be broken up. Since both the coupling strength $V_{\rm eff}$ and charge-carrier pair order parameter $\Delta_{\rm h}$ have been incorporated into the charge-carrier pair gap parameter $\bar{\Delta}_{\rm h}$, the strength $V_{\rm eff}$ of the charge-carrier attractive interaction mediated by spin excitations can be therefore obtained in terms of the ratio of $\bar{\Delta}_{\rm h}$ and $\Delta_{\rm h}$ as \cite{Feng12,Huang13},
\begin{eqnarray}\label{strength}
V_{\rm eff}={\bar{\Delta}_{\rm h}\over\Delta_{\rm h}},
\end{eqnarray}
where the charge-carrier pair order parameter $\Delta_{\rm h}$ can be evaluated explicitly from the charge-carrier off-diagonal Green's function (\ref{BCSHODGF}) as \cite{Feng03,Feng06,Feng06a,Feng12,Guo07,Huang13,Zhao12},
\begin{eqnarray}\label{CCDGP}
\Delta_{\rm h}={2\over N}\sum_{{\bf k}}[\gamma^{(\rm d)}_{{\bf k}}]^{2}{Z_{\rm hF}\bar{\Delta}_{\rm hZ}\over E_{{\rm h}{\bf k}}}{\rm tanh}\left({1\over 2}\beta E_{{\rm h}{\bf k}} \right).
\end{eqnarray}
In Fig. \ref{V-pair-order-parameter-doping}, we \cite{Feng12,Huang13} show (a) the coupling strength $V_{\rm eff}$ and (b) charge-carrier pair order parameter $\Delta_{\rm h}$ as a function of doping for $t/J=2.5$ and $t'/t=0.3$ with $T=0.002J$. For comparison, the corresponding experimental results of the coupling strength \cite{Johnson01} and pair order parameter \cite{He04} for Bi$_{2}$Sr$_{2}$Ca$_{2}$Cu$_{3}$O$_{8+\delta}$ are also shown in Fig. \ref{V-pair-order-parameter-doping}a and Fig. \ref{V-pair-order-parameter-doping}b (inset), respectively, where the pair order parameter is related to the SC peak ratio (SPR), which is defined as the SC peak intensity divided by the overall spectral weight in the antinodal point, and then the pair order parameter is obtained indirectly by the experimental measurements of SPR \cite{He04}. The result in Fig. \ref{V-pair-order-parameter-doping}a shows that the coupling strength $V_{\rm eff}$ smoothly decreases upon the increase of doping from a strong-coupling case in the underdoped regime to a weak-coupling side in the overdoped regime, and therefore is qualitatively consistent with the experimental result of cuprate superconductors \cite{Kordyuk10,Johnson01,Dahm09}. However, the charge-carrier pair order parameter $\Delta_{\rm h}$ increases with increasing doping in the lower doped regime, and reaches a maximum around the {\it critical doping} $\delta_{\rm critical}\approx 0.195$, then decreases with increasing doping in the higher doped regime. In comparison with the corresponding result of the charge-carrier pair gap parameter $\bar{\Delta}_{\rm h}$ in Fig. \ref{pair-gap-parameter-doping}, it is therefore found that the special doping dependence of the coupling strength $V_{\rm eff}$ in Fig. \ref{V-pair-order-parameter-doping}a induces an important shift from the {\it critical doping} $\delta_{\rm critical}\approx 0.195$ for the maximal $\Delta_{\rm h}$ to the {\it optimal doping} $\delta_{\rm optimal} \approx 0.15$ for the maximal $\bar{\Delta}_{\rm h}$.

\subsection{Doping dependence of $T_{\rm c}$}\label{DDTC}

\begin{figure}[h!]
\centering
\includegraphics[scale=0.55]{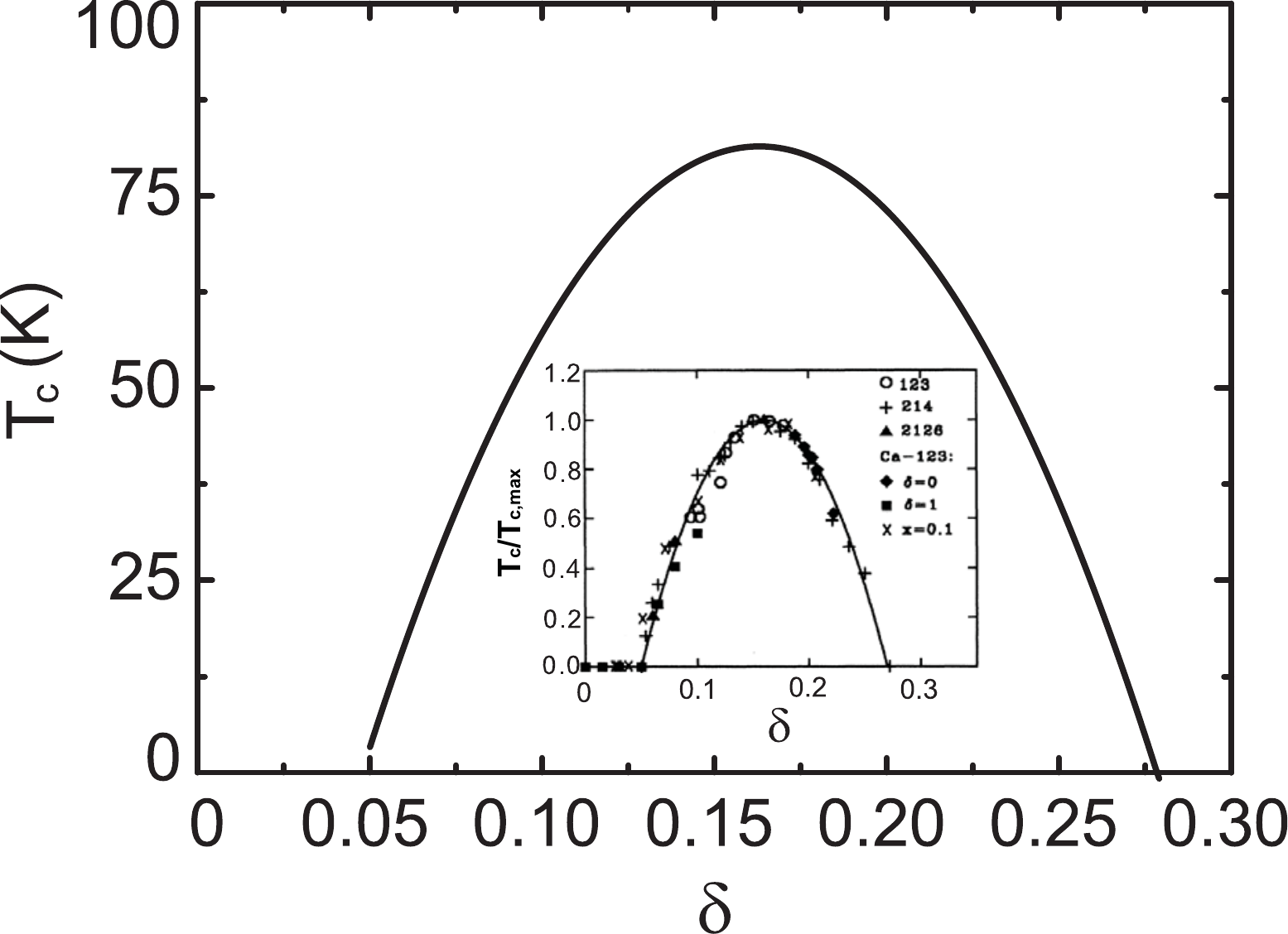}
\caption{$T_{\rm c}$ as a function of doping for $t/J=2.5$, $t'/t=0.3$, and $J=1000$K. Inset: the corresponding experimental result of cuprate superconductors taken from \cite{Tallon95}. [From Ref. \cite{Huang13}.] \label{Tc-doping}}
\end{figure}

$T_{\rm c}$ on the other hand can be obtained self-consistently from the self-consistent equations (\ref{SCE1}) and (\ref{SCE2}) by the condition ${\bar{\Delta}_{\rm h}}=0$, and the result \cite{Huang13} of $T_{\rm c}$ as a function of doping for $t/J=2.5$, $t'/t=0.3$, and $J=1000$K is shown in Fig. \ref{Tc-doping} in comparison with the corresponding experimental results \cite{Tallon95} (inset). Since the charge-carrier pair state (then the SC-state) of cuprate superconductors is controlled by both the charge-carrier pair gap and charge-carrier quasiparticle coherence, which leads to that the maximal $T_{\rm c}$ occurs around the {\it optimal doping}, and then decreases in both the underdoped and the overdoped regimes, in good agreement with the experimental results of cuprate superconductors \cite{Tallon95,Momono96,Presland91}. This calculated result also shows that the charge-carrier pairs (\ref{CCPGF}) are formed in momentum space, and then as in conventional superconductors \cite{Mourachkine04,Eliashberg60,Mahan81}, the pairing and onset of the phase coherence take place simultaneously at $T_{\rm c}$. In particular, $T_{\rm c}$ that is set by the pair gap and quasiparticle coherence has been observed experimentally in cuprate superconductors \cite{DLFeng00,Ding01}, which may be a common feature for all superconductors. This follows from a fact that in spite of the electron-phonon SC mechanism, $T_{\rm c}$ in conventional superconductors is also determined by the pair gap and quasiparticle coherence \cite{Eliashberg60,Mahan81}. Finally, it should be emphasized that except for the quasiparticle coherence, the superfluid density $\rho_{\rm s}$ in cuprate superconductors is closely related to the charge-carrier pair order parameter $\Delta_{\rm h}$, which is why $\rho_{\rm s}$ increases with increasing doping in the lower doped regime, and reaches a maximum around the {\it critical doping} $\delta_{\rm critical}\approx 0.195$, then decreases in the higher doped regime \cite{Huang13,Feng10,Bernhard01}. This is also why \cite{He04} the charge-carrier pair order parameter $\Delta_{\rm h}$ can be obtained indirectly in terms of the experimental measurement of the superfluid density $\rho_{\rm s}$. We shall come back to give an interpretation about this issue in the discussions of the doping dependence of the superfluid density (see subsection \ref{DTDSD}).

In the fermion-spin theory (\ref{CSS}), the electron is decoupled as the charge carrier and spin according to its charge and spin degrees of freedom. However, we work in the case $t\gg J$, so that charge carrier motions are much faster than spins. Although the charge carriers repel each other because of the Coulomb interaction, at low energies there is an effective attraction resulting directly from the interaction between the charge carriers and spins in the kinetic energy of the $t$-$J$ model (\ref{CSStJmodel}) by the exchange of spin excitations. This follows from a fact that a doped Mott insulator is formed by mobile charge carriers detaching themselves from spins that form the spin configuration. Such a mobile charge carrier is a magnetic dressing, and therefore its motion is strongly dependent on the spin configuration. When this charge carrier moves, the spin configuration rearrangements due to the presence of this charge carrier itself is left in its wake. This attracts a second charge carrier, leading to a net attraction between charge carriers. This kinetic-energy driven SC mechanism works because the spin dynamics is slow relative to charge carriers, which also is a natural consequence of the fact that the spin excitations are very local in cuprate superconductors \cite{Dean14}. In comparison with the conventional electron-phonon SC mechanism \cite{Schrieffer64,Bardeen57}, the spin excitation in cuprate superconductors plays a similar role to that of the phonon in conventional superconductors.

\begin{figure}[h!]
\centering
\includegraphics[scale=0.55]{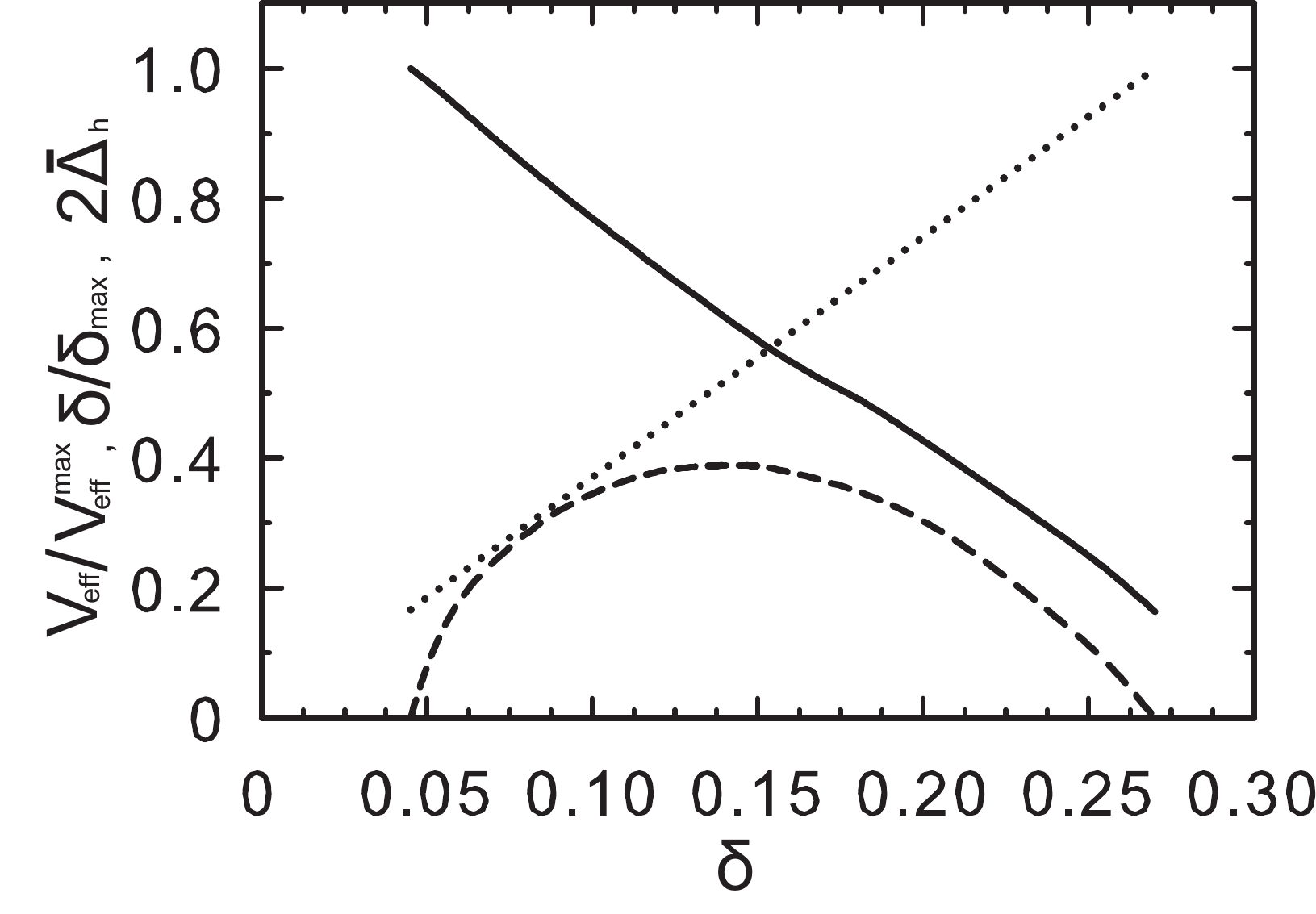}
\caption{The coupling strength (solid line), the doping concentration (dotted line), and the charge-carrier pair gap parameter ($2\bar{\Delta}_{\rm h}$) (dashed line) as a function of doping with $T=0.002J$ for $t/J=2.5$ and $t'/t=0.3$. [From Ref. \cite{Feng12}.] \label{V-doping}}
\end{figure}

On the other hand, the essential physics of a domelike shape of the doping dependence of $T_{\rm c}$ in the framework of the kinetic-energy driven SC mechanism can be attributed to a competition between the kinetic energy and magnetic energy in cuprate superconductors. The parent compounds of the cuprate superconductors are Mott insulators \cite{Anderson87,Phillips10}. When charge carriers are doped into a Mott insulator, there is a gain in the kinetic energy per charge carrier proportional to $t$ due to hopping, however, at the same time, the magnetic energy decreases, costing an energy of approximately $J$ per site \cite{Lee06}. This leads to that the spin excitation spectral intensity decreases with increasing doping. However, a decrease of the spin excitation spectral intensity with increasing doping also leads to a decrease of the coupling strength $V_{\rm eff}$ with increasing doping as shown in Fig. \ref{V-pair-order-parameter-doping}a. To see this competition clearly, $V_{\rm eff}/V^{\rm max}_{\rm eff}$ (solid line), $\delta/\delta_{\rm max}$ (dotted line), and $2\bar{\Delta}_{\rm h}$ (dashed line) as a function of doping with $T=0.002J$ for $t/J=2.5$ and $t'/t=0.3$ is shown in Fig. \ref{V-doping}, where $V^{\rm max}_{\rm eff}=V_{\rm eff}|_{\delta\approx 0.045}$ is the value of $V_{\rm eff}$ at the starting point of the SC dome, while $\delta_{\rm max}\approx 0.27$ is the doping concentration at the end point of the SC dome. In the underdoped regime, the coupling strength $V_{\rm eff}$ is very strong to bind the most charge carriers into charge-carrier pairs, and therefore the number of charge-carrier pairs increases with increasing doping, which leads to that the charge-carrier pair gap parameter $\bar{\Delta}_{\rm h}$ and $T_{\rm c}$ increase with increasing doping. However, in the overdoped regime, $V_{\rm eff}$ is relatively weak. In this case, not all charge carriers can be bound to form charge-carrier pairs by this weakly attractive interaction, and therefore the number of charge-carrier pairs decreases with increasing doping, which leads to that the charge-carrier pair gap parameter $\bar{\Delta}_{\rm h}$ and $T_{\rm c}$ decrease with increasing doping. In other words, in analogy to the electron-phonon SC mechanism for conventional superconductors, the reduction of $T_{\rm c}$ of cuprate superconductors in the overdoped side is driven by a reduction in the coupling strength $V_{\rm eff}$ of the pairing interaction. In particular, the optimal doping is a balanced point, where the number of charge-carrier pairs and coupling strength $V_{\rm eff}$ are optimally matched. This is why the maximal $\bar{\Delta}_{\rm h}$ and $T_{\rm c}$ occur around the optimal doping, and then decrease in both the underdoped and overdoped regimes.

\subsection{Summary}

We have reviewed briefly the theory of the kinetic-energy driven superconductivity. The kinetic-energy driven SC mechanism is purely electronic without phonons, where cuprate superconductors involve charge-carrier pairs bound together by the exchange of spin excitations. In particular, this kinetic-energy driven SC-state is conventional BCS-like with the d-wave symmetry, and then the obtained d-wave BCS type formalism for the charge-carrier pairing can be used to compute $T_{\rm c}$ and the related SC coherence of the low-energy excitations in cuprate superconductors on the first-principles basis much as can be done for conventional superconductors, although the pairing mechanism is driven by the kinetic energy by the exchange of spin excitations in the higher powers of the doping concentration, and other exotic magnetic scattering \cite{Fujita12,Dean14} is beyond the d-wave BCS type formalism. Furthermore, this kinetic-energy driven SC-state is controlled by both the charge-carrier pair gap and quasiparticle coherence, which leads to that the maximal $T_{\rm c}$ occurs around the optimal doping, and then decreases in both underdoped and overdoped regimes. The theory of kinetic-energy driven superconductivity also indicates that the strong electron correlation favors superconductivity, since the main ingredient is identified into a charge-carrier pairing mechanism not from the external degree of freedom such as the phonon but rather solely from the internal spin degree of freedom of the electron.

\section{Electromagnetic response}\label{Meissner-effect}

Superconductivity is characterized by exactly zero electrical resistance and expulsion of magnetic fields occurring in superconductors when cooled below $T_{\rm c}$. The later phenomenon is so-called Meissner effect \cite{Schrieffer64}, i.e., a superconductor is placed in an external magnetic field ${\rm B}$ smaller than the upper critical field ${\rm B}_{\rm c}$, the magnetic field ${\rm B}$ penetrates only to a penetration depth $\lambda$ (few hundred nm for cuprate superconductors at zero temperature) and is excluded from the main body of the system. This magnetic field penetration depth is a fundamental parameter of superconductors, and provides a rather direct measurement of the superfluid density $\rho_{\rm s}$ ($\rho_{\rm s}\equiv\lambda^{-2}$) \cite{Bonn96,Schrieffer64}, which is proportional to the squared amplitude of the macroscopic wave function. In particular, the variation of the magnetic field penetration depth (then the superfluid density) as a function of doping and temperature gives the information about the nature of quasiparticle excitations and their dynamics \cite{Bonn96}. Moreover, the magnetic field penetration depth can be also used as a probe of the pairing symmetry, since it can distinguish between a fully gapped and a nodal quasiparticle excitation spectrum \cite{Bonn96,Tsuei00}. The former results in the thermally activated (exponential) temperature dependence of the magnetic field penetration depth, whereas the latter one implies a power law behavior. This is why the first evidence of the d-wave pairing state in cuprate superconductors was obtained from the earlier experimental measurement of the magnetic field penetration depth \cite{Hardy93}.

\begin{figure}[h!]
\centering
\includegraphics[scale=0.8]{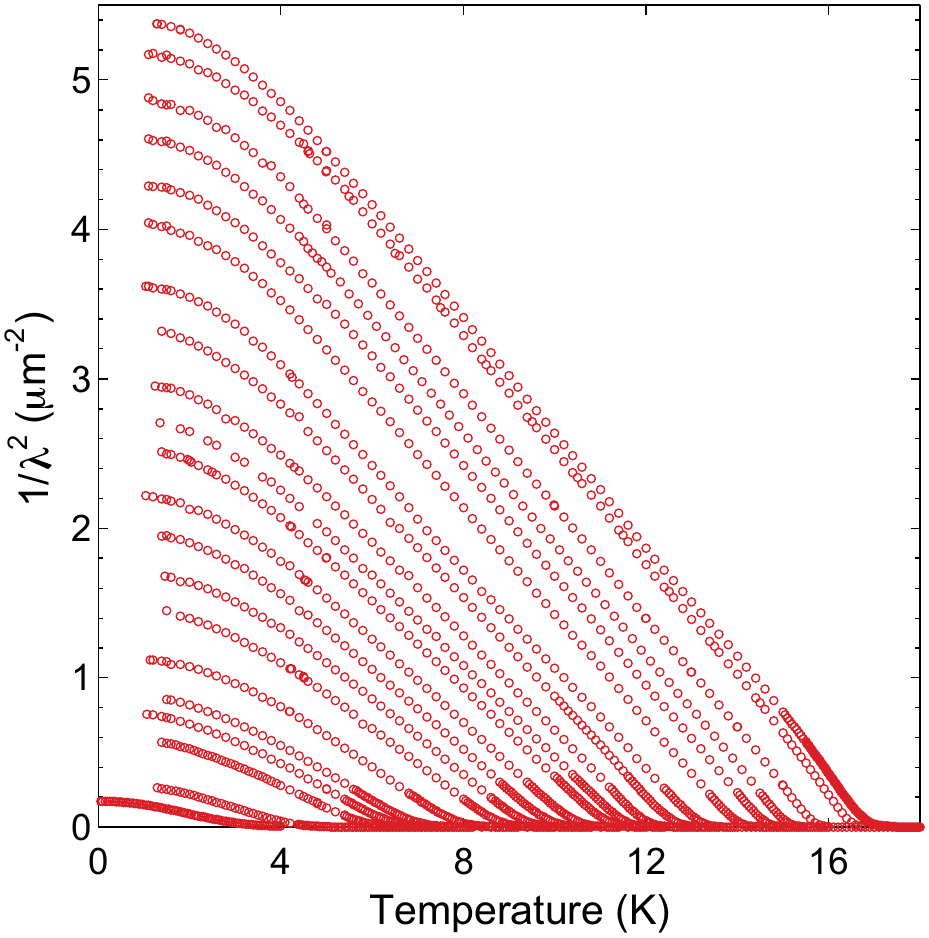}
\caption{(Color) The temperature dependence of the superfluid density measured for YBa$_{2}$Cu$_{3}$O$_{6+\delta}$ at different doping concentrations. [From Ref. \cite{Broun07}.] \label{superfluid-density-temp-exp}}
\end{figure}

\begin{figure}[h!]
\centering
\includegraphics[scale=0.8]{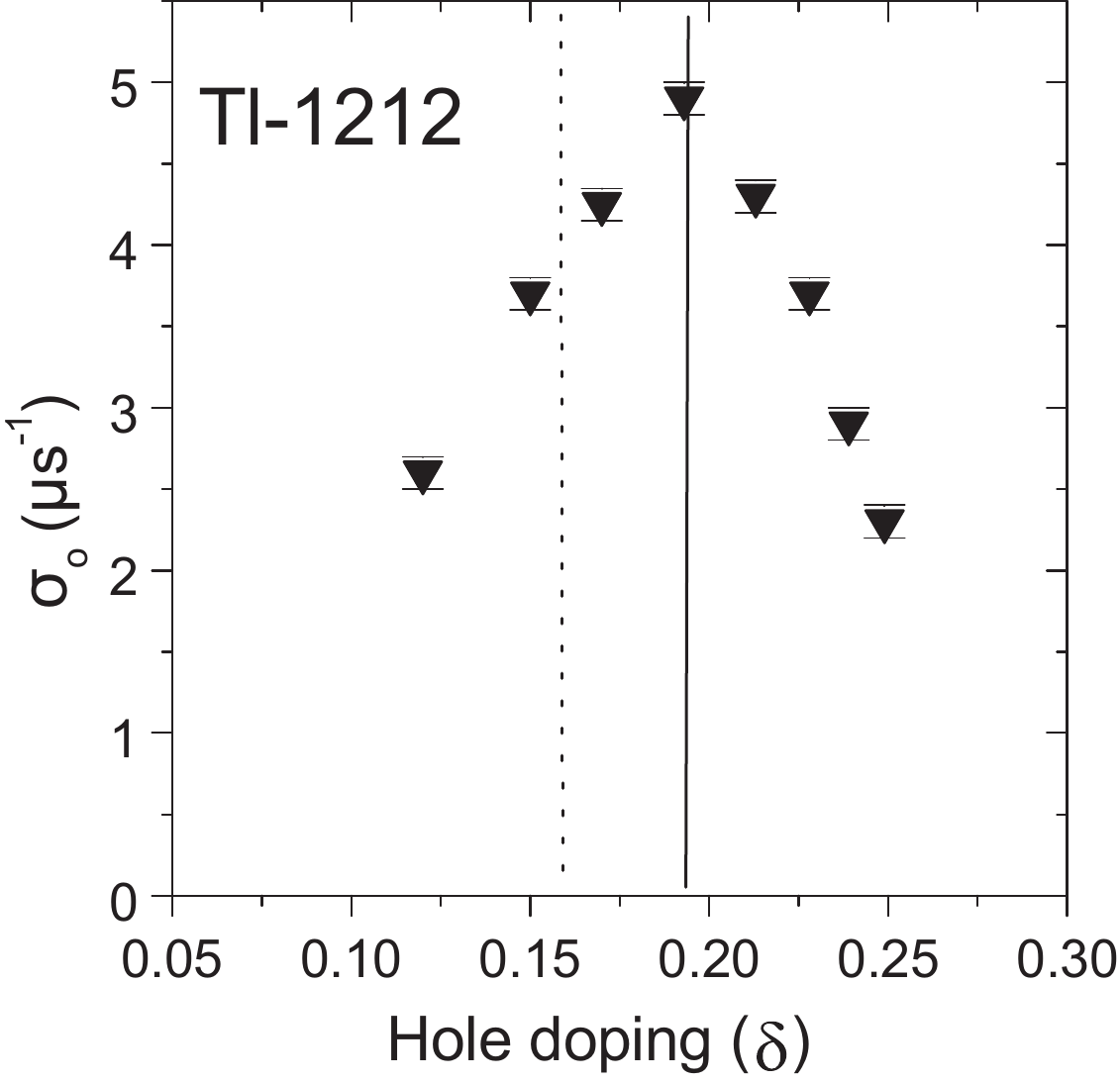}
\caption{The doping dependence of the low-temperature depolarization rate for underdoped to overdoped Tl$_{1-y}$Pb$_{y}$Sr$_{2}$Ca$_{1-x}$Y$_{x}$Cu$_{2}$O$_{7}$. The solid (dotted) line marks critical (optimal) doping. [From Ref. \cite{Bernhard01}.] \label{superfluid-density-doping-exp}}
\end{figure}

Experimentally, by virtue of systematic studies using the muon-spin-rotation measurement technique, the essential feature of the evolution of the magnetic field penetration depth and superfluid density in cuprate superconductors with doping and temperature has been established now for all the temperature $T\leq T_{\rm c}$ throughout the SC dome \cite{Bonn96,Hardy93,Bernhard01,Khasanov04,Suter04,Kamal98,Lee96,Broun07,Kim03,Lemberger11}: (a) the magnetic field screening is found to be of exponential character \cite{Khasanov04,Suter04}, in support of a local (London-type) nature of the electrodynamic response; (b) the magnetic field penetration depth is a linear temperature dependence at low temperatures except for the extremely low temperatures where a strong deviation from the linear characteristics (a nonlinear effect) appears \cite{Hardy93,Kamal98,Lee96}, which leads to that the superfluid density $\rho_{\rm s}$ shows a crossover from the linear temperature dependence at low temperatures to a nonlinear one at the extremely low temperatures as illustrated in Fig. \ref{superfluid-density-temp-exp} \cite{Kamal98,Lee96,Broun07,Kim03}; (c) the experimental measurements \cite{Bernhard01,Lemberger11} throughout the SC dome show that the superfluid density $\rho_{\rm s}$ appears from the starting point of the SC dome, and then increases with increasing doping in the lower doped regime. However, this $\rho_{\rm s}$ reaches its highest value around the {\it critical doping} $\delta_{\rm critical}\approx 0.19$, and then decreases at the higher doped regime, eventually disappearing together with superconductivity at the end of the SC dome (see Fig. \ref{superfluid-density-doping-exp}). This is different from the case of the doping dependence of $T_{\rm c}$, since the maximal $T_{\rm c}$ occurs around the the {\it optimal doping} $\delta_{\rm optimal}\approx 0.15$, and then decreases in both the underdoped and the overdoped regimes \cite{Tallon95,Momono96,Presland91}. In particular, it has been shown \cite{Bernhard01,Lemberger11,Tallon95,Momono96} that the maximal $T_{\rm c}$ around the {\it optimal doping} $\delta_{\rm optimal}\approx 0.15$ and the peak of $\rho_{\rm s}$ around the {\it critical doping} $\delta_{\rm critical}\approx 0.19$ are a common feature of the hole-doped cuprate superconductors. Thus both $T_{\rm c}$ and $\rho_{\rm s}$ variation as a function of doping provides important information crucial to understanding the details of the SC-state \cite{Bonn96}. Theoretically, the magnetic field penetration depth and the related superfluid density in cuprate superconductors have been discussed widely based on a phenomenological d-wave BCS formalism \cite{Yip92,Kosztin97,Franz97,Li00,Sheehy04}. In the local limit, where the magnetic field penetration depth $\lambda$ is much larger than the coherence length $\zeta$, i.e., $\lambda\gg\zeta$, it has been shown \cite{Kosztin97} that a simple d-wave pairing state gives the linear temperature dependence of the magnetic field penetration depth $\Delta\lambda(T)=\lambda(T)-\lambda(0)\propto T/\bar{\Delta}_{0}$ at low temperatures, where $\bar{\Delta}_{0}$ is the zero-temperature value of the d-wave gap amplitude. However, the characteristic feature of the d-wave energy gap is the existence of the gap nodes, which can lead to the nonlinear effect of field on the penetration depth (then superfluid density) at the extremely low temperatures \cite{Yip92,Kosztin97,Franz97,Li00,Sheehy04}. This follows from a fact that the nonlocal effect is closely related to the divergence of the coherence length $\zeta$ at the gap nodes, as the coherence length $\zeta$ varies in inverse proportion to the value of the energy gap, and then this nonlocal effect at the extremely low temperatures can lead to a nonlinear temperature dependence of the magnetic field penetration depth in the clean limit.

In this section, we show how the kinetic-energy driven SC mechanism can be applied to the discussions of the Meissner effect in cuprate superconductors \cite{Feng10,Krzyzosiak10}. For the discussions of the doping and temperature dependence of the electromagnetic response, the $t$-$J$ model (\ref{CSStJmodel}) can be extended by including the exponential Peierls factors as \cite{Feng10,Krzyzosiak10,Huang13},
\begin{eqnarray}\label{MtJmodel}
H&=&t\sum_{l\hat{\eta}}e^{-i(e/{\hbar}){\bf A}(l)\cdot\hat{\eta}}(h^{\dagger}_{l+\hat{\eta}\uparrow}h_{l\uparrow}S^{+}_{l}S^{-}_{l+\hat{\eta}}+h^{\dagger}_{l+\hat{\eta}\downarrow} h_{l\downarrow}S^{-}_{l}S^{+}_{l+\hat{\eta}}) \nonumber\\
&-&t'\sum_{l\hat{\tau}}e^{-i(e/{\hbar}){\bf A}(l)\cdot\hat{\tau}}(h^{\dagger}_{l+\hat{\tau}\uparrow}h_{l\uparrow}S^{+}_{l}S^{-}_{l+\hat{\tau}}+h^{\dagger}_{l+\hat{\tau}\downarrow} h_{l\downarrow}S^{-}_{l}S^{+}_{l+\hat{\tau}}) \nonumber\\
&-&\mu\sum_{l\sigma}h^{\dagger}_{l\sigma}h_{l\sigma}+J_{{\rm eff}}\sum_{l\hat{\eta}}{\bf S}_{l}\cdot {\bf S}_{l+\hat{\eta}},
\end{eqnarray}
where the exponential Peierls factors account for the coupling of the electron charge to an external magnetic field \cite{Hirsch92a,Misawa94} in terms of the vector potential ${\bf A}(l)$.

\subsection{Linear response approach}

In a superconductor, an external magnetic field generally represents a large perturbation on a superconductor, then the induced field arising from the superconductor cancels this external magnetic field over most of the system. In this case, the net field acts only near the surface on a scale of the magnetic field penetration depth, and then it can be treated as a weak perturbation on the system as a whole \cite{Schrieffer64}. This is why the Meissner effect can be successfully studied within the linear response approach \cite{Fukuyama69,Arseev06}, where the linear response current density $J_{\mu}$ and the vector potential $A_{\nu}$ are closely related by a kernel of the response function $K_{\mu\nu}$ as,
\begin{eqnarray}\label{linres}
J_{\mu}({\bf q},\omega)=-\sum\limits_{\nu=1}^{3}K_{\mu\nu}({\bf q},\omega)A_{\nu}({\bf q},\omega),
\end{eqnarray}
where the Greek indices label the axes of the Cartesian coordinate system. It should be noted that the relation (\ref{linres}), which is local in the reciprocal space, in general implies a nonlocal response in the coordinate space. The kernel of the response function plays a crucial role for the discussion of the electromagnetic response, and can be separated into two parts as $K_{\mu\nu}({\bf q},\omega)=K^{({\rm d})}_{\mu\nu}({\bf q},\omega)+K^{({\rm p})}_{\mu\nu}({\bf q},\omega)$, where $K^{({\rm d}) }_{\mu\nu}$ and $K^{({\rm p})}_{\mu\nu}$ are the corresponding diamagnetic and paramagnetic parts, respectively, and are related to the electron current density in the presence of the vector potential $A_{\nu}$. The evaluation of the diamagnetic contribution usually poses no difficulties since it is known almost immediately from the form of the diamagnetic current operator: it turns out to be diagonal and proportional to the average kinetic term. However, the paramagnetic part can only be calculated approximately since it involves evaluation of a retarded current-current correlation function (polarization bubble).

The vector potential ${\bf A}$ (then the external magnetic field ${\bf B}={\rm rot}\,{\bf A}$) has been coupled to the electrons, which are now represented by $C_{l\uparrow}= h^{\dagger}_{l\uparrow}S^{-}_{l}$ and $C_{l\downarrow}= h^{\dagger}_{l\downarrow}S^{+}_{l}$ in the fermion-spin representation (\ref{CSS}). However, in the CSS framework, the vector potential ${\bf A}$ is coupled to the electron charge $h^{\dagger}_{l\sigma}$ in the $t$-$J$ model (\ref{MtJmodel}), while the corresponding electron magnetic momentum can be coupled to the external magnetic field by including the Zeeman term in the $t$-$J$ model (\ref{MtJmodel}). However, for cuprate superconductors, the upper critical magnetic field is 50 Tesla or greater around the optimal doping \cite{Wang06,Wen08}, while we \cite{Feng10,Krzyzosiak10} mainly focus on the case where the applied external magnetic field ($B<10$ mT) is much less than the upper critical magnetic field, and then the Zeeman term in the $t$-$J$ model (\ref{MtJmodel}) has been dropped. We shall come back to give some comments about this issue in subsection \ref{Meissner-summary}. In this case, the electron current operator is obtained in terms of the electron polarization operator, which is a summation over all the particles and their positions, and can be expressed explicitly in the fermion-spin representation as \cite{Feng10,Krzyzosiak10},
\begin{eqnarray}\label{poloper}
{\bf P}=-e\sum\limits_{l\sigma}{\bf R}_{l}C^{\dagger}_{l\sigma}C_{l\sigma}=e\sum\limits_{l}{\bf R}_{l}h^{\dagger}_{l}h_{l},
\end{eqnarray}
then the electron current operator \cite{Mahan81} is obtained by evaluating the time-derivative of the polarization operator as ${\bf j}={\partial{\bf P}/\partial t}=(i/\hbar) [H,{\bf P}]$. In particular, in the linear response approach, this electron current operator is reduced as ${\bf j}={\bf j}^{(\rm d)}+{\bf j}^{(\rm p)}$, with the corresponding diamagnetic $({\rm d})$ and paramagnetic $({\rm p})$ components of the electron current operator that are given by \cite{Feng10,Krzyzosiak10},
\begin{subequations}
\begin{eqnarray}
{\bf j}^{(\rm d)}&=&{e^{2}t\over\hbar^{2}}\sum\limits_{l\hat{\eta}}\hat{\eta}{\bf A}(l)\cdot\hat{\eta}(h_{l\uparrow}h^{\dagger}_{l+\hat{\eta}\uparrow}S^{+}_{l}S^{-}_{l+\hat{\eta}} +h_{l\downarrow}h^{\dagger}_{l+\hat{\eta}\downarrow}S^{-}_{l}S^{+}_{l+\hat{\eta}})\nonumber\\
&-&{e^{2}t'\over\hbar^{2}}\sum\limits_{l\hat{\tau}}\hat{\tau}{\bf A}(l)\cdot\hat{\tau}(h_{l\uparrow}h^{\dagger}_{l+\hat{\tau}\uparrow}S^{+}_{l}S^{-}_{l+\hat{\tau}} +h_{l\downarrow}h^{\dagger}_{l+\hat{\tau}\downarrow}S^{-}_{l}S^{+}_{l+\hat{\tau}}),~~~~~~\label{tcurdia}\\
{\bf j}^{(\rm p)}&=&{iet\over\hbar}\sum\limits_{l\hat{\eta}}\hat{\eta}(h_{l\uparrow}h^{\dagger}_{l+\hat{\eta}\uparrow}S^{+}_{l}S^{-}_{l+\hat{\eta}}+h_{l\downarrow} h^{\dagger}_{l+\hat{\eta}\downarrow}S^{-}_{l}S^{+}_{l+\hat{\eta}})\nonumber\\
&-&{iet'\over\hbar}\sum\limits_{l\hat{\tau}}\hat{\tau}(h_{l\uparrow}h^{\dagger}_{l+\hat{\tau}\uparrow}S^{+}_{l}S^{-}_{l+\hat{\tau}}+h_{l\downarrow}
h^{\dagger}_{l+\hat{\tau}\downarrow}S^{-}_{l}S^{+}_{l+\hat{\tau}}),
\label{tcurpara9}
\end{eqnarray}
\end{subequations}
respectively. Since the diamagnetic component of the electron current operator (\ref{tcurdia}) is proportional to the vector potential, the diamagnetic part of the response kernel is obtained directly as,
\begin{eqnarray}\label{diakernel}
K_{\mu\nu}^{(\rm d)}({\bf q},\omega)=-{4e^{2}\over\hbar^{2}}(\chi_{1}\phi_{1}t-2\chi_{2}\phi_{2}t')\delta_{\mu\nu}={1\over\lambda^{2}_{L}}\delta_{\mu\nu},
\end{eqnarray}
where $\lambda^{-2}_{L}=-4e^{2}(\chi_{1}\phi_{1}t- 2\chi_{2} \phi_{2}t')/\hbar^{2}$ is the doping and temperature dependence of the London penetration depth.

However, the paramagnetic part of the response kernel is more complicated to calculate, since it involves evaluation of the following electron current-current correlation function,
\begin{equation}\label{corP}
P_{\mu\nu}({\bf q},\tau)=-\langle T_{\tau}\{j^{(\rm p)}_{\mu}({\bf q},\tau)j_{\nu}^{(\rm p)}(-{\bf q},0)\}\rangle ,
\end{equation}
then the paramagnetic part of the response kernel $K_{\mu\nu}^{(\rm p)}({\bf q},\omega)$ is obtained as $K_{\mu\nu}^{(\rm p)}({\bf q},\omega)=P_{\mu\nu}({\bf q},\omega)$. In the fermion-spin approach (\ref{CSS}), the paramagnetic component of the electron current operator (\ref{tcurpara9}) can be decoupled as,
\begin{eqnarray}
{\bf j}^{(\rm p)}&=&-{ie\chi_{1}t\over\hbar}\sum\limits_{l\hat{\eta}\sigma}\hat{\eta}h^{\dagger}_{l+\hat{\eta}\sigma}h_{l\sigma}+{ie\chi_{2}t'\over\hbar} \sum\limits_{l\hat{\tau}\sigma}\hat{\tau}h^{\dagger}_{l+\hat{\tau}\sigma}h_{l\sigma}\nonumber\\
&-&{ie\phi_{1}t\over\hbar}\sum\limits_{l\hat{\eta}}\hat{\eta}(S^{+}_{l}S^{-}_{l+\hat{\eta}}+S^{-}_{l}S^{+}_{l+\hat{\eta}})+{ie\phi_{2}t'\over\hbar} \sum\limits_{l\hat{\tau}}\hat{\tau}(S^{+}_{l}S^{-}_{l+\hat{\tau}}
+S^{-}_{l}S^{+}_{l+\hat{\tau}}),\label{tcurpara15}
\end{eqnarray}
where the third and fourth terms in the right-hand side refer to the contribution from the electron spin, and can be expressed explicitly as,
\begin{subequations}\label{tcurpara16}
\begin{eqnarray}
&~&-{ie\phi_{1}t\over\hbar}\sum\limits_{l\hat{a}=\hat{x},\hat{y}}\hat{a}[(S^{+}_{l}S^{-}_{l+\hat{a}}+S^{-}_{l}S^{+}_{l+\hat{a}})-(S^{+}_{l}S^{-}_{l-\hat{a}}
+S^{-}_{l}S^{+}_{l-\hat{a}})]\nonumber\\
&=&-{ie\phi_{1}t\over\hbar}\sum\limits_{l\hat{a}=\hat{x},\hat{y}}\hat{a}[(S^{+}_{l}S^{-}_{l+\hat{a}}+S^{-}_{l}S^{+}_{l+\hat{a}})
-(S^{+}_{l+\hat{a}}S^{-}_{l}+S^{-}_{l+\hat{a}}S^{+}_{l})]\equiv 0,\\
&~&{ie\phi_{2}t'\over\hbar}\sum\limits_{l}[(\hat{x}+\hat{y})(S^{+}_{l}S^{-}_{l+\hat{x}+\hat{y}}+S^{-}_{l}S^{+}_{l+\hat{x}+\hat{y}})-(\hat{x}+\hat{y})
(S^{+}_{l}S^{-}_{l-\hat{x}-\hat{y}}+S^{-}_{l}S^{+}_{l-\hat{x}-\hat{y}})\nonumber\\
&+&(\hat{x}-\hat{y})(S^{+}_{l}S^{-}_{l+\hat{x}-\hat{y}}+S^{-}_{l}S^{+}_{l+\hat{x}-\hat{y}})-(\hat{x}-\hat{y})(S^{+}_{l}
S^{-}_{l-\hat{x}+\hat{y}}+S^{-}_{l}S^{+}_{l-\hat{x}+\hat{y}})]\nonumber\\
&=&{ie\phi_{2}t'\over\hbar}\sum\limits_{l}[(\hat{x}+\hat{y})(S^{+}_{l}S^{-}_{l+\hat{x}+\hat{y}}+S^{-}_{l}
S^{+}_{l+\hat{x}+\hat{y}})-(\hat{x}+\hat{y})(S^{+}_{l+\hat{x}+\hat{y}}S^{-}_{l}+S^{-}_{l+\hat{x}+\hat{y}}S^{+}_{l})\nonumber\\
&+&(\hat{x}-\hat{y})(S^{+}_{l}S^{-}_{l+\hat{x}-\hat{y}}+S^{-}_{l}S^{+}_{l+\hat{x}-\hat{y}})-(\hat{x}-\hat{y})
(S^{+}_{l+\hat{x}-\hat{y}}S^{-}_{l}+S^{-}_{l+\hat{x}-\hat{y}}S^{+}_{l})]\equiv 0,
\end{eqnarray}
\end{subequations}
which shows that the majority contribution for the paramagnetic component of the electron current operator comes from the electron charge, however the strong interplay between charge carriers and spins has been considered through the spin's order parameters entering in the charge-carrier part of the contribution to the current-current correlation. In particular, if we want to keep the theory gauge invariant, it is crucial to approximate the correlation function in a way of maintaining local charge conservation \cite{Schrieffer64,Misawa94,Fukuyama69,Arseev06}. Since the calculations in this subsection will be worked on with a fixed gauge of the vector potential \cite{Feng10,Krzyzosiak10}, we postpone the detailed discussions of the gauge invariant problem until subsection  \ref{invapp}. Starting with the paramagnetic current operator (\ref{tcurpara15}), we can obtain its Fourier transform in the Nambu representation in terms of the charge-carrier Nambu operators $\Psi^{\dagger}_{\bf k}=(h^{\dagger}_{{\bf k}\uparrow}, h_{-{\bf k}\downarrow})$ and $\Psi_{{\bf k}+{\bf q}}=(h_{{\bf k}+{\bf q}\uparrow}, h^{\dagger}_{-{\bf k}-{\bf q} \downarrow})^{T}$. For the purpose of addressing the gauge invariance problem in subsection \ref{invapp}, it is convenient to find the charge-carrier Green's functions and electron density in the Nambu notation as well. From Eq. (\ref{BCSHGF}), the charge-carrier BCS-type Green's function with the d-wave symmetry can be expressed in the Nambu representation as,
\begin{eqnarray}\label{NPBCSHG}
\tilde{g}({\bf k},\omega)=Z_{\rm hF}{\omega\tau_{0}+\bar{\xi}_{{\bf k}}\tau_{3}-\bar{\Delta}_{\rm hZ} ({\bf k})\tau_{1}\over\omega^{2}- E^{2}_{{\rm h}{\bf k}}},
\end{eqnarray}
where $\tau_{0}$ is the unit matrix, $\tau_{1}$ and $\tau_{3}$ are the Pauli matrices. Since the density operator is summed over the position of all particles, its Fourier transform can be obtained as
$\rho({\bf q})=({e}/2)\sum_{{\bf k}\sigma}h^{\dagger}_{{\bf k}\sigma}h_{{\bf k}+{\bf q}\sigma}=({e}/2)\sum_{{\bf k}}\Psi^{\dagger}_{{\bf k}}\tau_{3}\Psi_{{\bf k}+{\bf q}}$, then the paramagnetic density-current operator can be represented in the Nambu representation as,
\begin{eqnarray}\label{curnambu}
j_{\mu}^{(\rm p)}({\bf q})={1\over N}\sum\limits_{{\bf k}\sigma}\Psi^{\dagger}_{\bf k}\gamma_{\mu}({\bf k},{\bf k}+{\bf q})\Psi_{{\bf k}+{\bf q}}.
\end{eqnarray}
with the bare current vertex,
\begin{eqnarray}
\gamma_{\mu}({\bf k}+{\bf q},{\bf k})=\left\{ \begin{array}{ll} -{2e\over\hbar}\, e^{{1\over 2}iq_{\mu}}\{\sin(k_{\mu}+{1\over 2}q_{\mu})[\chi_{1}t-2\chi_{2}t'
\sum\limits_{\nu\neq\mu}\cos({1\over 2}q_{\nu})\cos(k_{\nu}+{1\over 2}q_{\nu})]\\
-i(2\chi_{2}t')\cos(k_{\mu}+{1\over 2}q_{\mu})\sum\limits_{\nu\neq\mu}\sin q_{\nu}\sin(k_{\nu}+{1\over 2}q_{\nu})\}\tau_{0}
  & {\rm for}\ \mu\neq 0,~~\\
e\tau_3  & {\rm{for}}\ \mu=0.\\
\end{array}\right.
\label{barevertex}
\end{eqnarray}
It is necessary to be aware that we \cite{Feng10,Krzyzosiak10} are calculating the polarization bubble with the paramagnetic current operator (\ref{curnambu}), i.e., the bare current vertex (\ref{barevertex}), but the full charge-carrier Green's function (\ref{NPBCSHG}). Consequently, as in this scenario we do not take into account longitudinal excitations properly \cite{Schrieffer64,Misawa94}, the obtained results are valid only in the gauge, where the vector potential is purely transverse, e.g. in the Coulomb gauge. In this case, the correlation function (\ref{corP}) can be obtained in the Nambu representation as,
\begin{eqnarray}
P_{\mu\nu}({\bf q},i\omega_{n})&=&{1\over N}\sum\limits_{\bf k}\gamma_{\mu}({\bf k}+{\bf q},{\bf k})\gamma^{*}_{\nu}({\bf k}+{\bf q},{\bf k}){1\over\beta}
\sum\limits_{i\nu_{m}}{\rm Tr}[\tilde{g}({\bf k}+{\bf q},i\omega_{n}+i\nu_{m})\tilde{g}({\bf k},i\nu_{m})].~~~~~~~~\label{barepolmats}
\end{eqnarray}
Substituting the charge-carrier Green's function (\ref{NPBCSHG}) into Eq. (\ref{barepolmats}), the paramagnetic part of the response kernel in the static limit ($\omega\sim 0$) is obtained as,
\begin{eqnarray}
K_{\mu\nu}^{(\rm p)}({\bf q},0)&=&{1\over N}\sum\limits_{\bf k}\gamma_{\mu}({\bf k}+{\bf q},{\bf k})\gamma^{*}_{\nu}({\bf k}+{\bf q},{\bf k})[{\bar L}_{1}({\bf k},{\bf q})
+{\bar L}_{2}({\bf k},{\bf q})]=K_{\mu\mu}^{(\rm p)}({\bf q},0)\delta_{\mu\nu},~~~~~~~~~\label{parakernel}
\end{eqnarray}
where the functions ${\bar L}_{1}({\bf k},{\bf q})$ and ${\bar L}_{2}({\bf k},{\bf q})$ are given by,
\begin{subequations}\label{lfunctions}
\begin{eqnarray}
{\bar L}_{1}({\bf k},{\bf q})&=&Z^{2}_{\rm hF}\left (1+{\bar{\xi}_{{\bf k}+{\bf q}}\bar{\xi}_{\bf k}+\bar{\Delta}_{\rm hZ}({\bf k}+{\bf q})\bar{\Delta}_{\rm hZ}({\bf k})
\over E_{{\rm h}{\bf k}}E_{{\rm h}{{\bf k}+{\bf q}}}}\right){n_{\rm F}(E_{{\rm h}{\bf k}})-n_{\rm F}(E_{{\rm h}{{\bf k}+{\bf q}}})\over E_{{\rm h}{\bf k}}
-E_{{\rm h}{{\bf k}+{\bf q}}}},~~~~~\\
{\bar L}_{2}({\bf k},{\bf q})&=&Z^{2}_{\rm hF}\left (1-{\bar{\xi}_{{\bf k}+{\bf q}}\bar{\xi}_{\bf k}+\bar{\Delta}_{\rm hZ}({\bf k}+{\bf q})\bar{\Delta}_{\rm hZ}({\bf k})
\over E_{{\rm h}{\bf k}}E_{{\rm h}{{\bf k}+{\bf q}}}}\right){n_{\rm F}(E_{{\rm h}{\bf k}})+n_{\rm F}(E_{{\rm h}{{\bf k}+{\bf q}}})-1\over E_{{\rm h}{\bf k}}
+E_{{\rm h}{{\bf k}+{\bf q}}}},~~~~~~~~~~~~
\end{eqnarray}
\end{subequations}
respectively. Now the kernel of the response function is obtained from Eqs. (\ref{diakernel}) and (\ref{parakernel}) as,
\begin{eqnarray}\label{kernel1}
K_{\mu\nu}({\bf q},0)=\left [{1\over\lambda^{2}_{L}}+K_{\mu\mu}^{(\rm p)}({\bf q},0)\right ]\delta_{\mu\nu}.
\end{eqnarray}

\subsection{Doping dependence of Meissner effect in long wavelength limit}

\begin{figure}[h!]
\centering
\includegraphics[scale=0.35]{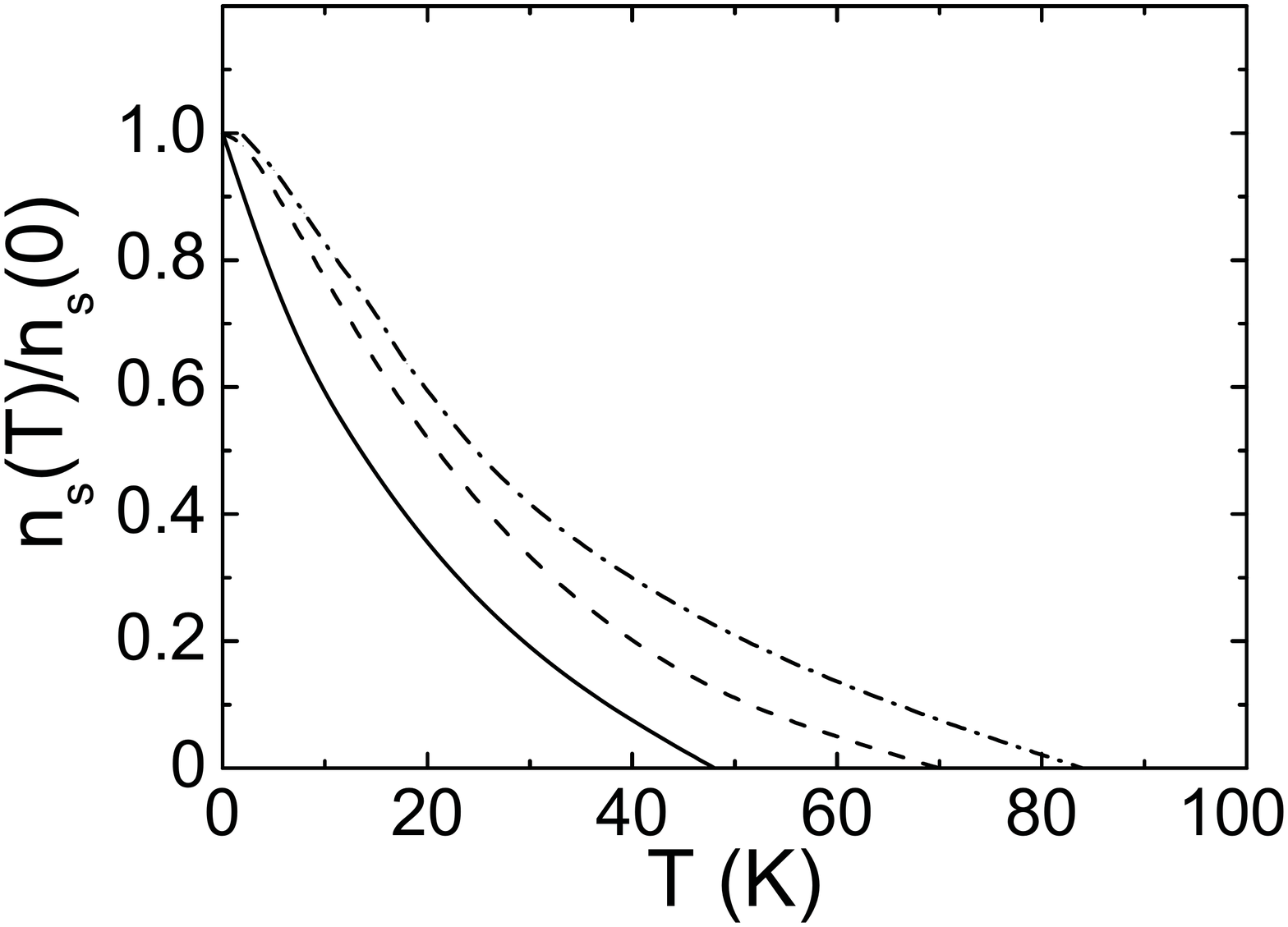}
\caption{The effective superfluid density as a function of temperature at $\delta=0.09$ (solid line), $\delta=0.12$ (dashed line), and $\delta=0.15$ (dash-dotted line) for $t/J=2.5$, $t'/t=0.3$, and $J=1000$K. [From Ref. \cite{Feng10}.] \label{effective-superfluid-density-doping}}
\end{figure}

In the long wavelength limit, i.e., $|{\bf q}|\to 0$, the function ${\bar L}_{2}({\bf k},{\bf q}\to 0)$ vanishes, and then the paramagnetic part of the response kernel can be obtained explicitly as \cite{Feng10},
\begin{eqnarray}
K_{\rm yy}^{(\rm p)}({\bf q}\to 0,0) &=& 2Z^{2}_{\rm hF}{4e^{2}\over\hbar^{2}}{1\over N}\sum\limits_{{\bf k}}\sin^{2}k_{y}[\chi_{1}t-2\chi_{2}t'\cos k_{x}]^{2}
\lim\limits_{{\bf q}\to 0}{n_{\rm F}(E_{{\rm h}{\bf k}})-n_{\rm F}(E_{{\rm h}{{\bf k}+{\bf q}}})\over E_{{\rm h}{\bf k}}-E_{{\rm h}{{\bf k}+{\bf q}}}}.~~~~~~ \label{kernel5}
\end{eqnarray}
At zero temperature $T=0$, it is found that $K_{\rm yy}^{(\rm p)}({\bf q}\to 0,0)|_{T=0}=0$, and then the long wavelength electromagnetic response is determined by the diamagnetic part of the kernel only. On the other hand, at $T=T_{\rm c}$, the charge-carrier gap parameter $\bar{\Delta}_{\rm h}|_{T=T_{\rm c}}=0$, and then the paramagnetic part of the response kernel is evaluated as \cite{Feng10},
\begin{eqnarray}
K_{\rm yy}^{(\rm p)}({\bf q}\to 0,0)|_{T=T_{\rm c}}&=&2Z^{2}_{\rm hF}{4e^{2}\over\hbar^{2}}{1\over N}\sum\limits_{{\bf k}}\sin^{2}k_{y}[\chi_{1}t-2\chi_{2}t'\cos k_{x}]^{2}\lim \limits_{{\bf q}\to 0}{n_{\rm F}(\bar{\xi}_{\bf k})-n_{\rm F}(\bar{\xi}_{{\bf k}+{\bf q}})\over \bar{\xi}_{\bf k}-\bar{\xi}_{{\bf k}+{\bf q}}}\nonumber\\
&=& -{1\over \lambda^{2}_{L}},~~~~~
\end{eqnarray}
which exactly cancels the diamagnetic part of the response kernel (\ref{diakernel}), and then the Meissner effect in cuprate superconductors is obtained for all the temperatures $T\leq T_{\rm c}$. To show this point clearly, the effective superfluid density $n_{\rm s}(T)$ at temperature $T$ is defined in terms of the paramagnetic part of the response kernel as,
\begin{eqnarray}
K_{\mu\nu}^{(\rm p)}({\bf q}\to 0,0)=-{1\over\lambda^{2}_{L}}\left [1-{n_{\rm s}(T)\over n_{\rm s}(0)}\right ]
\delta_{\mu\nu},
\end{eqnarray}
and then the kernel of the response function (\ref{kernel1}) can be rewritten as,
\begin{eqnarray}
K_{\mu\nu}({\bf q}\to 0,0)={1\over\lambda^{2}_{L}}{n_{\rm s}(T)\over n_{\rm s}(0)}\delta_{\mu\nu}, \label{kernel2}
\end{eqnarray}
where the ratio $n_{\rm s}(T)/n_{\rm s}(0)$ of the effective superfluid densities at temperature $T$ and zero temperature is given by,
\begin{eqnarray}
{n_{\rm s}(T)\over n_{\rm s}(0)}=1-2\lambda^{2}_{L}Z^{2}_{\rm hF}{4e^{2}\over\hbar^{2}}{1\over N}\sum\limits_{\bf k}\sin^{2}k_{y}[\chi_{1}t-2\chi_{2}t'\cos k_{x}]^{2}
{\beta e^{\beta E_{{\rm h}{\bf k}}}\over (e^{\beta E_{{\rm h}{\bf k}}}+1)^{2}}. \label{ratio2}
\end{eqnarray}
In Fig. \ref{effective-superfluid-density-doping}, we show the effective superfluid density $n_{\rm s}(T)/n_{\rm s}(0)$ as a function of temperature for $t/J=2.5$, $t'/t=0.3$, and $J=1000$K at $\delta=0.09$ (solid line), $\delta=0.12$ (dashed line), and $\delta=0.15$ (dash-dotted line), where the effective superfluid density diminishes with increasing temperatures, and disappears at $T_{\rm c}$, then all the charge carriers are in the normal fluid for the temperatures $T\geq T_{\rm c}$.

The main results obtained within the kinetic-energy driven SC mechanism are summarized as \cite{Feng10}: (a) the doping dependence of the Meissner effect in cuprate superconductors is obtained for all the temperatures $T\leq T_{\rm c}$ throughout the SC dome; (b) the electromagnetic response kernel goes to the London form in the long wavelength limit [see, e.g., Eq. (\ref{kernel2})]; (c) although the electromagnetic response kernel is not manifestly gauge invariant within the bare current vertex (\ref{barevertex}), however, we can keep the gauge invariance within the dressed current vertex \cite{Krzyzosiak10}, which will be proven clearly in subsection \ref{invapp}.

\subsection{Quantitative characteristics}

\begin{figure}[h!]
\centering
\includegraphics[scale=0.65]{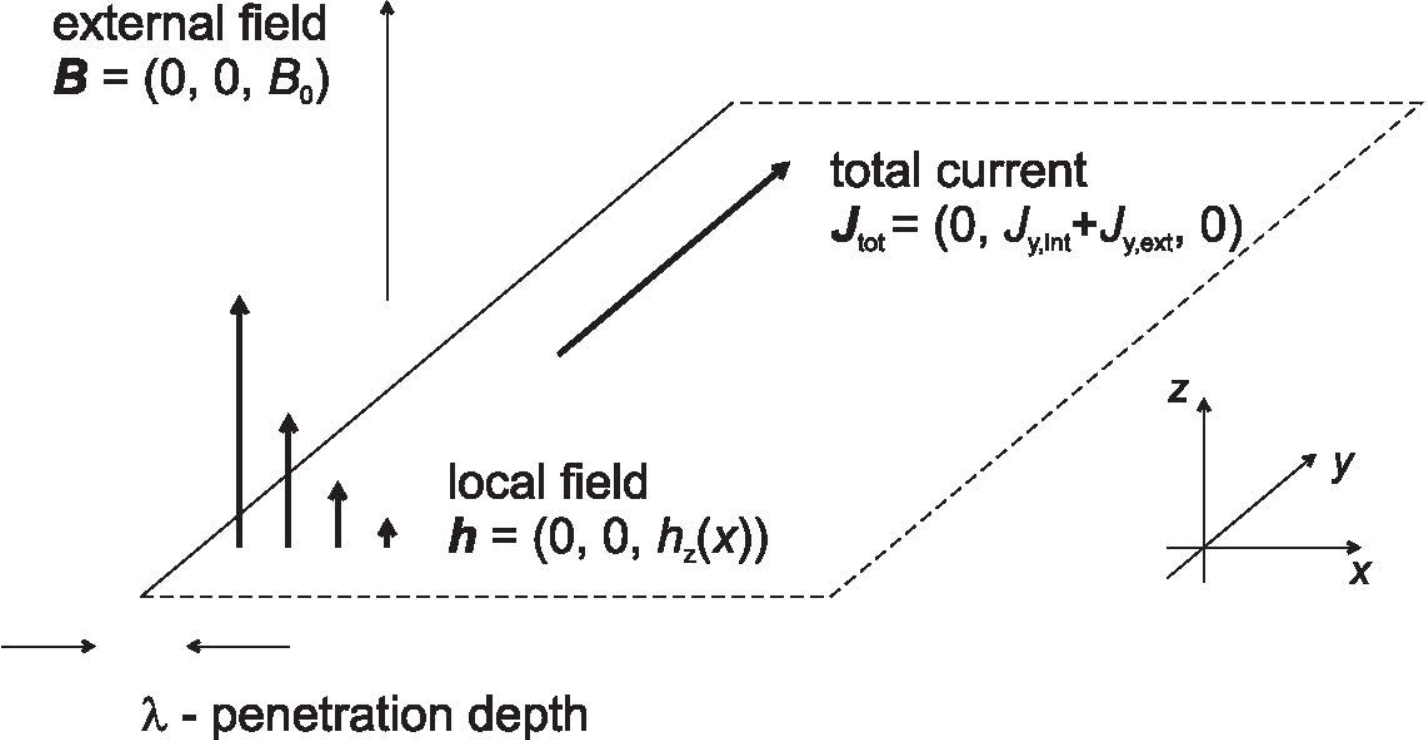}
\caption{The geometry of the specular reflection model. The current ${\bf J}_{\rm ext}$ simulates external magnetic field at the edge of the sample ($x=0$), whereas the induced supercurrent ${\bf J}_{\rm int}$ is the (linear) reaction of the system. [From Ref. \cite{Krzyzosiak10}.] \label{specular-reflection-model}}
\end{figure}

The way the system reacts to an external electromagnetic stimulus is entirely described by the linear response kernel (\ref{linres}), which is calculated within the kinetic-energy driven SC mechanism. Once the response kernel $K_{\mu\nu}$ is known, the effect of an electromagnetic field can be quantitatively characterized by experimentally measurable quantities such as the local magnetic field profile and the magnetic field penetration depth. Technically, it is needed to combine one of the Maxwell equations with the relation (\ref{linres}) describing the response of the system and solve them together for the vector potential. This is the step in which a particular gauge of the vector potential
--- usually implied by the geometry of the system --- is set. However, the result we have obtained the response kernel (\ref{kernel1}) can not be used for a direct comparison with the corresponding experimental data of cuprate superconductors because the kernel function derived within the linear response theory describes the response of an {\it infinite} system. In order to take into account the confined geometry of cuprate superconductors, it is necessary to introduce a surface being the boundary between the environment and the sample. This can be done within the standard specular reflection model \cite{Abrikosov88,Tinkham96} with a two-dimensional geometry of the SC plane, in the configuration with external magnetic field perpendicular to the {\rm ab} plane, as illustrated in Fig. \ref{specular-reflection-model}. In this subsection we study magnetic field penetration effects within the {\rm ab} plane only, so our goal is to find and discuss the in-plane magnetic field penetration depth.

In order to simulate an external magnetic field at the surface of a two-dimensional sample, we introduce an external current sheet
$J_{{\rm y},{\rm ext}}(x)=-2{\rm B}\delta(x)/\mu_{0}$ at the edge $x=0$, where $\mu_{0}$ is the magnetic permeability and ${\rm B}$ is the amplitude of the external magnetic field ${\bf B}$ at the surface ($x=0$). From the Maxwell equation for the curl of the local magnetic field
${\rm rot}\,{\bf h}=\mu_{0}({\bf J}_{\rm int}+{\bf J}_{\rm ext})=\mu_{0}{\bf J}_{\rm int}+[0,-2{\rm B}\delta(x),0]$ and the fact, that the induced supercurrent ${\bf J}_{\rm int}$ flows along the ${\rm y}$ axis, we can state that the local magnetic field is of the form ${\bf h}({\bf r})=[0,0,h_{\rm z}(x)]$. In order to discuss the magnetic field penetration effect, the spatial dependence of the local magnetic field has to be found. Let us begin with the identity
${\rm rot}\,{\rm rot}\,{\bf A}={\rm grad}\,{\rm div}\,{\bf A}-\nabla^2 {\bf A}$ and choose the vector potential as ${\bf A}({\bf r})=[0,A_{\rm y}(x),0]$ setting the Coulomb gauge. In this case, $q^{2}_{\rm x}A_{\rm y}({\bf q})=\mu_{0}[J_{{\rm y},{\rm int}}({\bf q})+J_{{\rm y},{\rm ext}}({\bf q})]$, because the vector potential has only non-zero ${\rm y}$ component. Finally, including the form of the external current, the linear relation (\ref{linres}) between the induced supercurrent and the vector potential
$J_{{\rm y},{\rm int}}({\bf q})=-K_{\rm yy}({\bf q})A_{\rm y}({\bf q})$, and solving for the vector potential we obtain,
\begin{equation}\label{aspec}
A_{\rm y}({\bf q})=-2{\rm B}{\delta(q_{\rm y})\delta(q_{\rm z})\over\mu_{0}K_{\rm yy}({\bf q})+q^{2}_{\rm x}}.
\end{equation}
Since the vector potential has only the ${\rm y}$ component, the only non-zero component of the local magnetic field ${\bf h}={\rm rot}\,{\bf A}$ is that along the ${\bf z}$ axis and $h_{\rm z}({\bf q})=iq_{\rm x}A_{\rm y}({\bf q})$. Substituting the derived form of the vector potential (\ref{aspec}), and taking the inverse Fourier transform, the local magnetic field profile can be obtained explicitly as,
\begin{equation}\label{profile}
h_{\rm z}(x)={{\rm B}\over\pi}\int\limits_{-\infty}^\infty {\rm d}q_{x}\,{q_{x}\sin(q_{x}x)\over\mu_{0}K_{\rm yy}(q_{x},0,0)+q_{x}^{2}}.
\end{equation}
This local magnetic field profiles can be measured experimentally, e.g. using the $\mu$SR measurement technique \cite{Khasanov04,Suter04}, providing an important tool to investigate the details of magnetic field screening inside the sample. For convenience in the following discussions at this section, we introduce a characteristic length scale $a_{0}=\sqrt{\hbar^{2}a/\mu_{0}e^{2}J}$. Using the lattice parameter $a\approx 0.383$nm for YBa$_2$Cu$_3$O$_{7-y}$, this characteristic length is obtain as $a_{0}\approx 97.8$nm. In this case, the local magnetic field profile (\ref{profile}) as a function of the distance from the surface has been studied at different doping levels \cite{Feng10,Krzyzosiak10}, and the theoretical results perfectly follow an exponential field decay as expected for the local electrodynamic response. In particular, this exponential character of the local magnetic field profile has been observed experimentally on different families of cuprate superconductors \cite{Khasanov04,Suter04}, in support of a local (London-type) nature of the electrodynamics.

\subsection{Doping and temperature dependence of magnetic field penetration depth}

\begin{figure}[h!]
\centering
\includegraphics[scale=0.4]{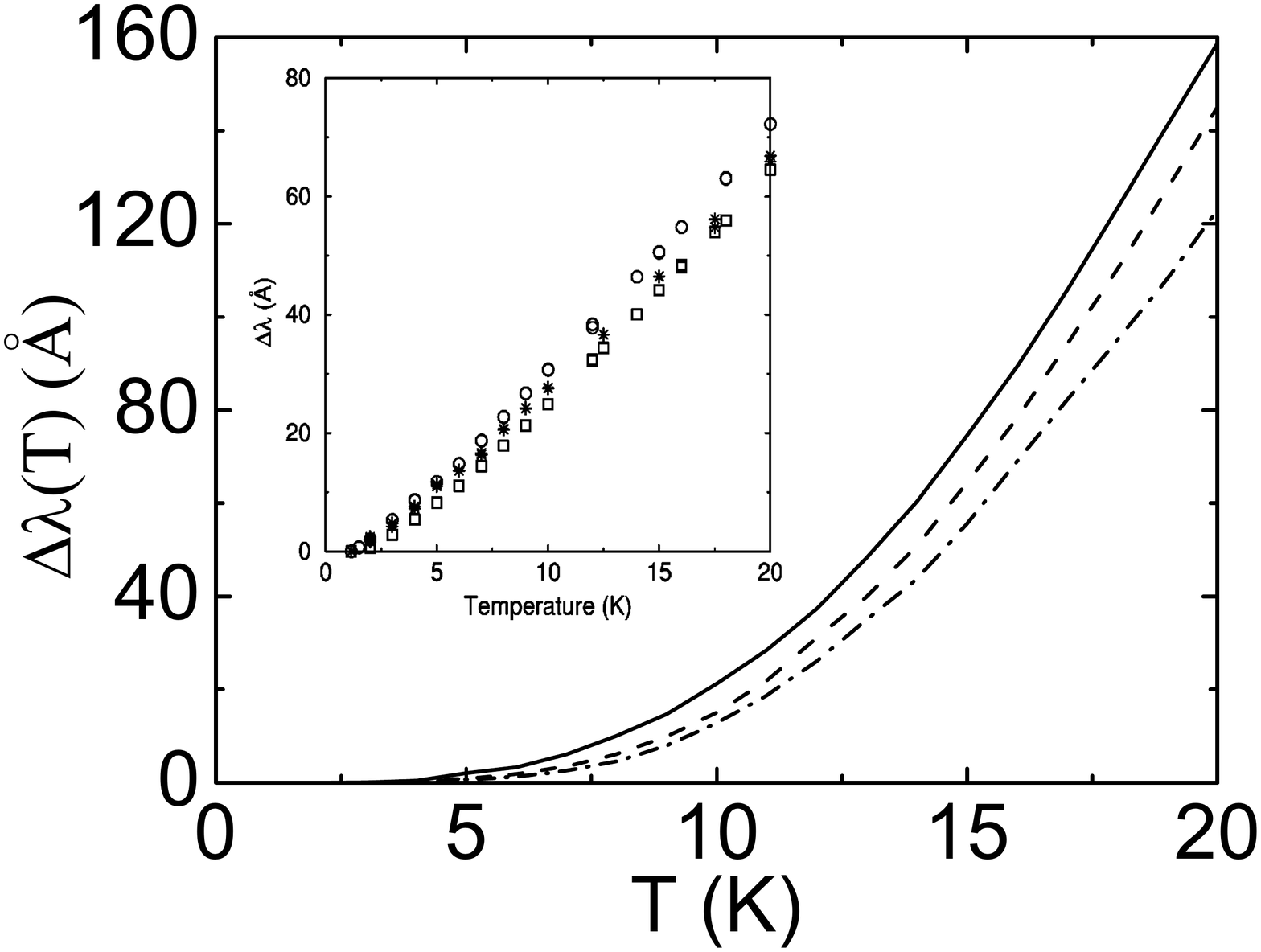}
\caption{The temperature dependence of the magnetic field penetration depth $\Delta\lambda(T)$ at $\delta=0.14$ (solid line), $\delta=0.15$ (dashed line), and $\delta=0.18$ (dash-dotted line) for $t/J=2.5$, $t'/t=0.3$, and $J=1000$K. Inset: the corresponding experimental data for YBa$_2$Cu$_3$O$_{7-y}$ taken from Ref. \cite{Kamal98}. [From Ref. \cite{Feng10}.] \label{penetration-depth-temp}}
\end{figure}

The local magnetic field profile $h_{\rm z}(x)$ in the Meissner state obtained in Eq. (\ref{profile}) allows us to determine the magnetic field penetration depth $\lambda(T)$ in a straightforward way. According to the definition $\lambda(T)={\rm B}^{-1}\int_{0}^\infty h_{\rm z}(x)\,{\rm d}x$, the magnetic field penetration depth can be evaluated as,
\begin{eqnarray}\label{lambda}
\lambda(T)={1\over {\rm B}}\int\limits_{0}^{\infty}h_{\rm z}(x)\,{\rm d}x={2\over\pi}\int\limits_{0}^{\infty}{{\rm d}q_{x}\over\mu_{0}K_{\rm yy}(q_{x},0,0)+q_{x}^{2}}.
\end{eqnarray}
At zero temperature, the calculated magnetic field penetration depths \cite{Feng10,Krzyzosiak10} are $\lambda(0)\approx 239.17$nm, $\lambda(0)\approx 234.76$nm, and $\lambda(0)\approx 224.44$nm at $\delta=0.14$, $\delta=0.15$, and $\delta=0.18$, respectively, which are qualitatively consistent with the values of the magnetic field penetration depth $\lambda\approx 156$nm $\sim 400$nm observed for different families of cuprate superconductors at different doping levels \cite{Bernhard01,Khasanov04,Broun07,Niedermayer93,Uemura93}. On the other hand, at $T=T_{\rm c}$, the kernel of the response function $K_{\mu\nu}({\rm q}\to 0,0)|_{T=T_{\rm c}}=0$, and then the magnetic field penetration depth from Eq. (\ref{lambda}) can be found as $\lambda(T_{\rm c})=\infty$, which reflects that in the normal-state, the external magnetic field can penetrate through the main body of the system, therefore there is no the Meissner effect in the normal-state. In Fig. \ref{penetration-depth-temp}, we \cite{Feng10} show the magnetic field penetration depth $\Delta\lambda(T)=\lambda(T)-\lambda(0)$ as a function of temperature for $t/J=2.5$, $t'/t=0.3$, and $J=1000$K at $\delta=0.14$ (solid line), $\delta=0.15$ (dashed line), and $\delta=0.18$ (dash-dotted line) in comparison with the corresponding experimental results \cite{Kamal98} of YBa$_2$Cu$_3$O$_{7-y}$ (inset). In low temperatures, the magnetic field penetration depth $\Delta\lambda(T)$ exhibits a linear temperature dependence, however, it crosses over to a nonlinear behavior in the extremely low temperatures, in good agreement with experimental observation in nominally clean crystals of cuprate superconductors \cite{Hardy93,Khasanov04,Suter04,Kamal98}. However, it should be emphasized that the result in Fig. \ref{penetration-depth-temp} for cuprate superconductors is different from that in conventional superconductors, where the characteristic feature is the existence of the isotropic SC gap $\bar{\Delta}_{\rm s}$, and then $\Delta\lambda(T)$ exhibits an exponential behavior as $\Delta\lambda(T)\propto {\rm exp}(-\bar{\Delta}_{\rm s}/T)$.

\subsection{Doping and temperature dependence of superfluid density}\label{DTDSD}

\begin{figure}[h!]
\centering
\includegraphics[scale=0.4]{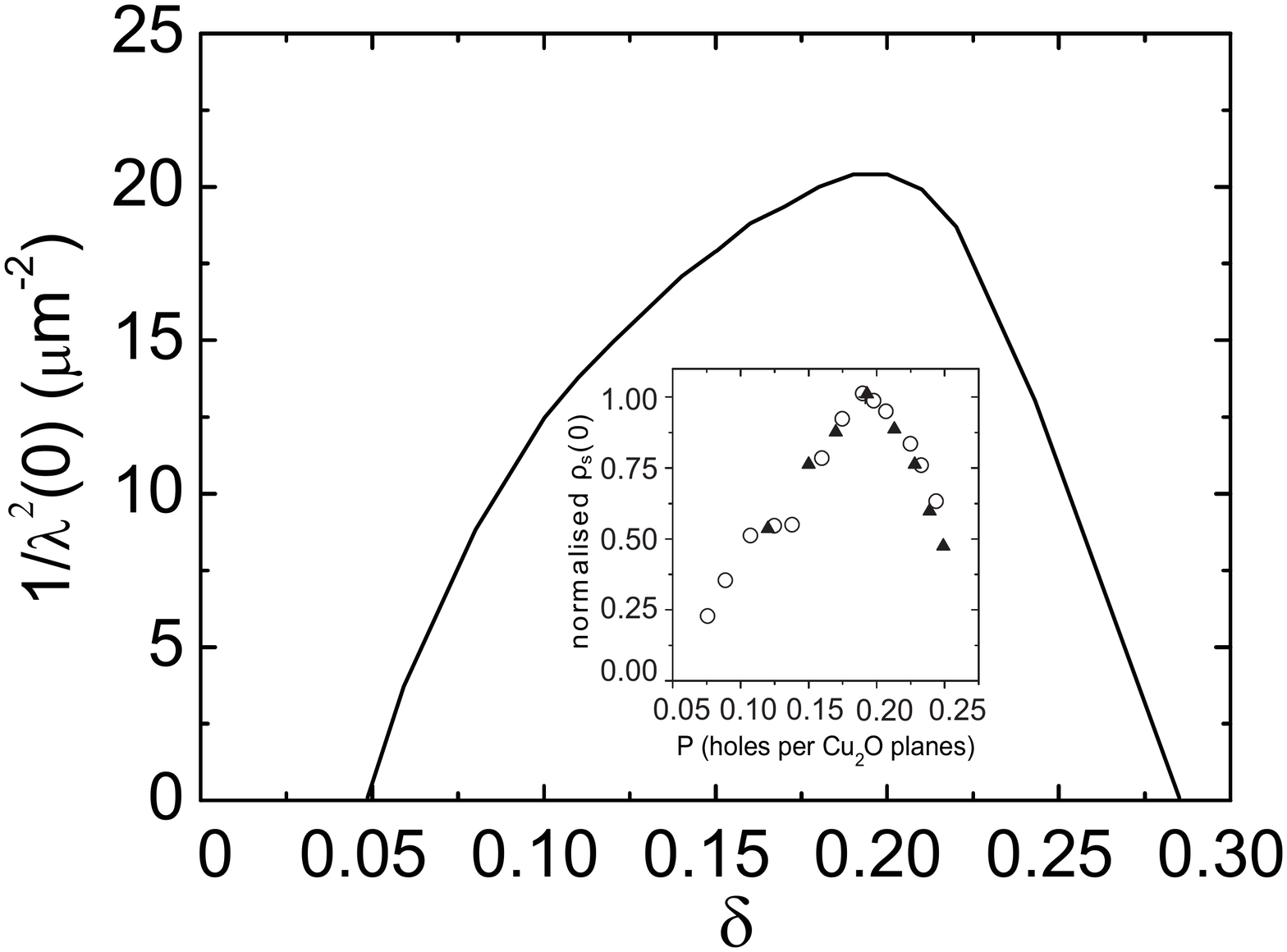}
\caption{The doping dependence of the superfluid density in $T=0$ for $t/J=2.5$, $t'/t=0.3$, and $J=1000$K. Inset: the corresponding experimental results for Y$_{0.8}$Ca$_{0.2}$Ba$_{2}$(Cu$_{1-z}$Zn$_{z}$)$_{3}$O$_{7-\delta}$ (open circles) and Tl$_{1-y}$Pb$_{y}$Sr$_{2}$Ca$_{1-x}$Y$_x$Cu$_2$O$_{7}$ (solid triangles) taken from Ref. \cite{Bernhard01}. [From Ref. \cite{Feng10}.] \label{superfluid-density-doping}}
\end{figure}

The superfluid density $\rho_{\rm s}(T)$ can be obtained directly from the magnetic field penetration depth $\lambda(T)$ as,
\begin{eqnarray}\label{rhodensity}
\rho_{\rm s}(T)\equiv {1\over\lambda^{2}(T)}.
\end{eqnarray}
The zero temperature superfluid density $\rho_{\rm s}(0)$ as a function of doping for $t/J=2.5$, $t'/t=0.3$, and $J=1000$K is shown in Fig. \ref{superfluid-density-doping} in comparison with the corresponding experimental data \cite{Bernhard01} for Y$_{0.8}$Ca$_{0.2}$Ba$_{2}$(Cu$_{1-z}$Zn$_{z}$)$_{3}$O$_{7-\delta}$ and
Tl$_{1-y}$Pb$_{y}$Sr$_{2}$Ca$_{1-x}$Y$_x$Cu$_2$O$_{7}$ (inset). This calculated result clearly shows that the superfluid density $\rho_{\rm s}(0)$ increases with increasing doping in the lower doped regime, and reaches a highest value (a peak) around the {\it critical doping} $\delta\approx 0.195$, then decreases in the higher doped regime. In particular, this anticipated value of the {\it critical doping} $\delta_{\rm critical}\approx 0.195$ is very close to the {\it critical doping} $\delta_{\rm critical}\approx 0.19$ observed experimentally for different families of cuprate superconductors \cite{Bernhard01,Broun07,Lemberger11,Niedermayer93}. The early experimental data observed from cuprate superconductors show that the superfluid density $\rho_{\rm s}(0)$ in the underdoped regime vanishes more or less linearly with decrease of the charge-carrier doping concentration $\delta$ \cite{Uemura89}. Later, a clear deviation from this linear relation between the superfluid density $\rho_{\rm s}(0)$ and charge-carrier doping concentration has been observed in the underdoped regime \cite{Bernhard01,Lee96}. However, the recent experimental measurement \cite{Broun07} on the cuprate superconductor YBa$_2$Cu$_3$O$_{7-y}$ indicate that the superfluid density $\rho_{\rm s}(0)$ is, in actual fact, linearly proportional to the charge-carrier doping concentration in the doped range $\delta\approx 0.054\sim 0.061$. The calculated result in Fig. \ref{superfluid-density-doping} in this doped range also is well consistent with the experimental observation \cite{Broun07}.

However, the {\it critical doping} $\delta_{\rm critical}\approx 0.195$ for the highest $\rho_{\rm s}$ is different from that in the doping dependence of $T_{\rm c}$ as shown in Fig. \ref{Tc-doping} and the doping dependence of the charge-carrier pair gap parameter $\bar{\Delta}_{\rm h}$ as shown in Fig. \ref{pair-gap-parameter-doping}, where the maximal $T_{\rm c}$ and $\bar{\Delta}_{\rm h}$ appear around the {\it optimal doping} $\delta_{\rm optimal}\approx 0.15$. This difference is a long-standing puzzle in cuprate superconductors, however, its interpretation is quite simple within the framework of the kinetic-energy driven SC mechanism \cite{Huang13,Feng10}. This follows from a fact that
the superfluid density $\rho_{\rm s}$ from Eqs. (\ref{rhodensity}) and (\ref{lambda}) is closely related to the kernel of the response function (\ref{kernel1}), and therefore the charge-carrier pair order parameter $\Delta_{\rm h}$, the coupling strength $V_{\rm eff}$, and all the other order parameters are relevant. In other words, the variation of the superfluid density $\rho_{\rm s}$ with doping and temperature is coupled to the doping and temperature dependence of $\Delta_{\rm h}$, $V_{\rm eff}$, and all the other order parameters \cite{Huang13,Feng10}. In particular, the doping-derivative of $\rho_{\rm s}$ at the {\it critical doping} $\delta_{\rm critical}\approx 0.195$ is obtained as
$({\rm d}\rho_{\rm s}/{\rm d}\delta)|_{\delta=\delta_{\rm critical}}=0$. Since $\rho_{\rm s}\equiv\lambda^{-2}$,
$({\rm d}\rho_{\rm s}/{\rm d}\delta)|_{\delta=\delta_{\rm critical}}=0$ is equivalent to $({\rm d}\lambda/{\rm d}\delta)|_{\delta=\delta_{\rm critical}}=0$. In this case,
$({\rm d}\lambda/{\rm d}\delta)|_{\delta=\delta_{\rm critical}}=0$ can be expressed in terms of Eq. (\ref{lambda}) as,
\begin{eqnarray}\label{dflambda}
\left [{{\rm d}\lambda\over {\rm d}\delta}\right ]_{\delta=\delta_{\rm critical}}=- {2\mu_{0}\over\pi}\int\limits_{0}^{\infty}\rm{d}q_{x}\left [{1\over [\mu_{0} K_{yy}(q_{x},0,0) +q_{x}^{2}]^{2}}{{\rm d}K_{yy}(q_{x},0,0)\over {\rm d}\delta}\right]_{\delta=\delta_{\rm critical}}=0,~~~~~~~
\end{eqnarray}
and then it is straightforward to obtain from Eq. (\ref{kernel1}) that when $({\rm d}\rho_{\rm s}/{\rm d}\delta)|_{\delta=\delta_{\rm critical}}=0$,
$({\rm d}\Delta_{\rm h}/{\rm d}\delta)|_{\delta=\delta_{\rm critical}}=0$, which shows that the doping effects from the coupling strength $V_{\rm eff}$ and all the other order parameters upon $\rho_{\rm s}$ are almost canceled each other, and then the behavior of the doping dependence of the superfluid density $\rho_{\rm s}$ is mainly dominated by the doping dependence of the charge-carrier pair order parameter $\Delta_{\rm h}$. However, the charge-carrier pair order parameter $\Delta_{\rm h}$ measures the strength of the binding of two charge carriers into a charge-carrier pair, and it has been shown in subsection \ref{KEDSQC} that $\Delta_{\rm h}$ has a domelike shape of the doping dependence with the maximal value appearing around the {\it critical doping} $\delta_{\rm critical}\approx 0.195$ (see Fig. \ref{V-pair-order-parameter-doping}b). In particular, the charge-carrier pair order parameter $\Delta_{\rm h}$ and the charge-carrier pair macroscopic wave functions in cuprate superconductors are the same as we have mentioned in section \ref{KEDM}, i.e., the charge-carrier pair order parameter is a {\it magnified} version of the charge-carrier pair macroscopic wave functions. On the other hand, the superfluid density $\rho_{\rm s}$ is a measurement of the phase stiffness \cite{Bonn96}, and is proportional to the squared amplitude of the charge-carrier pair macroscopic wave functions. Both $\rho_{\rm s}$ and $\Delta_{\rm h}$ thus describe the different aspects of the same charge-carrier pair macroscopic wave functions. In this case, the domelike shape of the doping dependence of $\rho_{\rm s}$ with the highest value appearing around the {\it critical doping} is a natural consequence of the domelike shape of the doping dependence of $\Delta_{\rm h}$ with the maximal value appearing around the same {\it critical doping}. In comparison with the results obtained in subsection \ref{DDTC}, it is therefore shown that except for the quasiparticle coherence, $\rho_{\rm s}$ is determined by the charge-carrier pair order parameter $\Delta_{\rm h}$, while $T_{\rm c}$ is set by the charge-carrier pair gap parameter ${\bar{\Delta}_{\rm h}}$, this is why there is a difference between the {\it optimal doping} for the maximal $T_{\rm c}$ and the {\it critical doping} for the highest $\rho_{\rm s}$ in cuprate superconductors.

\begin{figure}[h!]
\centering
\includegraphics[scale=0.4]{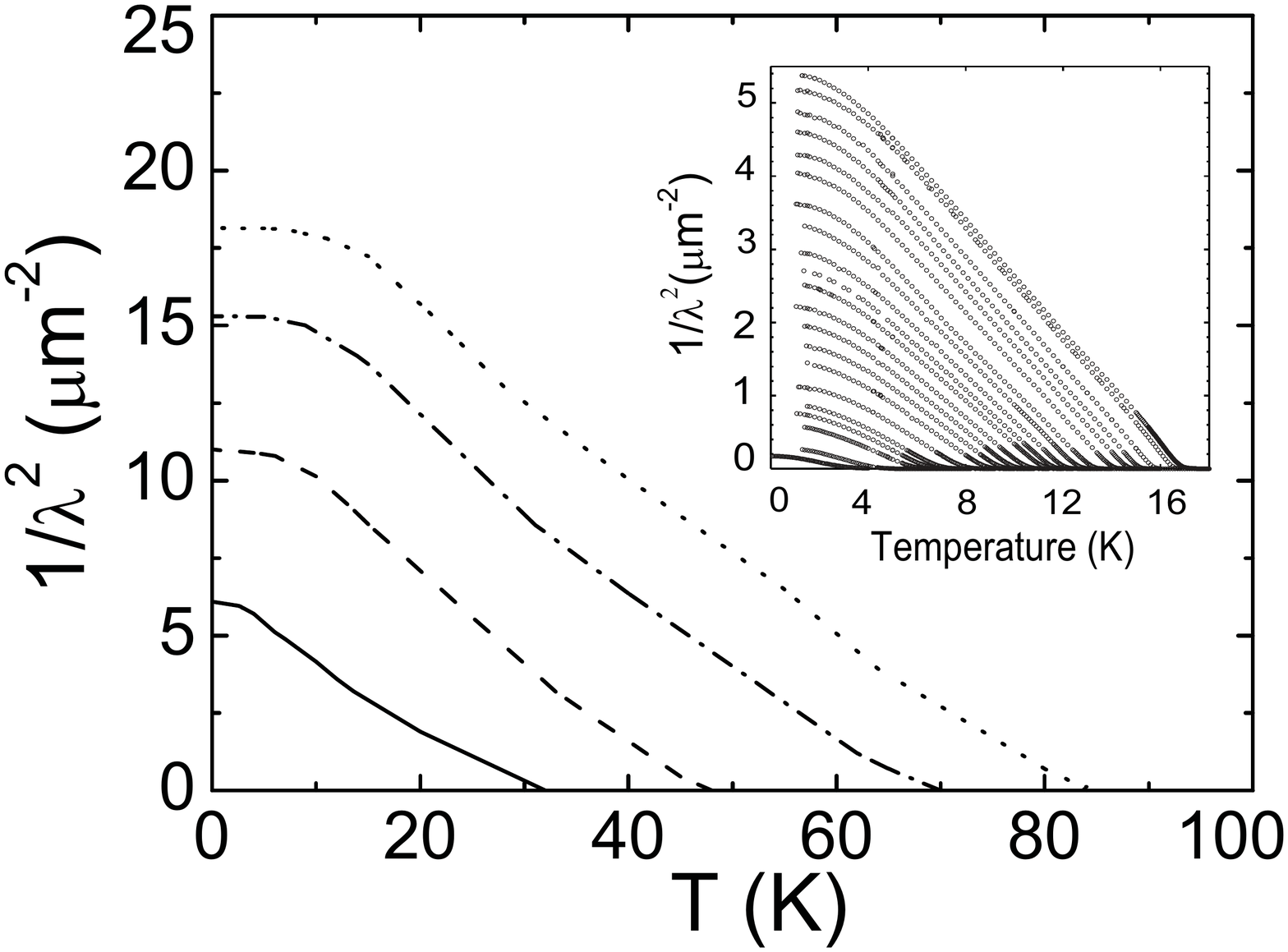}
\caption{The temperature dependence of the superfluid density at $\delta=0.06$ (solid line), $\delta=0.09$ (dashed line), $\delta=0.12$ (dash-dotted line), and $\delta=0.15$ (dotted line) for $t/J=2.5$, $t'/t=0.3$, and $J=1000$K. Inset: the corresponding experimental result for YBa$_2$Cu$_3$O$_{7-y}$ taken from Ref. \cite{Broun07}.[From Ref. \cite{Feng10}.] \label{superfluid-density-temp}}
\end{figure}

The doping dependence of the superfluid density shown in Fig. \ref{superfluid-density-doping} is also strongly temperature dependent. In particular, when the temperature $T=T_{\rm c}$, the kernel of the response function $K_{\mu\nu}({\bf q}\to 0,0)|_{T=T_{\rm c}}=0$, and then the magnetic field penetration depth $\lambda(T_{\rm c})=\infty$ as mentioned above, which leads to the superfluid density $\rho_{\rm s}(T_{c})=0$, which is consistent with the result of the effective superfluid density obtained from Eq. (\ref{ratio2}). In Fig. \ref{superfluid-density-temp}, we \cite{Feng10} show the superfluid density $\rho_{\rm s}(T)$ as a function of temperature for $t/J=2.5$, $t'/t=0.3$, and $J=1000$K at $\delta=0.06$ (solid line), $\delta=0.09$ (dashed line), $\delta=0.12$ (dash-dotted line), and $\delta=0.15$ (dotted line) in comparison with the corresponding experimental result \cite{Broun07} of YBa$_2$Cu$_3$O$_{7-y}$ (inset). $\rho_{\rm s}(T)$ decreases with increasing temperature, and vanishes at $T_{\rm c}$. In particular, the most striking feature of the results is the wide range of linear temperature dependence at low temperatures, extending from close to $T_{\rm c}$ to down to the temperatures $T\approx 4$K$\sim 8$K for different doping concentrations. However, in correspondence with the nonlinear temperature dependence of the magnetic field penetration depth at the extremely low temperatures shown in Fig. \ref{penetration-depth-temp}, the superfluid density $\rho_{\rm s}(T)$ crosses over to a nonlinear temperature behavior at the extremely low temperatures (below $T\approx 4$K$\sim 8$K for different doping concentrations). These calculated results are also qualitatively consistent with the corresponding experimental results \cite{Bernhard01,Broun07,Niedermayer93} of cuprate superconductors.

The explanation \cite{Feng10,Krzyzosiak10} for the nonlinearity in the temperature dependence of the penetration depth (then the superfluid density) at the extremely low temperatures is the same as the case based on the phenomenological d-wave BCS formalism \cite{Yip92,Kosztin97,Franz97,Li00,Sheehy04}, and can be found from the nonlocal effects induced by the gap nodes in a pure d-wave pairing state. An external magnetic field acts on the SC-state of cuprate superconductors as a perturbation. Within the linear response theory, one can find that the nonlocal relation between the supercurrent and the vector potential (\ref{linres}) in the coordinate space holds due to the finite size of charge-carrier pairs. In particular, in the kinetic-energy driven d-wave SC mechanism \cite{Feng03,Feng06,Feng06a}, the size of charge-carrier pairs in the clean limit is of the order of the coherence length $\zeta({\bf k})=\hbar v_{\rm F}/\pi\Delta_{\rm h}({\bf k})$, where $v_{\rm{F}}=\hbar^{-1}\partial\xi_{\bf k}/\partial {\bf k}|_{k_{\rm F}}$ is the charge-carrier velocity, which shows that the size of charge-carrier pairs is momentum dependent. In general, although the external magnetic field decays exponentially on the scale of the magnetic field penetration length $\lambda(T)$, any nonlocal contributions to measurable quantities are of the order of $\kappa^{-2}$, where the Ginzburg--Landau parameter $\kappa$
is the ratio of the magnetic field penetration depth $\lambda$ and the coherence length $\zeta$. However, for cuprate superconductors, because the pairing is d-wave as shown in Eq. (\ref{CCPGF}), the charge-carrier pair gap vanishes on the gap nodes, so that the quasiparticle excitations are gapless and therefore affect particularly the physical properties at the extremely low temperatures. This gapless quasiparticle excitation leads to a divergence of the coherence length $\zeta({\bf k})$ around the gap nodes, and then the behavior of the temperature dependence of the magnetic field penetration depth (then the superfluid density) depends sensitively on the quasiparticle scattering. At the extremely low temperatures, the quasiparticles selectively locate around the gap nodal region, and then the major contribution to measurable quantities comes from these quasiparticles. In this case, the Ginzburg--Landau ratio $\kappa({\bf k})$ around the gap nodal region is no longer large enough for the system to belong to the class of type-II superconductors, and the condition of the local limit is not satisfied \cite{Kosztin97}, which leads to the system in the extreme nonlocal limit, and therefore the nonlinear behavior in the temperature dependence of the magnetic field penetration depth (then superfluid density) is observed experimentally \cite{Bonn96,Khasanov04,Suter04}. On the other hand, with increasing temperatures, the quasiparticles around the gap nodal region become excited out of the condensate, and then the nonlocal effect fades away, where the momentum
dependent coherence length $\zeta({\bf k})$ can be replaced approximately with the isotropic one $\zeta_{0}=\hbar v_{\rm F}/\pi\Delta_{\rm h}$. In this case, the calculated Ginzburg--Landau parameters are $\kappa_{0}\approx\lambda(0)/\zeta_{0}\approx 166.29$, $\kappa_{0}\approx 175.55$, and $\kappa_{0}\approx 156.14$ for the doping concentrations $\delta=0.14$, $\delta=0.15$, and $\delta=0.18$, respectively, and then the condition for the local limit is satisfied. In particular, these calculated values of the Ginzburg--Landau parameter at different doping concentrations are very close to the range $\kappa_{0}\approx 150\sim 400$ estimated experimentally for different families of cuprate superconductors at different doping levels \cite{Bernhard01,Khasanov04,Broun07,Niedermayer93,Uemura93}. As a consequence, the study \cite{Feng10,Krzyzosiak10} based on the kinetic-energy driven SC mechanism shows that cuprate superconductors at moderately low temperatures turn out to be type-II superconductors, where nonlocal effects can be neglected, and then the electrodynamics is purely local and the magnetic field decays exponentially over a length of the order of a few hundreds nm.

\subsection{Gauge-invariant electromagnetic response}\label{invapp}

Although the electromagnetic response kernel is not manifestly gauge invariant within the bare current vertex (\ref{barevertex}), however, we can keep the theory gauge invariance within the dressed current vertex \cite{Krzyzosiak10}. It is well known that gauge invariance is a direct consequence of local charge conservation \cite{Fukuyama69,Schrieffer64}, which is mathematically expressed by the charge density-current continuity equation or its Green function analogue called the generalized Ward identity \cite{Fukuyama69,Schrieffer64,Misawa94,Arseev06}
\begin{equation}\label{GWI}
\sum\limits_{\mu=0}^{2}q_{\mu}\Gamma_{\mu}(k+q,k)=\tau_{3}\tilde{g}^{-1}(k)-\tilde{g}^{-1}(k+q)\tau_{3},
\end{equation}
where the charge $e$ and reduced Planck constant $\hbar$ have been set to the unity, $\Gamma_{\mu}$ is a dressed version of the density-current vertex function, and for convenience in the discussions at this subsection, the three-vector notation $q=({\bf q},q_{0}=i\omega_{n})$ along with the metric $(1,1,-1)$ has been introduced.

\begin{figure}[h!]
\centering
\includegraphics[scale=0.65]{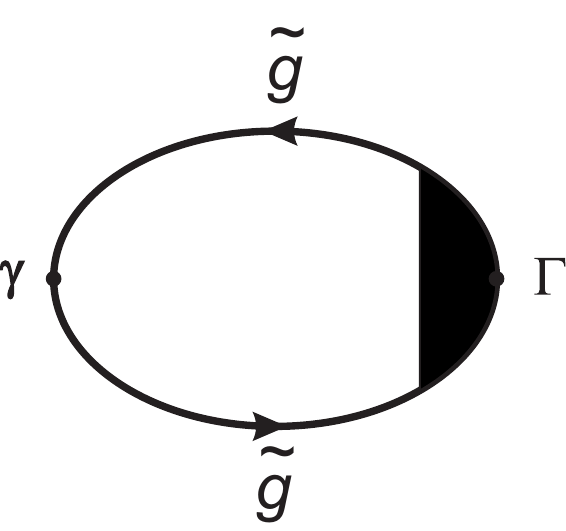}
\caption{Dressed polarization bubble (Nambu notation). Here both the charge-carrier Green function and the current vertex are dressed with the pairing interaction due to the spin bubble. [From Ref. \cite{Krzyzosiak10}.] \label{polarization-bubble}}
\end{figure}

Since the local charge conservation requirement is quite universal and fundamental, it should be inherent to any theory of the Meissner effect which is expected to be gauge invariant. The purpose of this subsection is to propose--within the formalism of kinetic-energy driven superconductivity--a method to dress the current vertex in a way, which does not violate the generalized Ward identity. Once such a method is found, the bare polarization bubble (\ref{corP}) can be replaced by its dressed version presented in Fig. \ref{polarization-bubble}, and the resulting kernel of the response function will provide correct results for any gauge of the vector potential.

In the first step we will note that \cite{Krzyzosiak10}
\begin{equation}\label{wardmean}
\sum\limits_{\mu=0}^{2}q_{\mu}\gamma_{\mu}(k+q,k)=\tau_{3}\tilde{g}^{(0)-1}(k)-\tilde{g}^{(0)-1}(k+q)\tau_{3},
\end{equation}
i.e. the generalized Ward identity for the bare current vertex is satisfied with the MF charge-carrier Green function (\ref{MFHGF})
$\tilde{g}^{(0)}(k)=[(i\omega_{n})^{2}-\xi_{\bf k}^{2}]^{-1}(i\omega_{n}\tau_{0}+\xi_{\bf k}\tau_{3})$. Substituting this MF charge-carrier Green function, the right-hand side of Eq. (\ref{wardmean}) turns into $\tau_{3}\tilde{g}^{(0)-1}(k)-\tilde{g}^{(0)-1}(k+q)\tau_{3}=(\xi_{\bf k+q}-\xi_{\bf k})\tau_{0}-q_{0}\tau_{3}$. Moreover, in the long wavelength limit, after including the explicit form of the MF charge-carrier dispersion relation (\ref{MFCCS}), it further simplifies to $\tau_{3}\tilde{g}^{(0)-1}(k)-\tilde{g}^{(0)-1}(k+q)\tau_{3}\approx [-q_{x}\sin k_{x}(2t\chi_{1}-Zt'\chi_{2}\cos k_{y})-q_{y}\sin k_{y}(2t\chi_{1}-Zt'\chi_{2}\cos k_{x})]\tau_{0}- q_{0}\tau_{3}$. Now, recalling the form of the bare vertex (\ref{barevertex}), it is easy to find that in the long wavelength limit the scalar product on the left-hand side of Eq. (\ref{wardmean}) is equal to the above obtained one of the right-hand side of Eq. (\ref{wardmean}), which proves the equality (\ref{wardmean}).

\begin{figure}[h!]
\centering
\includegraphics[scale=0.45]{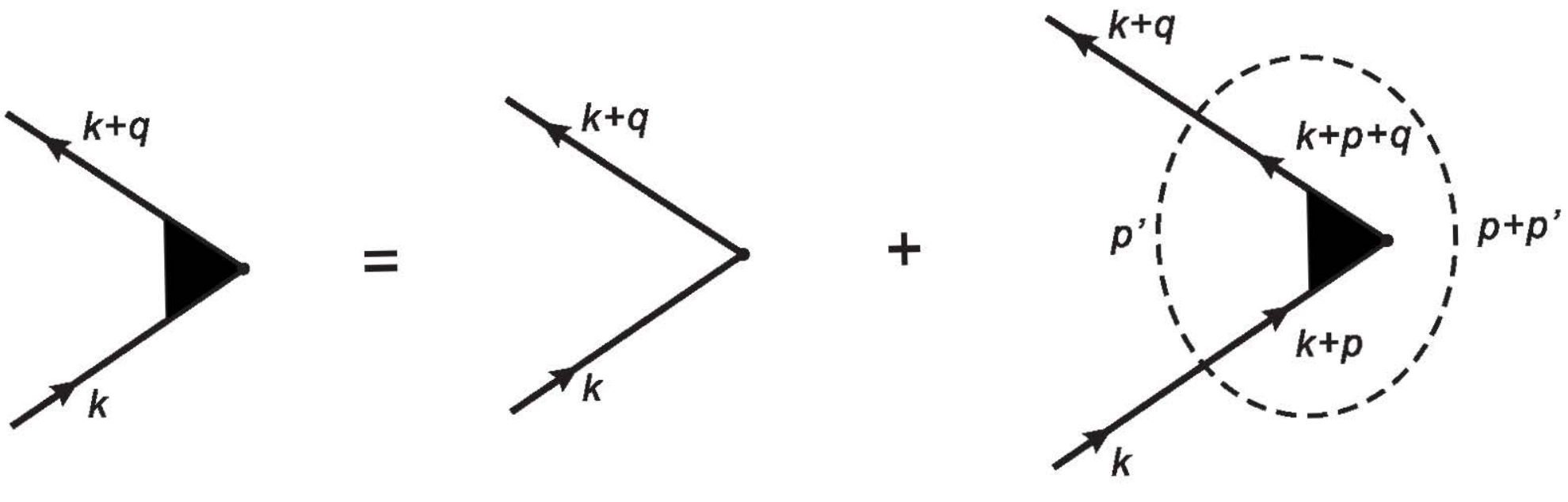}
\caption{Ladder-type approximation for the dressed vertex. [From Ref. \cite{Krzyzosiak10}.] \label{Ladder-diagram}}
\end{figure}

It is well known that in order to obtain a dressed vertex function, which does not violate the generalized Ward identity, a ladder-type approximation can be adapted \cite{Schrieffer64,Fukuyama69,Misawa94}. The nature of the pairing mechanism \cite{Feng03,Feng06}, which originates from the spin bubble (\ref{SB}), suggests a ladder-like approximation
of the form \cite{Krzyzosiak10},
\begin{eqnarray}
\Gamma_{\mu}(k+q,k)&=&\gamma_{\mu}(k+q,k)+{1\over N}{1\over\beta}\sum\limits_{p}\tau_{3}\tilde{g}(k+p+q)\Gamma_{\mu}(k+p+q,k+p){\tilde{g}}(k+p)\tau_{3}\nonumber\\
&\times& {1\over N}\sum\limits_{{\bf p}'}\Lambda^{2}_{{\bf p}+{\bf p}'+{\bf k}}\Pi({\bf p},{\bf p}',ip_{m}), \label{ladder}
\end{eqnarray}
which is graphically presented in Fig. \ref{Ladder-diagram}.

In order to prove that the approximation (\ref{ladder}) for the dressed vertex in fact implies a gauge invariant description of the Meissner effect, it is necessary and sufficient to check whether it does not violate the generalized Ward identity (\ref{GWI}). In order to prove it, we insert the dressed vertex function (\ref{ladder}) into the left-hand side of Eq. (\ref{GWI}) and use the identity $\sum_{\mu=0}^{2}q_{\mu}\Gamma_{\mu}(k+q,k)=\tau_{3}{\tilde{g}}^{-1}(k)-{\tilde{g}}^{-1}(k+q)\tau_{3}$ to obtain
\begin{eqnarray}\nonumber
\sum\limits_{\mu=0}^{2}q_{\mu}\Gamma_{\mu}(k+q,k)&=&\sum\limits_{\mu=0}^{2}q_{\mu}\gamma_{\mu}(k+q,k)+{1\over N}{1\over\beta}\sum\limits_{p} [\tau_{3}{\tilde{g}}(k+p+q) -{\tilde{g}}(k+p) \tau_{3}]\\
&\times& {1\over N}\sum\limits_{\bf p'}\Lambda^{2}_{{\bf p}+{\bf p}'+{\bf k}}\Pi({\bf p},{\bf p}',ip_{m}).\label{inveqn}
\end{eqnarray}
In the long wavelength limit  we use the approximation $\Lambda^{2}_{{\bf p}+{\bf p}'+{\bf k}}\approx\Lambda^{2}_{{\bf p}+{\bf p}'+{\bf k}+{\bf q}}$ for the second term of the right-hand side in Eq. (\ref{inveqn}). Then we can simplify Eq. (\ref{inveqn}) in terms of the self-energy (\ref{SE}) into $\sum_{\mu=0}^{2}q_{\mu}\Gamma_{\mu}(k+q,k)\approx\sum_{\mu=0}^{2}q_{\mu}\gamma_{\mu}(k+q,k)+\tilde{\Sigma}^{({\rm h})}(k+q)\tau_{3}-\tau_{3}\tilde{\Sigma}^{({\rm h})}(k)$. Using the fact that the bare vertex satisfies the generalized Ward identity with the MF charge-carrier Green function, as stated in Eq. (\ref{wardmean}), and arranging the terms with respect to the Pauli matrices, we \cite{Krzyzosiak10} have
\begin{eqnarray}
\sum\limits_{\mu=0}^{2}q_{\mu}\Gamma_{\mu}(k+q,k)&\approx&\tau_{3}[\tilde{g}^{(0)-1}(k)-\tilde{\Sigma}^{({\rm h})}(k)]-[\tilde{g}^{(0)-1}(k+q)-\tilde{\Sigma}^{({\rm h})}(k+q)] \tau_{3}. ~~~~~
\end{eqnarray}
Hence, identifying the terms in the square brackets as the full charge-carrier Green functions, we eventually obtain the generalized Ward identity (\ref{GWI}), which proves that the ladder-type approximation (\ref{ladder}) for the vertex function in the dressed polarization bubble in Fig. \ref{polarization-bubble} is consistent with the generalized Ward identity. Consequently, the kernel of the linear response calculated with the dressed polarization bubble is gauge invariant.

\subsection{Summary and discussions}\label{Meissner-summary}

Within the framework of the kinetic-energy driven SC mechanism, the doping dependence of the electromagnetic response is discussed. In the linear response approach, the electromagnetic response consists of two parts, the diamagnetic current, which is the acceleration in the magnetic field, and the paramagnetic current, which is a perturbation response of the excited quasiparticle and exactly cancels out the diamagnetic term in the normal state, then the Meissner effect is obtained for all the temperatures $T\leq T_{\rm c}$ throughout the SC dome. By considering the two-dimensional geometry of cuprate superconductors within the specular reflection model, the main features of the doping dependence of the local magnetic field profile, the magnetic field penetration depth, and the superfluid density are qualitatively reproduced. The local magnetic field profile follows an exponential law, while the magnetic field penetration depth shows a crossover from the linear temperature dependence at low temperatures to a nonlinear one at the extremely low temperatures. In particular, the domelike shape of the doping dependence of the superfluid density $\rho_{\rm s}$ with the highest value appearing around the {\it critical doping} $\delta_{\rm critical}\approx 0.195$ is a natural consequence of the domelike shape of the doping dependence of the charge-carrier pair order parameter $\Delta_{\rm h}$ with the maximal value appearing around the same {\it critical doping}.

\begin{figure}[h!]
\centering
\includegraphics[scale=0.7]{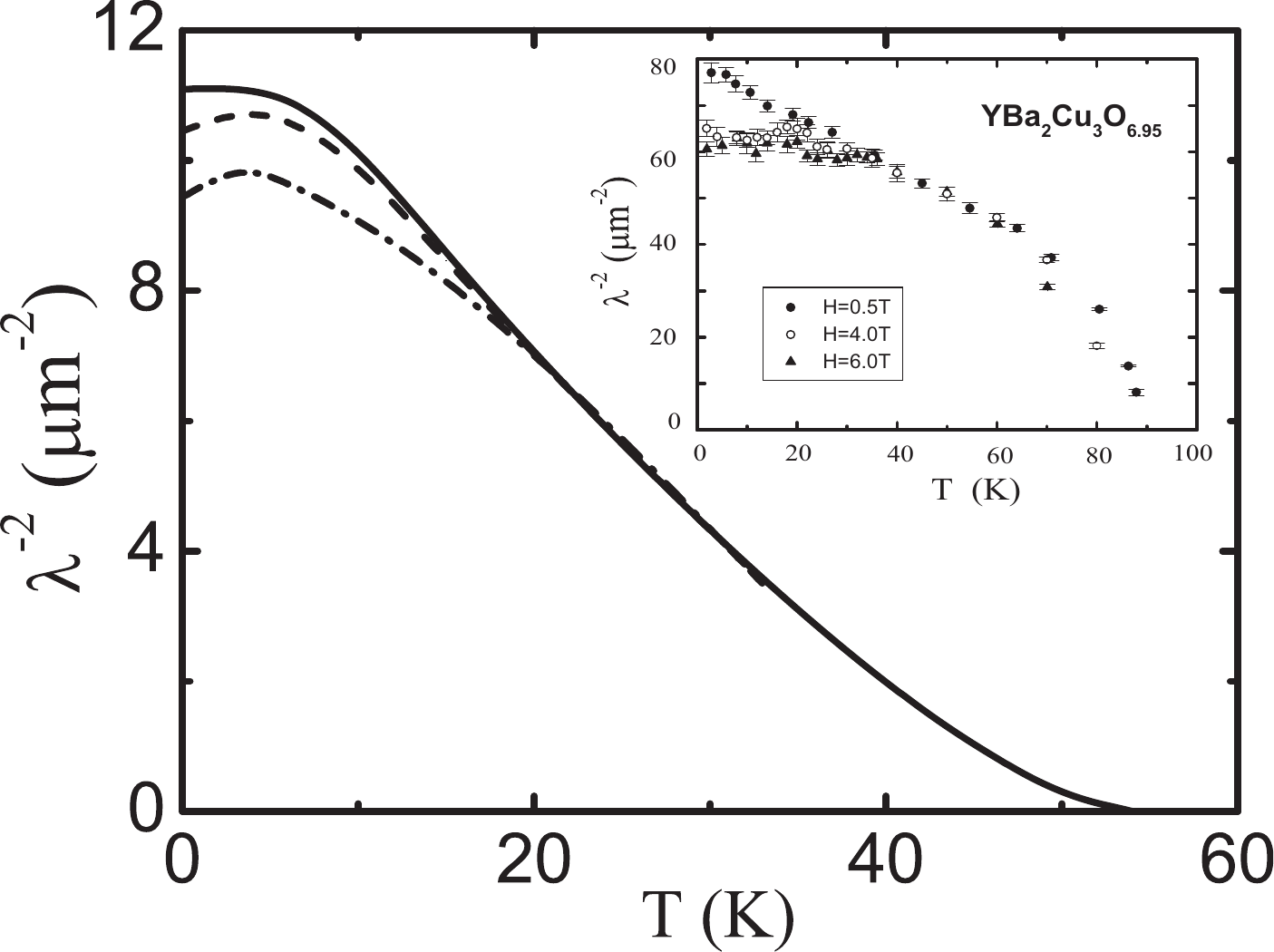}
\caption{The superfluid density as a function of temperature at $\delta=0.09$ with magnetic field $B=0$ (solid line), $B=0.5$ T (dashed line), and $B=1.0$ T (dash-dotted line) for $t/J=2.5$, $t'/t=0.3$, and $J=1000$K. Insets: the corresponding experimental results for YBa$_2$Cu$_3$O$_{6.95}$ taken from Ref. \cite{Sonier99}. [From Ref. \cite{Huang11}.] \label{superfluid-field}}
\end{figure}

Finally, it should be emphasized again that in the above discussions, the only coupling of the electron charge to an external magnetic field is considered in the $t$-$J$ model (\ref{MtJmodel}) in terms of the vector potential ${\bf A}$, while the coupling of the electron magnetic momentum with the external magnetic field in terms of the Zeeman mechanism has been dropped. However, the depairing due to the Pauli spin polarization is very important in the presence of a moderate or strong external magnetic field, since cuprate superconductors are doped Mott insulators with the strong AFSRO correlation dominating the entire SC phase \cite{Fujita12,Dean14}. In particular, within the framework of the kinetic-energy driven SC mechanism, a moderate or strong external magnetic field aligns the spins of the unpaired electrons, then the d-wave charge-carrier pairs in cuprate superconductors can not take advantage of the lower energy offered by a spin-polarized state \cite{Vorontsov10}. In this case, the magnetic field dependence of the superfluid density in cuprate superconductors has been studied \cite{Huang11} by considering both couplings of the electron charge and electron magnetic momentum with a weak magnetic field, and the calculated result of the superfluid density $\rho_{\rm s}$ as a function of temperature at $\delta=0.09$ with the magnetic field $B=0$ (solid line), $B=0.5$ T (dashed line), and $B=1.0$ T (dash-dotted line) for $t/J=2.5$, $t'/t=0.3$, and $J=1000$K is shown in Fig. \ref{superfluid-field} in comparison with the corresponding experimental data \cite{Sonier99} for YBa$_2$Cu$_3$O$_{6.95}$ (inset). Most importantly, the magnitude of $\rho_{\rm s}$ at the extremely low temperatures decreases with increasing magnetic field, and then it turns to be independent on a weak magnetic field away from the extremely low temperatures, in qualitative agreement with experimental data of cuprate superconductors \cite{Sonier99,Bidinosti99,Khasanov09,Serafin10}. The calculated result also indicates that the nature of the quasiparticle excitations at the extremely low temperatures is strongly influenced by a weak magnetic field. This weak magnetic field induced reduction of the superfluid density of cuprate superconductors at the extremely low temperatures contrasts with that observed from conventional superconductors \cite{Khasanov06}, where the curves of the temperature dependent superfluid density for differently weak magnetic fields were found to collapse onto a single curve since conventional superconductors are fully gaped.

\section{Dynamical spin response}\label{spin-response}

As illustrated in the schematic phase diagram in Fig. \ref{phase-diagram-exp}, cuprate superconductors exist on a continuum with a family of the Mott insulators. The undoped and extremely low-doped cuprates have an AFLRO at low temperatures. The spin excitation with AFLRO is called as magnon. However, in the doped regime of the SC dome, the AFLRO breaks down, but the AFSRO correlation remains. This AFSRO can still support spin waves, but the spin excitations with AFSRO are damped. The damped spin excitation is known as paramagnon \cite{Miller11}.
\begin{figure}[h!]
\centering
\includegraphics[scale=0.8]{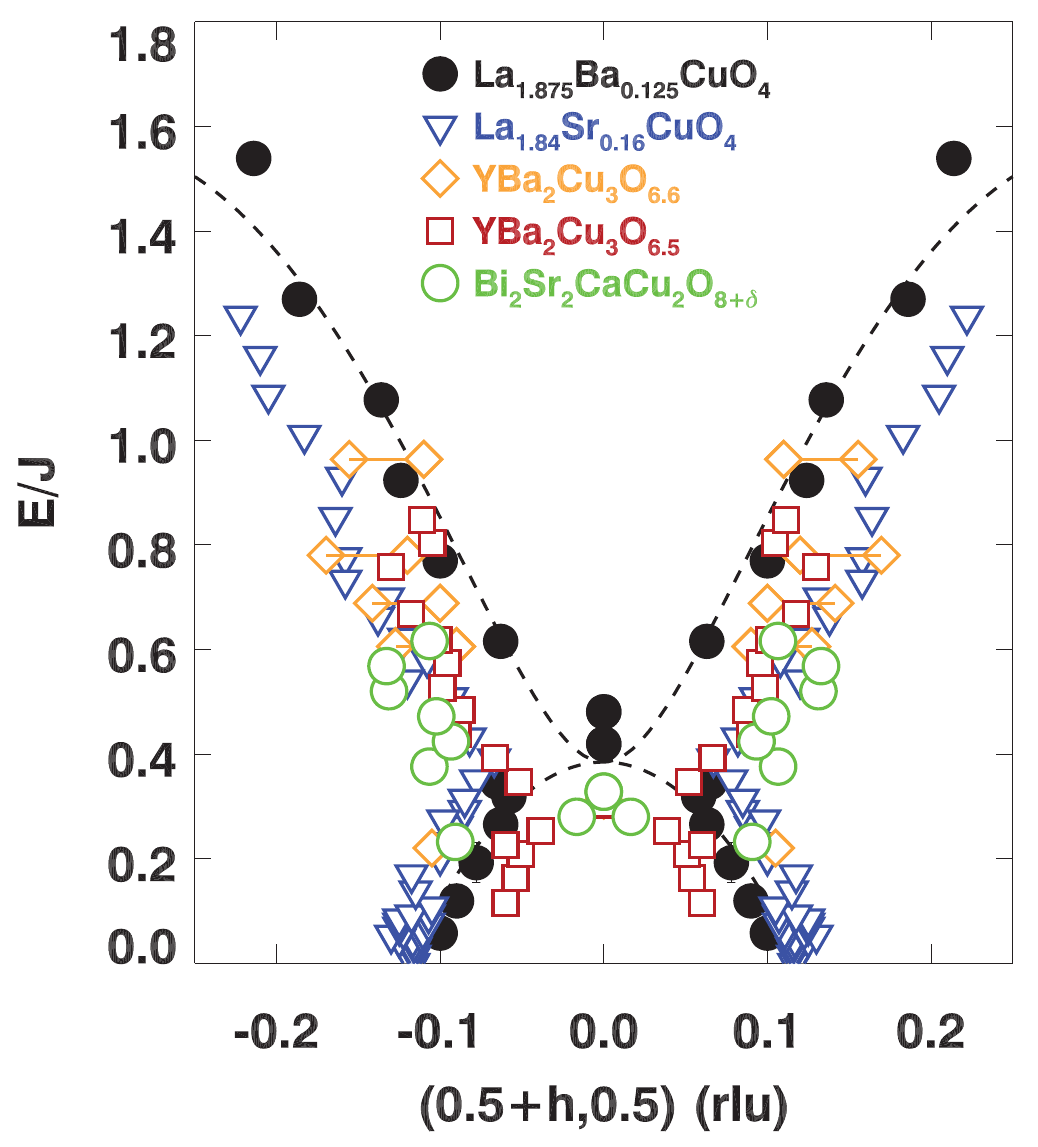}
\caption{(Color) The magnetic dispersion relation in different families of cuprate superconductors. [From Ref. \cite{Fujita12}.] \label{hour-glass}}
\end{figure}

\begin{figure}[h!]
\centering
\includegraphics[scale=0.5]{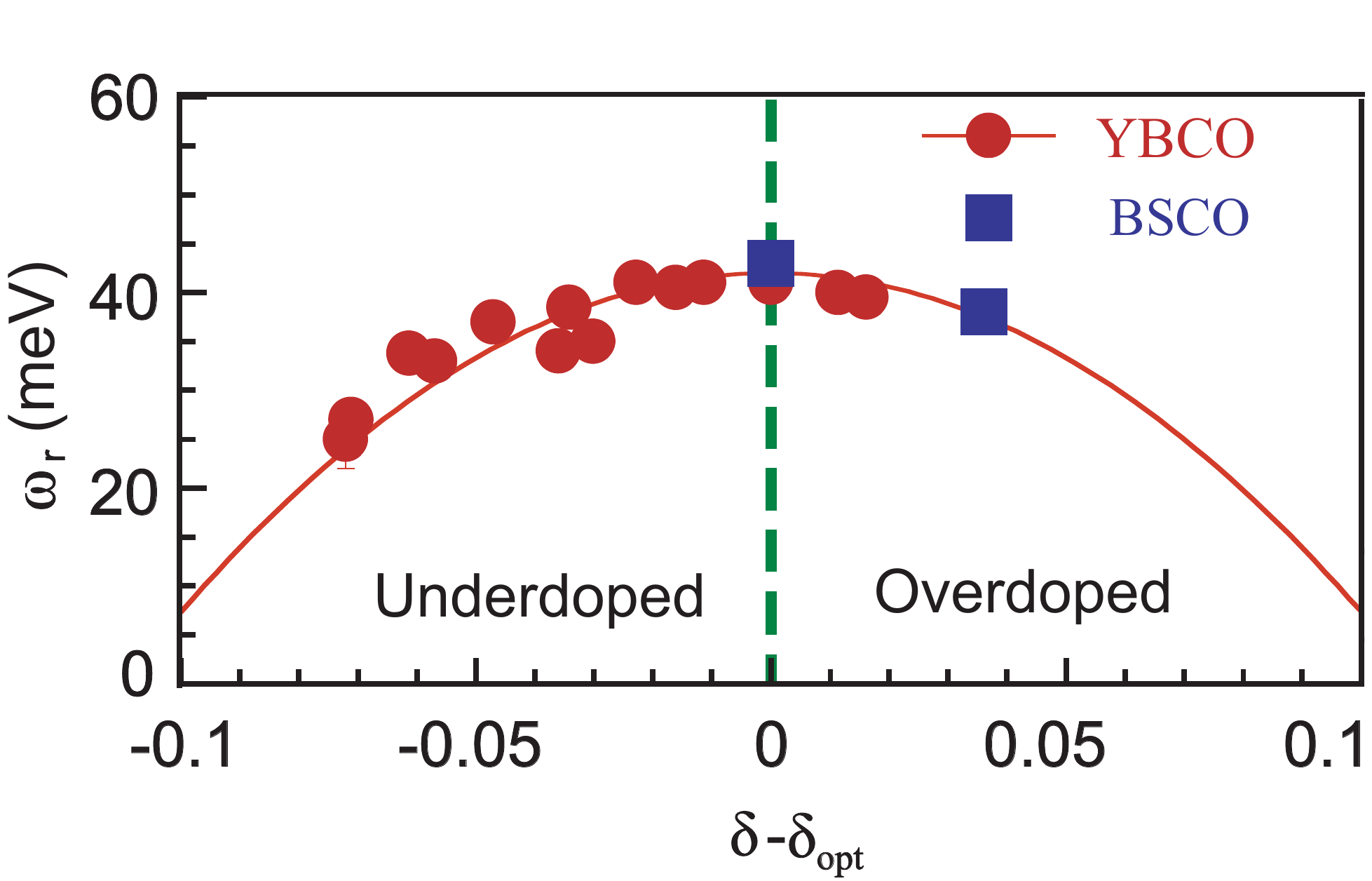}
\caption{(Color) The resonance energy as a function of doping referenced to the optimal doping, $\delta_{\rm opt}$, corresponding to $T^{\rm max}_{\rm c}$. The red full curve shows the doping dependence of $T_{\rm c}$ times 5.3. [From Ref. \cite{Bourges05}.] \label{resonance-doping}}
\end{figure}

\begin{figure}[h!]
\centering
\includegraphics[scale=0.6]{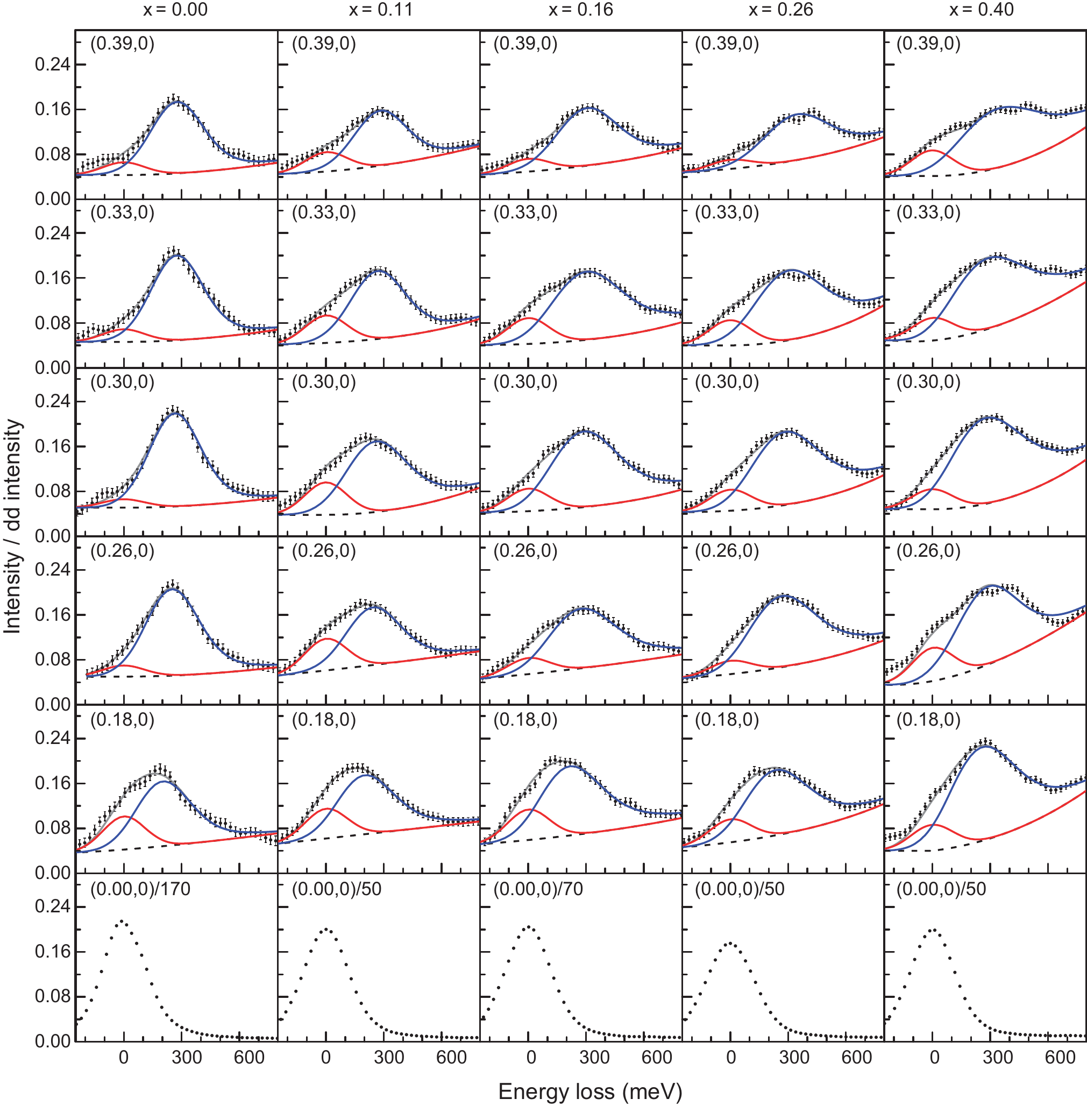}
\caption{(Color) The dispersion of the magnetic excitations in La$_{2-x}$Sr$_{x}$CuO$_{4}$ as function of {\bf Q} and $x$. [From Ref. \cite{Dean13}.] \label{RIXS}}
\end{figure}

The early INS measurements \cite{Fujita12,Eschrig06,Fong95,Birgeneau89,Rossat91,Cheong91,Yamada98,Dai01,Wakimoto04,Hayden04,Tranquada04} on cuprate superconductors have demonstrated that the doped charge carriers cause substantial changes to the low-energy spin excitation spectrum, and a consistent pattern has been identified as the {\it hour-glass-shaped} dispersion as illustrated in Fig. \ref{hour-glass}. This hour-glass-shaped dispersion was first observed in the spin excitations of YBa$_{2}$Cu$_{3}$O$_{6.6}$ \cite{Hayden04} and La$_{1.875}$Ba$_{0.125}$CuO$_{4}$ \cite{Tranquada04}, where two IC components of the low-energy spin excitation spectrum are separated by a commensurate resonance energy $\omega_{\rm r}$ at the waist of the hour glass. In the upward component, above the commensurate resonance energy $\omega_{\rm r}$, the spin excitation spectrum is similar to what one would expect from AF spin fluctuations with a finite gap, and is relevant to the results for different families of cuprate superconductors that appear to scale with the magnetic exchange coupling constant $J$ for the undoped parent compounds of cuprate superconductors. In particular, for a given excitation energy, the magnetic scattering peaks lie on a circle of radius of $\bar{\delta}_{\rm IC}'$, with the incommensurability parameter $\bar{\delta}_{\rm IC}'$ that is defined as a deviation of the peak position from the AF wave vector $[1/2,1/2]$ (for convenience, in this section we use the units of $[2\pi,2\pi]$) in the Brillouin zone (BZ), and then the distribution of the spectral weight of IC magnetic scattering peaks is rather isotropic. On the other hand, in the downward component, below $\omega_{\rm r}$, the distribution of the spectral weight of IC magnetic scattering peaks is quite anisotropic \cite{Fujita12,Eschrig06,Fong95,Birgeneau89,Rossat91,Cheong91,Yamada98,Dai01,Wakimoto04,Hayden04,Tranquada04}. In particular, it is remarkable \cite{Fujita12,Eschrig06,Fong95,Birgeneau89,Rossat91,Cheong91,Yamada98,Dai01,Wakimoto04,Hayden04,Tranquada04,Hinkov04,Bourges05,Sidis04,He01,Bourges00,Arai99} that in analogy to the domelike shape of the doping dependence of $T_{\rm c}$, the commensurate resonance energy $\omega_{\rm r}$ increases with increasing doping in the underdoped regime, and reaches a maximum around the optimal doping, then decreases in the overdoped regime as illustrated in Fig. \ref{resonance-doping}, reflecting a intrinsical relationship between $\omega_{\rm r}$ and $T_{\rm c}$. Although the IC magnetic scattering has been also observed in the normal-state, the commensurate resonance is a new feature that appears in the SC-state {\it only} \cite{Fujita12,Eschrig06,Fong95,Birgeneau89,Rossat91,Cheong91,Yamada98,Dai01,Wakimoto04,Hayden04,Tranquada04,Hinkov04,Bourges05,Sidis04,He01,Bourges00,Arai99}. Later, this hour-glass-shaped dispersion was found in several different families of cuprate superconductors \cite{Vignolle07,Stock05,Stock10,Xu09}. However, because of technical limitations, only the low-energy ($E\sim 10-80$ meV) spin excitations in a small range of momentum space around the AF wave vector are detected by INS measurements \cite{Eschrig06,Fong95,Birgeneau89,Rossat91,Cheong91,Yamada98,Dai01,Wakimoto04,Hayden04,Tranquada04,Hinkov04,Bourges05,Sidis04,He01,Bourges00,Arai99,Vignolle07,Stock05,Stock10,Xu09}. In recent years, instrumentation for RIXS with both soft and hard X-rays has improved dramatically, allowing this technique to directly measure the high-energy ($E\sim 80-500$ meV) spin excitations of cuprate superconductors in the wide energy-momentum window that cannot be detected by INS measurements \cite{Dean14,Vojta11}. In this case, as a compensation for the miss of a significant part of the spectral weight of spin excitations in INS studies \cite{Eschrig06,Fong95,Birgeneau89,Rossat91,Cheong91,Yamada98,Dai01,Wakimoto04,Hayden04,Tranquada04,Hinkov04,Bourges05,Sidis04,He01,Bourges00,Arai99,Vignolle07,Stock05,Stock10,Xu09}, the RIXS technique has been used to measure the high-energy spin excitations of cuprate superconductors in the whole doping range, and the experimental data \cite{Dean14,Tacon11,Dean13,Dean13a,Tacon13} indicates that the key feature of the high-energy spin excitations even in the overdoped regime is strikingly similar to that of the undoped parent compounds \cite{Braicovich10,Guarise10,Piazza12} (see Fig. \ref{RIXS}).

Within the framework of the kinetic-energy driven SC mechanism, the dynamical spin response of cuprate superconductors from low-energy to high-energy has been studied \cite{Kuang15,Yuan01,Feng98,Feng04}, where one of the main results is that both the damped but well-defined dispersive low-energy and high-energy spin excitations exist across the whole doping phase diagram. In the SC-state \cite{Kuang15}, the low-energy spin excitations are strongly renormalized due to the interaction between charge carriers and spins to form an hour-glass-shaped dispersion. In particular, the commensurate resonance is closely related to the process of the creation of charge-carrier pairs, and appears in the SC-state only, while the low-energy IC magnetic scattering is mainly associated with mobile charge-carrier quasiparticles, and therefore persists into the normal-state. On the other hand, the charge-carrier doping has a more modest effect on the high-energy spin excitations \cite{Kuang15}, and the high-energy spin fluctuations bear a striking resemblance to those found in the undoped parent compounds \cite{Braicovich10,Guarise10,Piazza12}. In this section, we summarize a few calculated results for the dynamical spin response of cuprate superconductors \cite{Kuang15,Yuan01,Feng98,Feng04}.

\subsection{Dynamical spin structure factor in superconducting-state}\label{SC-state-DSSF}

In the framework of the CSS fermion-spin theory (\ref{CSS}), the scattering of spins due to the charge-carrier fluctuation is responsible to the spin dynamics. For the discussions of the dynamical spin response in cuprate superconductors, it is needed to calculate the full spin Green's function, which can be expressed as \cite{Kuang15,Yuan01,Feng98,Feng04},
\begin{eqnarray}\label{FSGF}
D({\bf k},\omega)={1\over D^{(0)-1}({\bf k},\omega)-\Sigma^{({\rm s})}({\bf k},\omega)}.
\end{eqnarray}
In the SC-state, the spin fluctuation occurs in the charge-carrier quasiparticle background, and then the spin self-energy can be obtained within the framework of the equation of motion method in terms of the collective charge-carrier modes in the particle-hole and particle-particle channels \cite{Kuang15} as,
\begin{eqnarray}\label{SSF1}
\Sigma^{({\rm s})}({\bf k},ip_{m})&=&-{1\over N^{2}}\sum_{\bf pq}(\Lambda^{2}_{{\bf k}-{\bf p}}+\Lambda^{2}_{{\bf p}+{\bf q}+{\bf k}}){1\over\beta}\sum_{iq_{m}}
D^{(0)}({\bf q}+{\bf k},iq_{m}+ip_{m})\nonumber\\
&\times& [\Pi^{(s)}_{gg}({\bf p},{\bf q},iq_{m})-\Pi^{(s)}_{\Gamma\Gamma}({\bf p},{\bf q},iq_{m})],~~~~~~~
\end{eqnarray}
where the charge-carrier bubble $\Pi^{(s)}_{gg}({\bf p},{\bf q},iq_{m})$ in the particle-hole channel is obtained from the full charge-carrier diagonal Green's function (\ref{BCSHDGF}) as,
\begin{eqnarray}\label{SBPH}
\Pi^{(s)}_{gg}({\bf p},{\bf q},iq_{m})&=&{1\over\beta}\sum_{i\omega_{n}}g({\bf p},i\omega_{n})g({\bf p}+{\bf q},i\omega_{n}+iq_{m}), ~~~~~~~
\end{eqnarray}
and is closely related to mobile charge-carrier quasiparticles, while the charge-carrier bubble $\Pi^{(s)}_{\Gamma\Gamma}({\bf p},{\bf q},iq_{m})$ in the particle-particle channel is obtained from the full charge-carrier off-diagonal Green's function (\ref{BCSHODGF}) as,
\begin{eqnarray}\label{SBPP}
\Pi^{(s)}_{\Gamma\Gamma}({\bf p},{\bf q},iq_{m})&=&{1\over\beta}\sum_{i\omega_{n}}\Gamma^{\dagger}({\bf p},i\omega_{n})\Gamma({\bf p}+{\bf q},i\omega_{n}+iq_{m}),~~~~~~~
\end{eqnarray}
and therefore is directly associated with the creation of charge-carrier pairs. Substituting the full charge-carrier Green's function (\ref{BCSHGF}) and the MF spin Green's function (\ref{MFSGF}) into Eqs. (\ref{SBPH}), (\ref{SBPP}), and (\ref{SSF1}), the spin self-energy can be evaluated explicitly as \cite{Kuang15},
\begin{eqnarray}\label{SSF}
\Sigma^{({\rm s})}({\bf k},\omega)&=&-{1\over 2N^{2}}\sum_{{\bf pq},\nu}(-1)^{\nu+1}\Omega^{(\rm s)}_{\bf kpq}\left ({I_{+}({\bf p},{\bf q})F^{\rm (s)}_{\nu+}({\bf k},{\bf p}, {\bf q})\over\omega^{2}-[\omega_{{\bf q}+ {\bf k}}- (-1)^{\nu+1}(E_{{\rm h}{\bf p}+{\bf q}}-E_{{\rm h}{\bf p}})]^{2}}\right .\nonumber\\
&+& \left . {I_{-}({\bf p},{\bf q})F^{\rm (s)}_{\nu-}({\bf k},{\bf p},{\bf q})\over\omega^{2}-[\omega_{{\bf q}+{\bf k}}- (-1)^{\nu+1}(E_{{\rm h}{\bf p}+{\bf q}}+E_{{\rm h}{\bf p}}) ]^{2}} \right ),
\end{eqnarray}
where $\nu=1,2$, $\Omega^{(\rm s)}_{\bf kpq}=Z^{2}_{\rm hF}(\Lambda^{2}_{{\bf k}-{\bf p}}+\Lambda^{2}_{{\bf p}+{\bf q}+{\bf k}})B_{{\bf q}+{\bf k}}/(2\omega_{{\bf q}+{\bf k}})$, the charge-carrier coherence factors for the processes,
\begin{subequations}\label{CFS}
\begin{eqnarray}
I_{+}({\bf p},{\bf q})&=&1+{\bar{\xi}_{\bf p}\bar{\xi}_{{\bf p}+{\bf q}}-\bar{\Delta}_{\rm hZ}({\bf p})\bar{\Delta}_{\rm hZ}({\bf p}+{\bf q})\over E_{{\rm h}{\bf p}}E_{{\rm h}{\bf p} +{\bf q}}},\label{CFS1}\\
I_{-}({\bf p},{\bf q})&=&1-{\bar{\xi}_{\bf p}\bar{\xi}_{{\bf p}+{\bf q}}-\bar{\Delta}_{\rm hZ}({\bf p})\bar{\Delta}_{\rm hZ}({\bf p}+{\bf q})\over E_{{\rm h}{\bf p}}E_{{\rm h}{\bf p} +{\bf q}}},\label{CFS2}
\end{eqnarray}
\end{subequations}
and the functions,
\begin{subequations}
\begin{eqnarray}
F^{\rm (s)}_{\nu+}({\bf k},{\bf p},{\bf q})&=&[\omega_{{\bf q}+{\bf k}}-(-1)^{\nu+1}(E_{{\rm h}{\bf p}+{\bf q}}-E_{{\rm h}{\bf p}})]\{n_{\rm B}(\omega_{{\bf q}+{\bf k}})[n_{\rm F} (E_{{\rm h}{\bf p}})-n_{\rm F}(E_{{\rm h}{\bf p}+{\bf q}})]\nonumber\\
&-&(-1)^{\nu+1}n_{\rm F}[(-1)^{\nu}E_{{\rm h}{\bf p}}]n_{\rm F}[(-1)^{\nu+1}E_{{\rm h}{\bf p}+{\bf q}}]\},\\
F^{\rm (s)}_{\nu-}({\bf k},{\bf p},{\bf q})&=& [\omega_{{\bf q}+{\bf k}}-(-1)^{\nu+1}(E_{{\rm h}{\bf p}+{\bf q}}+E_{{\rm h}{\bf p}})]\{n_{\rm B}(\omega_{{\bf q}+{\bf k}})[1-n_{\rm F} (E_{{\rm h}{\bf p}})-n_{\rm F}(E_{{\rm h}{\bf p}+{\bf q}})]\nonumber\\
&-&(-1)^{\nu+1}n_{\rm F}[(-1)^{\nu+1}E_{{\rm h}{\bf p}}]n_{\rm F}[(-1)^{\nu+1}E_{{\rm h}{\bf p}+{\bf q}}]\}.
\end{eqnarray}
\end{subequations}
With the help of the full spin Green's function (\ref{FSGF}), the dynamical spin structure factor of cuprate superconductors is obtained as \cite{Kuang15},
\begin{eqnarray}\label{DSSF}
S({\bf k},\omega)&=&-2[1+n_{\rm B}(\omega)]{\rm Im}D({\bf k},\omega)\nonumber\\
&=&-{2[1+n_{\rm B}(\omega)]B^{2}_{{\bf k}}{\rm Im}\Sigma^{({\rm s})}({\bf k},\omega)\over [\omega^{2}-\omega^{2}_{\bf k}-B_{{\bf k}}{\rm Re}\Sigma^{({\rm s})}({\bf k},\omega)]^{2} +[B_{{\bf k}}{\rm Im}\Sigma^{({\rm s})}({\bf k},\omega)]^{2}}, ~~~~~
\end{eqnarray}
where ${\rm Im}\Sigma^{({\rm s})}({\bf k},\omega)$ and ${\rm Re}\Sigma^{({\rm s})}({\bf k},\omega)$ are the corresponding imaginary and real parts of the spin self-energy (\ref{SSF}), respectively.

\subsection{Universal Low-energy spin excitation spectrum}

\begin{figure}[h!]
\centering
\includegraphics[scale=0.52]{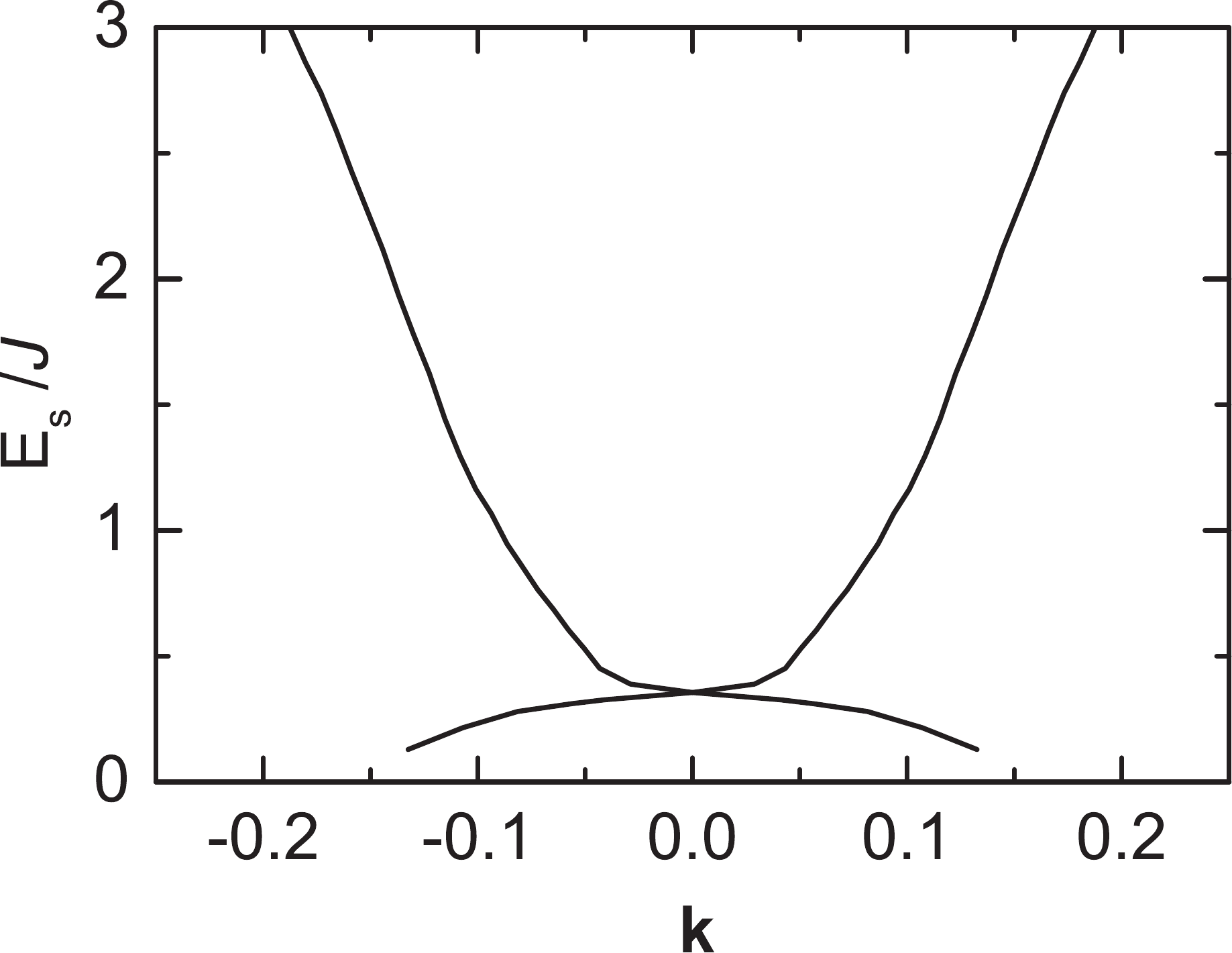}
\caption{The energy dependence of the position of magnetic scattering peaks at $\delta=0.21$ with $T=0.002J$ for $t/J=2.5$ and $t'/t=0.3$. [From Ref. \cite{Kuang15}.] \label{peak-position}}
\end{figure}

First of all, we discuss unusual feature of the low-energy magnetic scattering. Of course, at half-filling, the undoped parent compounds of cuprate superconductors are Mott insulators, and then AFLRO gives rise to a commensurate peak at $[1/2,1/2]$. However, the calculated dynamical spin structure factor spectrum shows that when AFLRO is suppressed with doping, two IC magnetic scattering modes separated by a commensurate resonance energy $\omega_{\rm r}$ are developed \cite{Kuang15}. Well above the magnetic resonance energy $\omega_{\rm r}$, the IC magnetic scattering peaks lie uniformly on a circle of radius of $\bar{\delta}_{\rm IC}'$, and then the distribution of the spectral weight of IC magnetic scattering peaks is quite isotropic. However, the geometry of the magnetic scattering is energy dependent. In particular, below $\omega_{\rm r}$, although some IC satellite peaks appear along the diagonal direction of BZ, the main weight of IC magnetic scattering peaks is in the parallel direction, which leads to an rather anisotropic distribution of the spectral weight of IC magnetic scattering peaks below $\omega_{\rm r}$. To show the energy dependence of the position of the low-energy magnetic scattering peaks clearly, the evolution of magnetic scattering peaks with energy \cite{Kuang15} at $\delta=0.21$ with $T=0.002J$ for $t/J=2.5$ and $t'/t=0.3$ is shown in Fig. \ref{peak-position}, where the hour-glass-shaped dispersion of the low-energy magnetic scattering peaks observed from different families of cuprate superconductors is qualitatively reproduced \cite{Fujita12,Eschrig06,Fong95,Birgeneau89,Rossat91,Cheong91,Yamada98,Dai01,Wakimoto04,Hayden04,Tranquada04,Hinkov04,Bourges05,Sidis04,He01,Bourges00,Arai99,Vignolle07,Stock05,Stock10,Xu09}. In particular, in contrast to the case at energies below $\omega_{\rm r}$, the spin excitations at the energies above $\omega_{\rm r}$ disperse almost linearly with energy, which has been observed experimentally on cuprate superconductors \cite{Fujita12,Eschrig06,Fong95,Birgeneau89,Rossat91,Cheong91,Yamada98,Dai01,Wakimoto04,Hayden04,Tranquada04,Hinkov04,Bourges05,Sidis04,He01,Bourges00,Arai99,Vignolle07,Stock05,Stock10,Xu09}.

\subsection{Doping dependence of commensurate resonance}

\begin{figure}[h!]
\centering
\includegraphics[scale=0.5]{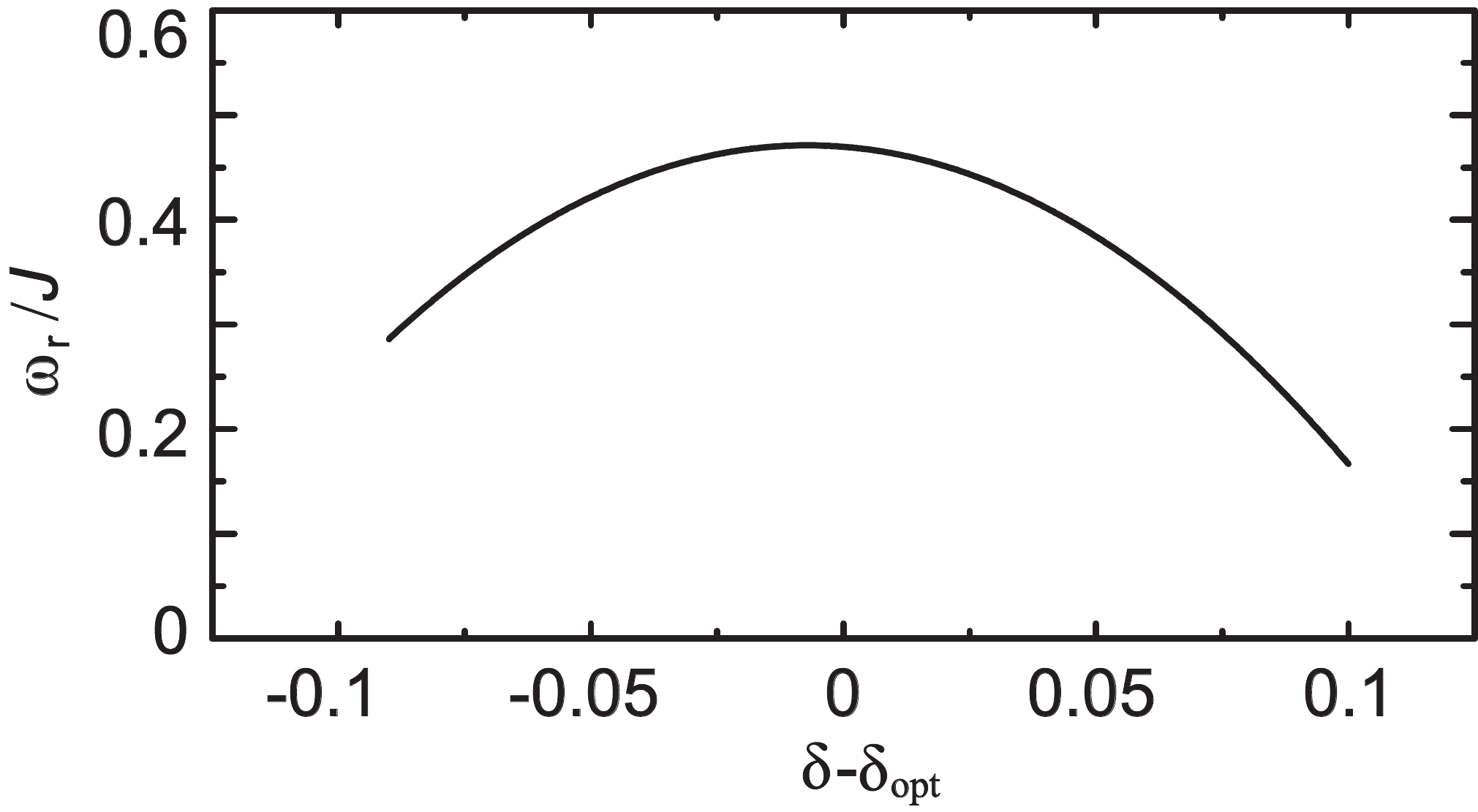}
\caption{The magnetic resonance energy $\omega_{\rm r}$ as a function of doping with $T=0.002J$ for $t/J=2.5$ and $t'/t=0.3$. [From Ref. \cite{Kuang15}.] \label{resonance-energy}}
\end{figure}

The commensurate resonance energy $\omega_{\rm r}$ is strongly doping dependent. In Fig. \ref{resonance-energy}, we \cite{Kuang15} show the commensurate resonance energy $\omega_{\rm r}$ as a function of doping with $T=0.002J$ for $t/J=2.5$ and $t'/t=0.3$, where in analogy to the domelike shape of the doping dependence of $T_{\rm c}$, the maximal $\omega_{\rm r}$ occurs around the optimal doping, and then decreases in both the underdoped and the overdoped regimes, also in qualitative agreement with the experimental results \cite{Fujita12,Bourges05,Sidis04}. In particular, using a reasonably estimative value of $J\sim 100$ meV, the anticipated resonance energy $\omega_{\rm r}=0.46J\approx 46$ meV in the optimal doping $\delta_{\rm opt}=0.15$ is not too far from the resonance energy $\omega_{\rm r}\approx 41$ meV observed in the optimally doped YBa$_{2}$Cu$_{3}$O$_{6+\delta}$  \cite{Fujita12,Fong95,Dai01,Bourges05,Sidis04}. This commensurate resonance has been also discussed based on the slave-boson approach at {\it zero} temperature \cite{Brinckmann01}.

\subsection{Evolution of high-energy spin excitations with doping}

\begin{figure}[t!]
\centering
\includegraphics[scale=0.72]{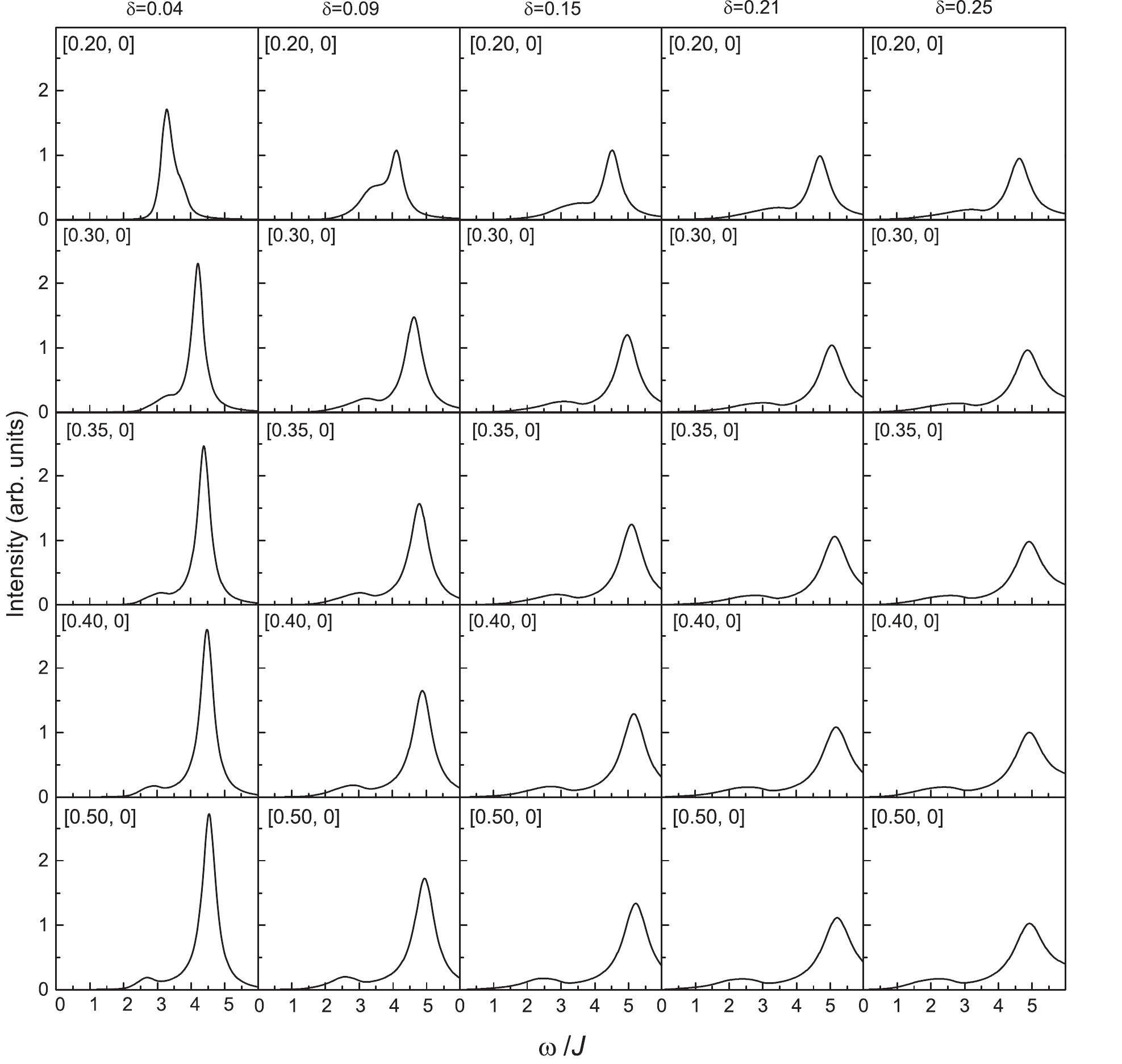}
\caption{The dynamical spin structure factor $S({\bf k},\omega)$ as a function of energy along the ${\bf k}=[0,0]$ to ${\bf k}=[0.5,0]$ direction of the Brillouin zone at $\delta=0.04$, $\delta=0.09$, $\delta=0.15$, $\delta=0.21$, and $\delta=0.25$ with $T=0.002J$ for $t/J=2.5$ and $t'/t=0.3$. [From Ref. \cite{Kuang15}.] \label{high-energy-excitations}}
\end{figure}

For a comparison of the high-energy spin excitations at different doping levels just as it has been done in RIXS experiments \cite{Dean14,Tacon11,Dean13,Dean13a,Tacon13}, the dynamical spin structure factor $S({\bf k},\omega)$ as a function of energy along the ${\bf k}=[0,0]$ to ${\bf k}=[0.5,0]$ direction of BZ with $T=0.002J$ for $t/J=2.5$ and $t'/t=0.3$ at $\delta=0.04$ (non-SC regime), $\delta=0.09$, $\delta=0.15$, $\delta=0.21$, and $\delta=0.25$ is shown in Fig. \ref{high-energy-excitations}. These calculated results \cite{Kuang15} capture the qualitative feature of the high-energy spin excitations observed experimentally on cuprate superconductors \cite{Dean14,Tacon11,Dean13,Dean13a,Tacon13}. The high-energy spin excitations persist across the whole doping phase diagram with comparable spectral weight and similar energies, i.e., in contrast to the dramatic change of the low-energy spin excitations with doping, the high-energy spin excitations retain roughly constant energy as a function of doping, and the shapes of the high-energy magnetic scattering peaks in the heavily overdoped regime are very similar to those in the extremely low-doped and underdoped regimes, although the width of the high-energy spin excitations increases continuously with doping, consistent with the spin excitation being damped by the increasing doping. Furthermore, for example, the magnetic scattering peak on energy scale of $4.7J$ appears in the ${\bf k}=[0.2,0]$ point at $\delta=0.21$, while the peak on the energy scale of $5.0J$ emerges in the ${\bf k}=[0.3,0]$ point, reflecting the dispersive nature of the high-energy spin excitations along the ${\bf k}=[0,0]$ to ${\bf k}=[0.5,0]$ direction. In particular, this dispersion relation of the high-energy spin excitations in the overdoped regime resembles those in the extremely low-doped and underdoped regimes.

\subsection{Dispersion of spin excitations}

\begin{figure}[h!]
\centering
\includegraphics[scale=0.5]{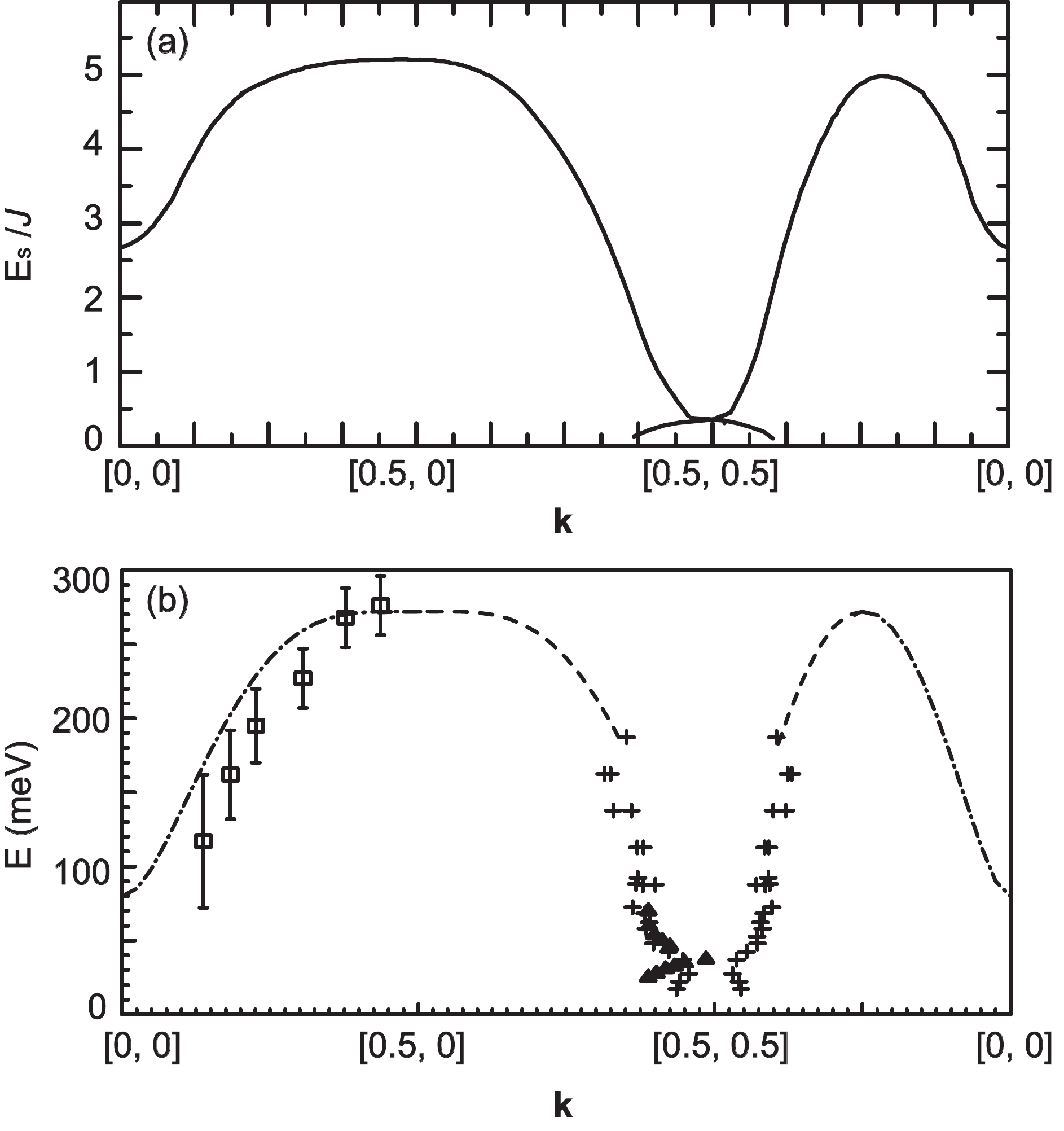}
\caption{(a) The dispersion of the spin excitations along the high symmetry directions of the Brillouin zone at $\delta=0.21$ for $t/J=2.5$ and $t'/J=0.3$ with $T=0.002J$. (b) The experimental result of the dispersion of the spin excitations along the high symmetry directions of the Brillouin zone for YBa$_{2}$Cu$_{3}$O$_{6+\delta}$ taken from Ref. \cite{Vojta11}. [From Ref. \cite{Kuang15}.] \label{dispersion}}
\end{figure}

To determine the overall spin excitation spectrum $E_{\rm s}({\bf k})$ in Eq. (\ref{DSSF}), spin excitations at different momenta in the whole doping phase diagram have been determined by a {\it self-consistent} calculation \cite{Kuang15},
\begin{eqnarray}
\omega^{2}=\omega^{2}_{\bf k}+B_{{\bf k}}{\rm Re}\Sigma^{({\bf s})}({\bf k},\omega),
\end{eqnarray}
and the result shows that spin excitations are well defined at all momenta. In Fig. \ref{dispersion}a, we \cite{Kuang15} show $E_{\rm s}({\bf k})$ as a function of momentum along the high symmetry directions of BZ at $\delta=0.21$ with $T=0.002J$ for $t/J=2.5$ and $t'/t=0.3$. For comparison, the experimental result \cite{Vojta11} of $E_{\rm s}({\bf k})$ along the high symmetry directions of BZ for the cuprate superconductor YBa$_{2}$Cu$_{3}$O$_{6+\delta}$ is shown in Fig. \ref{dispersion}b. It is shown that these theoretical calculations reproduce qualitatively the overall dispersion of spin excitations in cuprate superconductors \cite{Vojta11}. In comparison with the spin excitation spectrum (the spin wave) of the parent compounds of cuprate superconductors \cite{Fujita12,Hayden91,Hayden96,Coldea01,Braicovich10,Guarise10,Piazza12}, the spin excitation spectrum in the doped regime has been renormalized due to the presence of the interaction between charge carriers and spins directly from the kinetic energy of the $t$-$J$ model (\ref{CSStJmodel}). However, the charge-carrier doping does not uniformly renormalizes the dispersion of spin excitations. In particular, the low-energy magnetic correlation is strongly reorganized, where two IC components of the spin excitation spectrum are separated by the commensurate resonance energy $\omega_{\rm r}$, which therefore leads to an hour-glass-shaped dispersion of magnetic scattering peaks \cite{Fujita12,Eschrig06,Fong95,Birgeneau89,Rossat91,Cheong91,Yamada98,Dai01,Wakimoto04,Hayden04,Tranquada04,Hinkov04,Bourges05,Sidis04,He01,Bourges00,Arai99,Vignolle07,Stock05,Stock10,Xu09} as shown in Fig. \ref{peak-position}. Moreover, the spin excitations at energies well above $\omega_{\rm r}$ disperse almost linearly with energy, which is similar to spin wave with a finite gap, reflecting a fact that the charge carrier doping strongly renormalizes the spin excitations at energies below $\omega_{\rm r}$, but has a modest effect on the spin excitation dispersion at energies above $\omega_{\rm r}$ \cite{Fujita12}. However, in contrast to the case of the low-energy spin excitations, the dispersion of the high-energy spin excitations in cuprate superconductors in the overdoped regime is strikingly similar to that of their parent compounds \cite{Vojta11,Dean14,Braicovich10,Guarise10,Piazza12}.

The physical interpretation \cite{Kuang15} to the above results can be found from the special property of the spin self-energy $\Sigma^{({\rm s})}({\bf k}, \omega)$ in Eq. (\ref{SSF}) obtained directly from the interaction between charge carriers and spins in the kinetic energy of the $t$-$J$ model (\ref{CSStJmodel}). This follows from a fact that the dynamical spin structure factor $S({\bf k},\omega)$ in Eq. (\ref{DSSF}) has a well-defined resonance character, where $S({\bf k},\omega)$ exhibits peaks when the incoming neutron energy $\omega$ is equal to the spin excitation energy $E_{\rm s}({\bf k})$, i.e.,
\begin{eqnarray}
\omega^{2}-\omega_{{\bf k}_{\rm c}}^{2}-B_{{\bf k}_{\rm c}}{\rm Re}\Sigma^{({\rm s})}({\bf k}_{\rm c},\omega)=\omega^{2}-E^{2}_{\rm s}({\bf k}_{\rm c})\sim 0,
\end{eqnarray}
for certain critical wave vectors ${\bf k}_{\bf c}$, the magnetic scattering peaks appear, and then the weights of these peaks are dominated by the inverse of the imaginary part of the spin self-energy $1/{\rm Im}\Sigma^{({\rm s})}({\bf k}_{\rm c},\omega)$. In other words, the positions of magnetic scattering peaks are determined by both the spin excitation energy $E_{\rm s}({\bf k})$ and the imaginary part of the spin self-energy ${\rm Im} \Sigma^{({\rm s})}({\bf k}_{\rm c},\omega)$. At half-filling, the low-energy magnetic scattering peak locates at the AF wave vector $[1/2,1/2]$, so the commensurate AF peak appears there. However, away from half-filling, the doped charge carriers disturb the AF background. In particular, within the framework of the kinetic-energy driven SC mechanism, as a result of the self-consistent interplay between charge carriers and spins, the unusual magnetic correlation is developed. As mentioned in subsection \ref{SC-state-DSSF}, the spin self-energy $\Sigma^{({\rm s})}({\bf k},\omega)$ in Eq. (\ref{SSF}) is obtained in terms of the full charge-carrier diagonal Green's function (\ref{BCSHDGF}) and off-diagonal Green's function (\ref{BCSHODGF}), and renormalizes spin excitations. However, in the charge-carrier quasiparticle spectrum $E_{{\rm h}{\bf k}}=\sqrt{\bar{\xi}^{2}_{{\bf k}}+\mid\bar{\Delta}_{\rm hZ} ({\bf k}) \mid^{2}}$ in the full charge-carrier diagonal Green's function (\ref{BCSHDGF}) and off-diagonal Green's function (\ref{BCSHODGF}), the maximal $|\bar{\xi}_{{\bf k}}|$ appears around the nodal region, and $\bar{\xi}_{{\bf k}}$ has an effective band width $W_{\rm h}\sim 2J$ at the end of the SC dome \cite{Kuang15}. In particular, this band width decreases with decreasing doping. However, the d-wave charge-carrier pair gap $\bar{\Delta}_{\rm h}({\bf k})$ vanishes on the gap nodes, while the d-wave charge-carrier pair gap parameter $\bar{\Delta}_{\rm h}$ has a domelike shape of the doping dependence with the maximal $\bar{\Delta}_{\rm h}\sim 0.2J$ appearing around the optimal doping. These properties of $\bar{\xi}_{{\bf k}}$ and $\bar{\Delta}_{\rm h}({\bf k})$ lead to that the effective band width of the charge-carrier quasiparticle spectrum $E_{{\rm h}{\bf k}}$ is almost the same as $\bar{\xi}_{{\bf k}}$, and then the spin self-energy in Eq. (\ref{SSF}) strongly renormalizes the spin excitation at the energies below $W_{\rm h}$, but has a weak effect on the spin excitation at the energies above $W_{\rm h}$. This is why the magnetic correlation at the energies below $W_{\rm h}$ is strongly reorganized, while the high-energy spin fluctuation bears a striking resemblance to those found in the undoped parent compounds. Furthermore, as seen from the spin self-energy (\ref{SSF}), there are two parts of charge-carrier quasiparticles contribution to the spin self-energy renormalization. The contribution from the first term of the right-hand side in Eq. (\ref{SSF}) mainly comes from mobile charge-carrier quasiparticles, and the coherence factor for this process is given in Eq. (\ref{CFS1}). This process mainly leads to the low-energy IC magnetic scattering, which can persist into the normal-state. However, the additional contribution from the second term of the right-hand side in Eq. (\ref{SSF}) originates from the creation of charge-carrier pairs, and the coherence factor for this additional process is given in Eq. (\ref{CFS2}). This additional process occurs in the SC-state {\it only}, and gives a dominant contribution to the commensurate resonance \cite{Fong95}, reflecting that the commensurate resonance is intimately related to superconductivity, and then appears in the SC-state {\it only}. In particular, it is easy to find from Eqs. (\ref{DSSF}) and (\ref{SSF}) when the incoming neutron energy $\omega\sim 2\bar{\Delta}_{\rm h}$ at the AF wave vector $[1/2,1/2]$, the commensurate resonance peak appears, which leads to that $\omega_{\rm r}$ show the same domelike shape of the doping dependence as $T_{\rm c}$. This universal relationship $\omega_{\rm r}\sim 2\bar{\Delta}_{\rm h}$ between the resonance energy and charge-carrier pair gap parameter in cuprate superconductors has been confirmed by the experimental data \cite{Yu09,Fujita12,Fong95,Dai01,Bourges05,Sidis04}. The theory of the kinetic-energy driven SC mechanism thus naturally explains why the commensurate resonance is seen below $T_{\rm c}$ only, and why its doping and temperature dependences scale with the SC order parameter.

\subsection{Dynamical spin response in normal-state}\label{normal-state-DSSF}

At the temperature $T>T_{\rm c}$, the charge-carrier pair gap parameter $\bar{\Delta}_{\rm h}=0$, and superconductivity disappears, then the system is in the normal-state. In this case, the dynamical spin structure factor $S({\bf k},\omega)$ in Eq. (\ref{DSSF}) in the SC-state is reduced to that in the normal-state, where the spin self-energy is obtained in terms of the collective charge-carrier mode in the particle-hole channel {\it only}, and the calculated results are summarized as \cite{Kuang15,Yuan01,Feng98,Feng04}: (a) the commensurate resonance that originates from the creation of charge-carrier pairs and appears in the SC-state \cite{Fujita12,Fong95,Rossat91,Dai01,Bourges05,Sidis04} is absent from the normal-state, while only the low-energy IC spin fluctuation in the SC-state persists into the normal-state. Moreover, the low-energy IC magnetic scattering peaks lie on a circle of radius $\bar{\delta}_{\rm IC}$. Although some IC satellite peaks along the diagonal direction appear, the main weight of IC magnetic scattering peaks is in the parallel direction as in the case of the SC-state. Since the height of IC magnetic scattering peaks is determined by damping, the IC magnetic scattering peaks broaden and weaken in amplitude as the energy increases. In particular, the dynamical spin structure factor spectrum has been used to extract the doping dependence of the incommensurability parameter $\bar{\delta}_{\rm IC}$, and the results show clearly that $\bar{\delta}_{\rm IC}$ increases progressively with doping at the lower doped regime, but saturates at the higher doped regime, in qualitative agreement with experiments \cite{Yamada98,Enoki13}; (b) Although the high-energy spin excitations retain roughly constant energy as a function of doping, the width of these high-energy spin excitations increases with increasing doping. In particular, the high-energy spin excitations, in their overall dispersion, their spectral weight, and the shapes of the magnetic scattering peaks, are strikingly similar to those in the corresponding SC-state, although the magnetic scattering peak in the normal-state is softening and broadening, also in qualitative agreement with the experimental results \cite{Dean14,Tacon11,Dean13,Dean13a,Tacon13}; (c) the integrated dynamical spin susceptibility appears to be particularly universal, and is scaled approximately as $\propto {\rm arctan}[a_{1}\omega/T+a_{3}(\omega/T)^{3}]$, where $a_{1}$ and $a_{3}$ are constants, which is qualitatively consistent with the experiments \cite{Keimer92,Sternlieb92}.

\subsection{Summary}

In this section, we have summarized some calculated results of the dynamical spin response of cuprate superconductors obtained based on the kinetic-energy driven SC mechanism. The spin self-energy in the SC-state is evaluated explicitly in terms of the collective charge-carrier modes in the particle-hole and particle-particle channels, and then employed to calculate the dynamical spin structure factor. The calculated results show the existence of damped but well-defined dispersive spin excitations in the whole doping phase diagram. In particular, the low-energy spin excitations in the SC-state have an hour-glass-shaped dispersion, with commensurate resonance that originates from the process of the creation of charge-carrier pairs, and appears in the SC-state only, while the low-energy IC spin fluctuation is dominated by the process from mobile charge-carrier quasiparticles, and therefore can persist into the normal-state. The high-energy spin excitations in the SC-state on the other hand retain roughly constant energy as a function of doping, with spectral weights and dispersion relations comparable to those in the corresponding normal-state, although the magnetic scattering peak in the normal-state is softening and broadening. The dynamical spin response probes the local magnetic fluctuation and is a very detailed and stringent test of the microscopic theory of superconductivity. The calculated results \cite{Kuang15,Yuan01,Feng98} based on the kinetic-energy driven SC mechanism lead to the behaviors similar to that seen in the experiments.

\section{Theory of normal-state pseudogap state}\label{SC-Pseudogap}

The discovery of superconductivity in cuprate superconductors has been underlined by two salient phenomena, their high temperature superconductivity and the occurrence of the normal-state pseudogap \cite{Timusk99,Norman05,Hufner08,Hufner08b,Batlogg94,Loeser96,Warren89,Johnston89,Alloul89}. In particular, the normal-state pseudogap is particularly obvious in the underdoped regime, where the charge-carrier concentration is too low for the optimal superconductivity. A number of experimental probes \cite{Timusk99,Norman05,Hufner08,Hufner08b,Batlogg94,Loeser96,Warren89,Johnston89,Alloul89} show that below a characteristic temperature $T^{*}$, which can be well above $T_{\rm c}$ in the underdoped regime, the physical response of cuprate superconductors can be interpreted in terms of the formation of a normal-state pseudogap by which it means a suppression of the spectral weight of the low-energy excitation spectrum. Moreover, this normal-state pseudogap crossover temperature $T^{*}$ decreases with increasing doping in the underdoped regime and since $T_{\rm c}$ rises with doping, then $T^{*}$ seems to merge with $T_{\rm c}$ in the overdoped regime, eventually disappearing together with superconductivity at the end of the SC dome \cite{Timusk99,Norman05,Hufner08}, which leads to an anomalous normal-state pseudogap state at the underdoped regime and eventually a crossover to the normal-metal phase at the heavily overdoped regime.

In the early days of Mott insulators, Mott \cite{Mott68} introduced the term pseudogap to indicate a minimum in the density of states at the Fermi energy, resulting from the Coulomb repulsion between electrons at the same site. During the last two decades, several scenarios were proposed to explain the formation of the normal-state pseudogap in cuprate superconductors. In particular, it has been argued \cite{Emery95,Chen05} that the normal-state pseudogap originates from preformed pairs at $T^{*}$, which would then condense (that is, become phase coherent) at $T_{\rm c}$. On the other hand, it has been suggested that the normal-state pseudogap is distinct from the SC gap and related with a certain order which competes with superconductivity \cite{Kondo11,Wise08,Kondo09}. Furthermore, it has been proposed that the normal-state pseudogap is a combination of a quantum disordered d-wave superconductor and an entirely different form of competing order, originating from the particle-hole channel \cite{Tesanovic08}. However, up to now, no general consensus for the normal-state pseudogap has been reached yet on its origin, its role in the onset of superconductivity itself, and not even on its evolution across the phase diagram of cuprate superconductors \cite{Timusk99,Norman05,Hufner08}. In particular, the key questions surrounding the normal-state pseudogap phenomenon and its relevance to superconductivity have been raised \cite{Norman05,Hufner08}: (a) What phase diagram is the correct phase diagram with respect to the normal-state pseudogap line? (b) Is the normal-state pseudogap the result of some one-particle band structure effect? (c) Is there a true order parameter defining the existence of a normal-state pseudogap phase? (d) Do the normal-state pseudogap and SC gap coexist? (e) Is the normal-state pseudogap a necessary ingredient for superconductivity?

In section \ref{KEDM}, we \cite{Feng03,Feng06,Feng06a} have discussed the kinetic-energy driven SC mechanism, and shown that the interaction between charge carriers and spins directly from the kinetic energy by the exchange of spin excitations in higher powers of the doping concentration generates SC-state in the particle-particle channel. Based on this kinetic-energy driven SC mechanism, we \cite{Feng12,Lan13} have developed a microscopic theory of the normal-state pseudogap state, and shown that the same charge-carrier interaction arising through the exchange of spin excitations that generates the SC-state in the particle-particle channel also induces the normal-state pseudogap state in the particle-hole channel, indicating that the spin excitation plays a decisive role in formation of both the SC-state and normal-state pseudogap state. In this section, we review briefly this microscopic theory of the normal-state pseudogap state.


\subsection{Relationship between normal-state pseudogap and quasiparticle coherence}\label{SC-Pseudogap-1}

Within the framework of the kinetic-energy driven SC mechanism, the normal-state pseudogap opens due to the strong electron correlation without symmetry breaking. This follows from the fact that the charge-carrier self-energy $\Sigma^{({\rm h})}_{1}({\bf k},\omega)$ in the particle-hole channel in Eq. (\ref{PHSE}) also can be rewritten approximately as \cite{Feng12},
\begin{eqnarray}\label{PHSE-1}
\Sigma^{({\rm }h)}_{1}({\bf k},\omega)&\approx&{[2\bar{\Delta}_{\rm pg}({\bf k})]^{2}\over \omega+M_{\bf k}},
\end{eqnarray}
where $M_{\bf k}$ is the energy spectrum of $\Sigma^{({\rm h})}_{1}({\bf k},\omega)$. As in the case of the charge-carrier pair gap mentioned in subsection \ref{KEDS}, the interaction force and {\it order parameter} in the characterization of the normal-state pseudogap state have been incorporated into $\bar{\Delta}_{\rm pg}({\bf k})$, then in this case, $\bar{\Delta}_{\rm pg}({\bf k})$ is so-called the normal-state pseudogap. It should be emphasized that the equation (\ref{PHSE-1}) is an identity only in the case of $\omega =0$, however, it is a proper approximation for the case of $\omega\neq 0$. Since the normal-state pseudogap $\Sigma^{({\rm h})}_{1}({\bf k},\omega)$ originates from the charge-carrier self-energy $\Sigma^{({\rm h})}_{1}({\bf k},\omega)$ in the particle-hole channel, it can be identified as being a region of the self-energy effect in the particle-hole channel in which the normal-state pseudogap suppresses the spectral weight of the low-energy excitation spectrum. In particular, this normal-state pseudogap is directly related to the charge-carrier quasiparticle coherent weight (\ref{CCQCW}) as,
\begin{eqnarray}\label{CCQCW1}
{1\over Z_{\rm hF}}=1-{\rm Re}\Sigma^{({\rm h})}_{\rm 1o}({\bf k},\omega=0)\mid_{{\bf k}= [\pi,0]}=1+{[2\bar{\Delta}_{\rm pg}({\bf k})]^{2}\over M^{2}_{\bf k}}\mid_{{\bf k}= [\pi,0]},
\end{eqnarray}
which shows that the main effect of the normal-state pseudogap has been contained in the quasiparticle coherent weight. As a byproduct, it is therefore established a relationship (\ref{CCQCW1}) between the normal-state pseudogap $\bar{\Delta}_{\rm pg}$ and charge-carrier quasiparticle coherent weight $Z_{\rm hF}$. Since the SC-state in the kinetic-energy driven SC mechanism is controlled by both the SC gap and quasiparticle coherence as mentioned in subsection \ref{KEDSQC}, in this sense, the normal-state pseudogap is a necessary ingredient for superconductivity. However, there is a strong competition between the charge-carrier quasiparticle coherence and charge-carrier pairs \cite{Feng03,Feng06}, which leads $T_{\rm c}$ in cuprate superconductors to be reduced to lower temperatures, indicating that the normal-state pseudogap has a competitive role in engendering superconductivity.

\subsection{Interplay between superconductivity and normal-state pseudogap state}\label{SC-Pseudogap-1}

Substituting the self-energy $\Sigma^{({\rm h})}_{1}({\bf k},\omega)$ in Eq. (\ref{PHSE-1}) into Eq. (\ref{HGF}), the full charge-carrier diagonal and off-diagonal Green's functions can be obtained by considering the interplay between the SC gap and normal-state pseudogap as,
\begin{subequations}\label{HGF-1}
\begin{eqnarray}
g({\bf k},\omega)&=&{1\over\omega-\xi_{\bf k}-\Sigma^{({\rm h})}_{1}({\bf k},\omega)-\bar{\Delta}^{2}_{\rm h}({\bf k})/[\omega+\xi_{\bf k}
+\Sigma^{({\rm h})}_{1}({\bf k},-\omega)]}\nonumber\\
&=&{U^{2}_{1{\rm h}{\bf k}}\over\omega-E_{1{\rm h}{\bf k}}}+{V^{2}_{1{\rm h}{\bf k}}\over\omega+E_{1{\rm h}{\bf k}}}+{U^{2}_{2{\rm h}{\bf k}}\over\omega
-E_{2{\rm h}{\bf k}}}+{V^{2}_{2{\rm h}{\bf k}}\over\omega+E_{2{\rm h}{\bf k}}}, ~~~~~~~~\label{HDGF-1}\\
\Gamma^{\dagger}({\bf k},\omega)&=&-{\bar{\Delta}_{\rm h}({\bf k})\over [\omega-\xi_{\bf k}-\Sigma^{({\rm h})}_{1}
({\bf k},\omega)][\omega+\xi_{\bf k}+\Sigma^{({\rm h})}_{1}({\bf k},-\omega)]-\bar{\Delta}^{2}_{\rm h}({\bf k})}\nonumber\\
&=&-{\alpha_{1{\bf k}}\bar{\Delta}_{\rm h}({\bf k})\over 2 E_{1{\rm h}{\bf k}}}\left ({1\over\omega-E_{1{\rm h}{\bf k}}}-{1\over\omega+E_{1{\rm h}{\bf k}}}\right )\nonumber\\
&+&{\alpha_{2{\bf k}}\bar{\Delta}_{\rm h}({\bf k})\over 2 E_{2{\rm h}{\bf k}}}\left ({1\over\omega-E_{2{\rm h}{\bf k}}}-{1\over\omega+E_{2{\rm h}{\bf k}}}\right ), \label{HODGF-1}
\end{eqnarray}
\end{subequations}
where $\alpha_{1{\bf k}}=(E^{2}_{1{\rm h}{\bf k}}-M^{2}_{\bf k})/(E^{2}_{1{\rm h}{\bf k}}-E^{2}_{2{\rm h}{\bf k}})$,
$\alpha_{2{\bf k}}=(E^{2}_{2{\rm h}{\bf k}}-M^{2}_{\bf k})/(E^{2}_{1{\rm h}{\bf k}}-E^{2}_{2{\rm h}{\bf k}})$, and there are four branches of the charge-carrier quasiparticle spectrum due to the presence of both the normal-state pseudogap and SC gap, $E_{1{\rm h}{\bf k}}$, $-E_{1{\rm h}{\bf k}}$, $E_{2{\rm h}{\bf k}}$, and $-E_{2{\rm h}{\bf k}}$, with
$E_{1{\rm h}{\bf k}}=\sqrt{[\Omega_{\bf k}+\Theta_{\bf k}]/2}$, $E_{2{\rm h}{\bf k}}=\sqrt{[\Omega_{\bf k}-\Theta_{\bf k}]/2}$, and the functions,
\begin{subequations}
\begin{eqnarray}
\Omega_{\bf k}&=&\xi^{2}_{\bf k}+M^{2}_{\bf k}+8\bar{\Delta}^{2}_{\rm pg}({\bf k})+\bar{\Delta}^{2}_{\rm h}({\bf k}),\\
\Theta_{\bf k}&=&\sqrt{(\xi^{2}_{\bf k}-M^{2}_{\bf k})\beta_{1{\bf k}}+16\bar{\Delta}^{2}_{\rm pg}({\bf k})\beta_{2{\bf k}}+\bar{\Delta}^{4}_{\rm h}({\bf k})},
\end{eqnarray}
\end{subequations}
with $\beta_{1{\bf k}}=\xi^{2}_{\bf k}-M^{2}_{\bf k}+2\bar{\Delta}^{2}_{\rm h}({\bf k})$, $\beta_{2{\bf k}}=(\xi_{\bf k}-M_{\bf k})^{2}+\bar{\Delta}^{2}_{\rm h}({\bf k})$, while the charge-carrier coherence factors,
\begin{subequations}\label{coherence-factors}
\begin{eqnarray}
U^{2}_{1{\rm h}{\bf k}}&=&{1\over 2}\left [\alpha_{1{\bf k}}\left (1+{\xi_{\bf k}\over E_{1{\rm h}{\bf k}}}\right )-
\alpha_{3{\bf k}}\left (1+{M_{\bf k}\over E_{1{\rm h}{\bf k}}}\right )\right ],\\
V^{2}_{1{\rm h}{\bf k}}&=&{1\over 2}\left [\alpha_{1{\bf k}}\left (1-{\xi_{\bf k}\over E_{1{\rm h}{\bf k}}}\right )-
\alpha_{3{\bf k}}\left (1-{M_{\bf k}\over E_{1{\rm h}{\bf k}}}\right )\right ],\\
U^{2}_{2{\rm h}{\bf k}}&=&-{1\over 2}\left [\alpha_{2{\bf k}}\left (1+{\xi_{\bf k}\over E_{2{\rm h}{\bf k}}}\right )-
\alpha_{3{\bf k}}\left (1+{M_{\bf k}\over E_{2{\rm h}{\bf k}}}\right )\right ],\\
V^{2}_{2{\rm h}{\bf k}}&=&-{1\over 2}\left [\alpha_{2{\bf k}}\left (1-{\xi_{\bf k}\over E_{2{\rm h}{\bf k}}}\right )-
\alpha_{3{\bf k}}\left (1-{M_{\bf k}\over E_{2{\rm h}{\bf k}}}\right )\right ],
\end{eqnarray}
\end{subequations}
satisfy the sum rule $U^{2}_{1{\rm h}{\bf k}}+V^{2}_{1{\rm h}{\bf k}}+U^{2}_{2{\rm h}{\bf k}}+V^{2}_{2{\rm h}{\bf k}}=1$ for any wave vector ${\bf k}$ (normalization), where
$\alpha_{3{\bf k}}=[2\bar{\Delta}_{\rm pg}({\bf k})]^{2}/(E^{2}_{1{\rm h}{\bf k}}-E^{2}_{2{\rm h}{\bf k}})$, and the corresponding normal-state pseudogap
$\bar{\Delta}_{\rm pg}({\bf k})$ and energy spectrum $M_{\bf k}$ in Eq. (\ref{PHSE-1}) can be obtained explicitly in terms of the self-energy $\Sigma^{({\rm h})}_{1}({\bf k},\omega)$ in Eq. (\ref{PHSE-1}) as,
\begin{subequations}\label{PG}
\begin{eqnarray}
\bar{\Delta}_{\rm pg}({\bf k})&=&{L_{2}({\bf k})\over 2\sqrt{L_{1}({\bf k})}},\\
M_{\bf k}&=&{L_{2}({\bf k})\over L_{1}({\bf k})},
\end{eqnarray}
\end{subequations}
with the functions $L_{1}({\bf k})=-\Sigma^{({\rm }h)}_{\rm 1o}({\bf k},\omega=0)$ and $L_{2}({\bf k})=\Sigma^{({\rm }h)}_{1}({\bf k},\omega=0)$, and can be obtained directly from the charge-carrier self-energy $\Sigma^{({\rm h})}_{1}({\bf k}, \omega)$ in Eq. (\ref{PHSE}). In this case, it is then straightforward to obtain the normal-state pseudogap parameter from Eq. (\ref{PG}) as \cite{Feng12},
\begin{eqnarray}\label{PG-parameter}
\bar{\Delta}_{\rm pg}={1\over N}\sum_{\bf k}\bar{\Delta}_{\rm pg}({\bf k}).
\end{eqnarray}
In comparison with the charge-carrier Green's functions in Eq. (\ref{BCSHGF}), the charge-carrier Green's functions in Eq. (\ref{HGF-1}) are still BCS-like with the d-wave symmetry, although the charge-carrier quasiparticle spectrum has been split further due to the presence of the normal-state pseudogap. However, it should be noted that in spite of the origins of the normal-state pseudogap state, the main feature of the charge-carrier propagators in Eq. (\ref{HGF-1}) is very similar to these proposed from the preformed pair theory \cite{Chen05}, where the pseudogap state is associated with the preformed pairs, and then the calculated result \cite{Dan12} of the conductivity in the underdoped cuprates is consistent with the experimental observations, or from several phenomenological theories \cite{Benfatto00,Cho06,Millis06} of the normal-state pseudogap state based on the d-wave BCS formalism, where by introducing a phenomenological doping and temperature dependence of the normal-state pseudogap, the two-gap feature in cuprate superconductors is reproduced, or from a phenomenological theory \cite{Yang06,Rice12} of the normal-state pseudogap state based on the RVB theory, where an ansatz is proposed for the coherent part of the single particle Green's function in a doped RVB state, and then the calculated result of the electronic properties in the normal-state pseudogap phase is in qualitative agreement with the experimental data.

\subsection{Doping and temperature dependence of normal-state pseudogap}\label{SC-Pseudogap-2}

\begin{figure}[h!]
\centering
\includegraphics[scale=0.6]{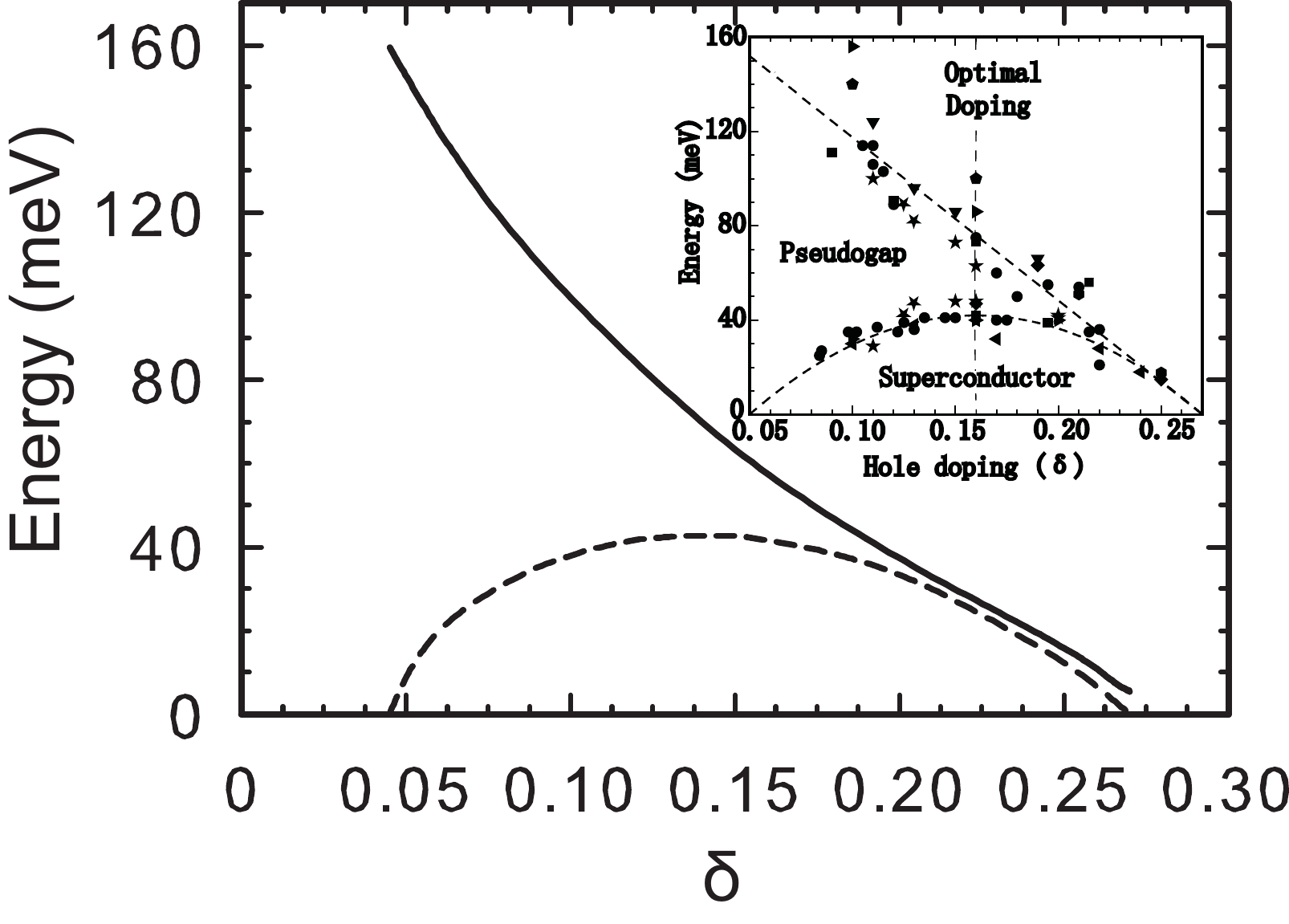}
\caption{The normal-state pseudogap parameter (2$\bar{\Delta}_{\rm pg}$) (solid line) and charge-carrier pair gap parameter ($2\bar{\Delta}_{\rm h}$) (dashed line) as a function of doping with $T=0.002J$ for $t/J=2.5$, $t'/t=0.3$, and $J=110$meV. Inset: the corresponding experimental data of cuprate superconductors taken from Ref. \cite{Hufner08}. [From Ref. \cite{Feng12}.] \label{phase-diagram-theor1}}
\end{figure}

The normal-state pseudogap parameter $\bar{\Delta}_{\rm pg}$ in Eq. (\ref{PG-parameter}) has been calculated \cite{Feng12,Lan13}, and the results of 2$\bar{\Delta}_{\rm pg}$ (solid line) and charge-carrier pair gap parameter $2\bar{\Delta}_{\rm h}$ (dashed line) as a function of doping with $T=0.002J$ for $t/J=2.5$, $t'/t=0.3$, and $J=110$meV are shown in Fig. \ref{phase-diagram-theor1} in comparison with the corresponding experimental data \cite{Hufner08} of cuprate superconductors (inset). Obviously, the two-gap feature observed on different families of cuprate superconductors \cite{Timusk99,Norman05,Hufner08,Hufner08b,Batlogg94,Loeser96,Warren89,Johnston89,Alloul89,Fischer07} is qualitatively reproduced. Although the charge-carrier pair gap parameter $\bar{\Delta}_{\rm h}$ has a domelike shape of the doping dependence, the magnitude of the normal-state pseudogap parameter $\bar{\Delta}_{\rm pg}$ is particularly large in the underdoped regime, and then smoothly decreases with increasing doping, eventually disappearing together with superconductivity at the end of the SC dome. In particular, since this normal-state pseudogap $\bar{\Delta}_{\rm pg}$ is directly related to the charge-carrier quasiparticle coherent weight $Z_{\rm hF}$ as shown in Eq. (\ref{CCQCW1}), the decrease of $\bar{\Delta}_{\rm pg}$ with increasing doping leads to that the charge-carrier quasiparticle coherent weight near the charge-carrier Fermi surface grows linearly with doping, which together with the charge-carrier pair gap parameter $\bar{\Delta}_{\rm h}$ shows that only $\delta$ number of the coherent doped carriers are recovered in the SC-state, consistent with the picture of a doped Mott insulator with $\delta$ charge carriers \cite{Anderson87,Phillips10,Johnson01,Randeria04,Fournier10}. This is much different from the case in conventional superconductors \cite{Bardeen57,Schrieffer64}, where the charge-carrier coherent weight $Z_{\rm F}\sim 1$ near the Fermi surface, since the normal-state of conventional superconductors is a standard Landau Fermi-liquid.

\begin{figure}[h!]
\centering
\includegraphics[scale=0.7]{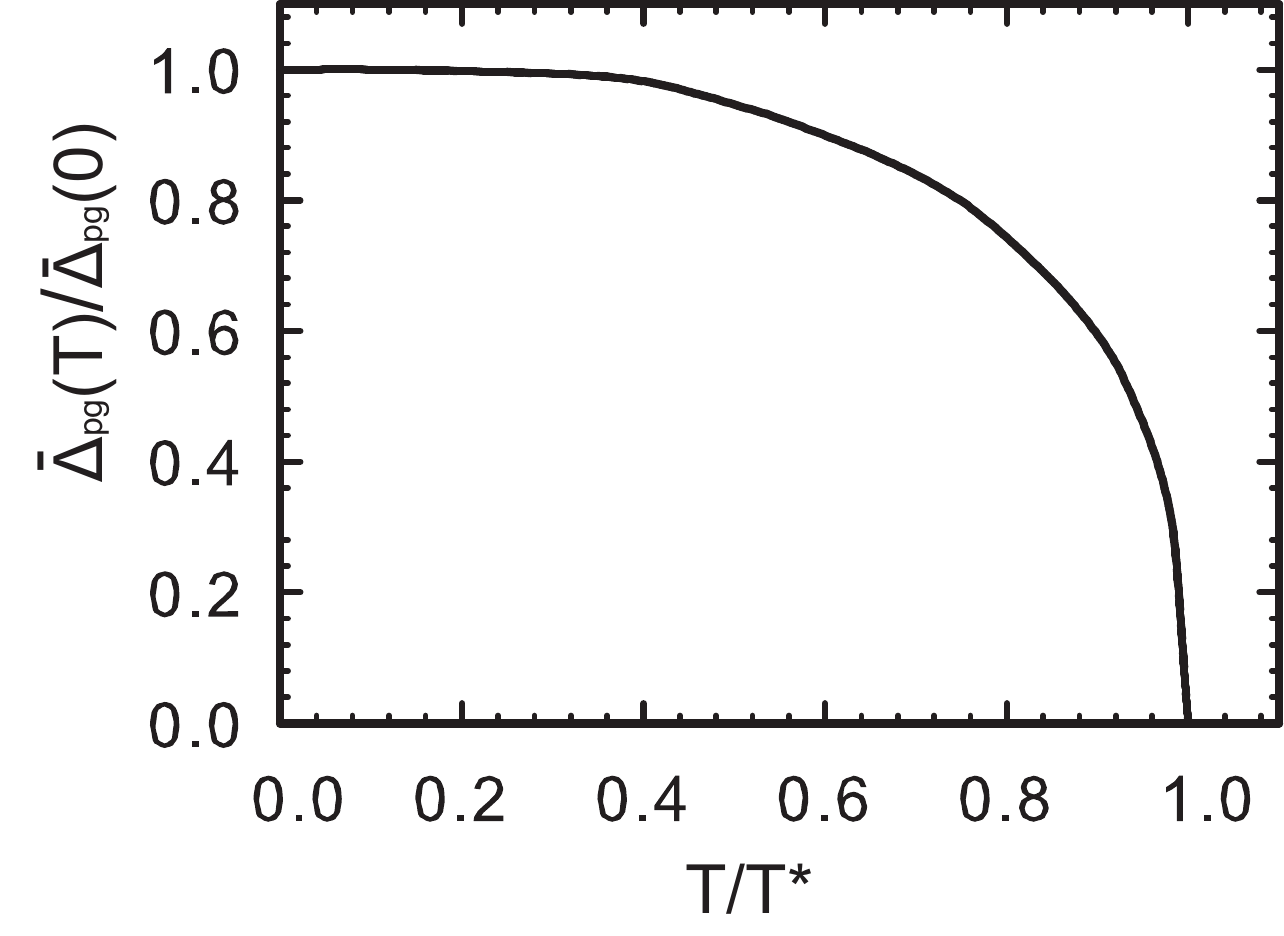}
\caption{The normal-state pseudogap parameter as a function of temperature at $\delta=0.09$ for $t/J=2.5$ and $t'/t=0.3$. [From Ref. \cite{Zhao12}.] \label{pseudogap-temp}}
\end{figure}

\begin{figure}[h!]
\centering
\includegraphics[scale=0.6]{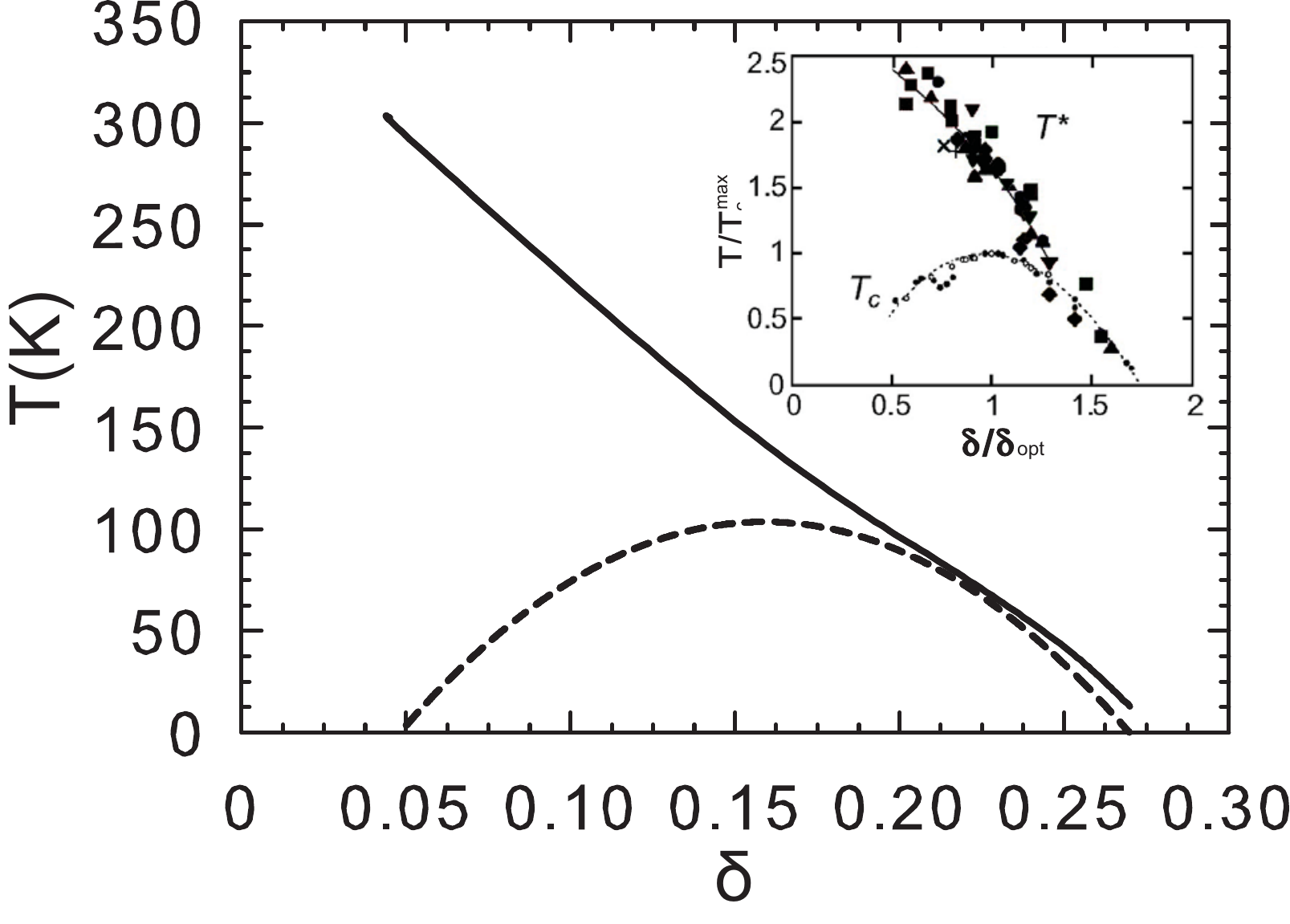}
\caption{The normal-state pseudogap crossover temperature $T^{*}$ (solid line) and superconducting transition temperature $T_{\rm c}$ (dashed line) as a function of doping for $t/J=2.5$, $t'/t=0.3$, and $J=110$meV. Inset: the corresponding experimental data of cuprate superconductors taken from Ref. \cite{Fischer07}. [From Ref. \cite{Feng12}.] \label{phase-diagram-theor2}}
\end{figure}

As in the temperature dependence of the charge-carrier pair gap (see Fig. \ref{pair-gap-parameter-temp}), the normal-state pseudogap is also temperature dependent. To show this point clearly, the result \cite{Zhao12} of the normal-state pseudogap parameter as a function of temperature at $\delta=0.09$ for $t/J=2.5$ and $t'/t=0.3$ is shown in Fig. \ref{pseudogap-temp}, where in analogy to the temperature dependence of the charge-carrier pair gap parameter (see Fig. \ref{pair-gap-parameter-temp}), the normal-state pseudogap parameter decreases with increasing temperatures, and vanishes when temperature reaches the normal-state pseudogap crossover temperature $T^{*}$, then the system crossovers from the strongly correlated normal-state pseudogap state at the temperatures $T<T^{*}$ to the normal-metal phase with largely transport properties at the temperatures $T>T^{*}$. Furthermore, this normal-state pseudogap crossover temperature $T^{*}$ is also doping dependent, and the results \cite{Feng12} of $T^{*}$ (solid line) and $T_{\rm c}$ (dashed line) as a function of doping for $t/J=2.5$, $t'/t=0.3$, and $J=110$meV are shown in Fig. \ref{phase-diagram-theor2} in comparison with the experimental data obtained from different families of cuprate superconductors \cite{Fischer07}. In corresponding to the results of the doping dependence of $\bar{\Delta}_{\rm pg}$ and $\bar{\Delta}_{\rm h}$ in Fig. \ref{phase-diagram-theor1}, $T^{*}$ is much larger than $T_{c}$ in the underdoped regime, and then smoothly decreases with increasing doping. Moreover, both $T^{*}$ and $T_{c}$ converge to the end of the SC dome, in qualitative agreement with the experimental results \cite{Timusk99,Norman05,Hufner08,Hufner08b,Batlogg94,Loeser96,Warren89,Johnston89,Alloul89,Fischer07}. In particular, in comparison with the doping dependence of the coupling (interaction) strength $V_{\rm eff}$ in Fig. \ref{V-pair-order-parameter-doping}(a), it is therefore found \cite{Feng12} that all $T^{*}$, $\bar{\Delta}_{\rm pg}$, and $V_{\rm eff}$ show the same doping dependence, i.e., $T^{*}\sim \bar{\Delta}_{\rm pg}\sim V_{\rm eff}$, while such a relationship among $T^{*}$, $\bar{\Delta}_{\rm pg}$, and $V_{\rm eff}$ has been confirmed experimentally on cuprate superconductors \cite{Kordyuk10,Johnson01}.

The results \cite{Feng12,Lan13} in Fig. \ref{phase-diagram-theor1} and Fig. \ref{phase-diagram-theor2} show clearly that there are two coexisting energy gaps in the whole SC dome: one associated with a direct measurement of the binding energy of the two charge carriers forming a charge-carrier pair, while the other with a suppression of the spectral weight of the low-energy excitation spectrum. The charge carriers interact by the exchange of spin excitations in both the particle-particle and particle-hole channels, with the coupling (interaction) strength $V_{\rm eff}$ that falls linearly with doping from a strong-coupling case in the underdoped regime to a weak-coupling side in the overdoped regime (see Fig. \ref{V-pair-order-parameter-doping}a). On the other hand, the normal-state pseudogap $\bar{\Delta}_{\rm pg}({\bf k})$ originates from the self-energy in the particle-hole channel due to the charge-carrier interaction, which leads to that the normal-state pseudogap parameter $\bar{\Delta}_{\rm pg}$ and normal-state pseudogap crossover temperature $T^{*}$ follow qualitatively a doping dependence in like manner of the coupling strength $V_{\rm eff}$. In particular, since the charge-carrier interactions in both the particle-hole and particle-particle channels are mediated by the same spin excitations as we have shown in Eq. (\ref{Interaction}), all these charge-carrier interactions in the particle-hole and particle-particle channels are controlled by the same magnetic exchange coupling $J$. In this sense, both the normal-state pseudogap and SC gap in the phase diagram of the cuprate superconductors are dominated by one energy scale. Moreover, the theory of the kinetic-energy driven superconductivity starting from the $t$-$J$ model (\ref{CSStJmodel}) also shows that both the normal-state pseudogap and SC gap in the cuprate superconductors are the result of the strong electron correlation, a generic consequence of the strong Coulomb interaction in the large-$U$ Hubbard model (\ref{Hubbard}), and they have the same magnetic origin.

\subsection{Summary}

In this section, we have given a brief review of the microscopic theory of the normal-state pseudogap state in cuprate superconductors. The charge-carrier interaction arising through the exchange of spin excitations that generates charge-carrier pair state in the particle-particle channel also induces the normal-state pseudogap state in the particle-hole channel, therefore the kinetic-energy driven SC mechanism provides a natural explanation of both the origin of the normal-state pseudogap state and the SC mechanism for superconductivity in cuprate superconductors. In this microscopic theory of the normal-state pseudogap state, the normal-state pseudogap has been identified as being a region of the self-energy effect in the particle-hole channel in which the normal-state pseudogap suppresses the spectral weight of the low-energy excitation spectrum. Furthermore, this microscopic theory of the normal-state pseudogap state also indicates that (a) there is a true {\it order parameter} defining the existence of the normal-state pseudogap phase; (b) there is a coexistence of the SC gap and normal-state pseudogap in the whole SC dome; (c) both the normal-state pseudogap and the SC gap are dominated by one energy scale, and they are the result of the strong electron correlation; (d) the normal-state pseudogap is directly related to the quasiparticle coherence, and therefore antagonizes superconductivity; (e) the correct phase diagram with respect to the normal-state pseudogap line is that the normal-state pseudogap is much larger than that of the SC gap in the underdoped regime, and then it merges gradually with the SC gap in the overdoped regime, eventually disappearing together with superconductivity at the end of the SC dome.

\section{Charge transport}\label{charge-transport}

Since the discovery of superconductivity in cuprte superconductors \cite{Bednorz86,Wu87,Schilling93}, a significant body of reliable and reproducible data has been accumulated by using many probes  \cite{Kastner98,Timusk99,Norman05,Hufner08,Hufner08b,Batlogg94,Loeser96,Warren89,Johnston89,Alloul89}, which shows that the most remarkable expression of the nonconventional physics is found in the normal-state. The normal-state properties in the underdoped and optimally doped regimes exhibit a number of anomalous properties in sense that they do not fit in with the standard Landau Fermi-liquid theory. However, there is mounting evidence that the anomalous normal-state properties in the underdoped and optimally doped regimes is dominated by the normal-state pseudogap \cite{Kastner98,Timusk99,Norman05,Hufner08,Hufner08b,Batlogg94,Loeser96,Warren89,Johnston89,Alloul89}. Among the striking features of the normal-state properties in the underdoped and optimally doped regimes, the physical quantity which most evidently displays the signature for the normal-state pseudogap is the charge transport \cite{Timusk99,Homes04,Schlesinger90,Orenstein90,Uchida91,Puchkov96,Puchkov96a,Basov96,Lee05,Basov05,Hwang07,Mirzaei13,Cooper09}, which is manifested by the conductivity and resistivity. The optical studies of the quasiparticle excitations have revealed much about the nature of the charge carriers in cuprate superconductors. In particular, the normal-state pseudogap can be observed directly by the infrared measurements of the conductivity. Experimentally, it has been shown in terms of the Kramers-Kronig analysis of the reflectance that the conductivity is rather universal within the whole cuprate superconductors \cite{Timusk99,Homes04,Schlesinger90,Orenstein90,Uchida91,Puchkov96,Puchkov96a,Basov96,Lee05,Basov05,Hwang07,Mirzaei13,Cooper09}, where a key feature is the two-component conductivity: a narrow band centered around energy $\omega\sim 0$ followed by a broadband centered in the midinfrared region in the underdoped and optimally doped regimes. The conductivity in the underdoped and optimally doped regimes shows a non-Drude behavior (the conductivity decays as $\rightarrow 1/\omega$) at low energies, and is carried by $\delta$ charge carriers, while the midinfrared spectral weight is biased towards the low-energy region with the increase of doping. In particular, this two-component conductivity extends to the normal-state pseudogap boundary in the phase diagram at $T^{*}$ \cite{Lee05,Basov05,Hwang07,Mirzaei13}.

Within the framework of the kinetic-energy driven SC mechanism, the doping and temperature dependence of the conductivity have been studied in the whole doping range from the underdoped to heavily overdoped by considering the effect of the normal-state pseudogap \cite{Feng04,Feng97,Yuan03,Qin14,Qin14a}, and the result shows that the part of the low-energy spectral weight of the conductivity spectrum in the underdoped and optimally doped regimes is transferred to the higher energy region to form the unusual midinfrared band, however, the onset of the region to which the spectral weight is transferred, is always close to the normal-state pseudogap.

\subsection{Linear response theory}

Through the standard linear response theory \cite{Mahan81}, the finite-frequency conductivity of cuprate superconductors can be expressed as \cite{Feng97,Yuan03,Qin14,Qin14a},
\begin{eqnarray}\label{conductivity-1}
\sigma(\omega)=-{{\rm Im}\Pi(\omega)\over\omega},
\end{eqnarray}
with the electron current-current correlation function,
\begin{eqnarray}\label{correlation}
\Pi(\tau-\tau')=-\langle T_{\tau}{\bf j}(\tau)\cdot {\bf j}(\tau')\rangle,
\end{eqnarray}
where the electron current operator ${\bf j}$ is obtained by evaluating the time-derivative of the polarization operator (\ref{poloper}), and has been given explicitly in Eqs. (\ref{tcurpara15}) and (\ref{tcurpara16}) as,
\begin{eqnarray}\label{current}
{\bf j}=-{ie\chi_{1}t\over\hbar}\sum\limits_{l\hat{\eta}\sigma}\hat{\eta}h^{\dagger}_{l+\hat{\eta}\sigma}h_{l\sigma}+{ie\chi_{2}t'\over\hbar}\sum\limits_{l\hat{\tau}\sigma}
\hat{\tau}h^{\dagger}_{l+\hat{\tau}\sigma}h_{l\sigma}.  \label{current}
\end{eqnarray}
In the normal-state, the electron current-current correlation function is evaluated in terms of the full charge-carrier Green's function as,
\begin{eqnarray}\label{correlation-1}
\Pi(i\omega_{n})&=&-{1\over 2}(Ze)^{2}{1\over N}\sum_{\bf k}\gamma^{2}_{{\rm s}{\bf k}}{1\over\beta}\sum_{i\omega_{n'}}g({\bf k},i\omega_{n'}+i\omega_{n})g({\bf k},i\omega_{n'}),
\end{eqnarray}
with the current vertex,
\begin{eqnarray}
\gamma^{2}_{{\rm s}{\bf k}}={1\over 4}[(\chi_{1}t-2\chi_{2}t'\cos k_{y})^{2}\sin^{2}k_{x}+(\chi_{1}t-2\chi_{2}t'\cos k_{x})^{2}\sin^{2}k_{y}],
\end{eqnarray}
while the full charge-carrier Green's function in the normal-state can be obtained in terms of the full charge-carrier diagonal Green's function (\ref{HDGF-1}) in the condition of the charge-carrier pair gap $\bar{\Delta}_{\rm h}=0$, and has been evaluated explicitly as \cite{Qin14},
\begin{eqnarray}\label{normal-state-HGF}
g({\bf k},\omega)={\alpha^{\rm (n)}_{1{\bf k}}\over\omega-E^{+}_{{\rm h}{\bf k}}}+{\alpha^{\rm (n)}_{2{\bf k}}\over\omega -E^{-}_{{\rm h}{\bf k}}},
\end{eqnarray}
where there are two branches of the charge-carrier quasiparticle spectrum due to the presence of the normal-state pseudogap,
\begin{subequations}
\begin{eqnarray}
E^{+}_{{\rm h}{\bf k}}={1\over 2}\left [\xi_{{\bf k}}-M_{\bf k}+ \sqrt{(\xi_{{\bf k}}+M_{\bf k})^{2}+16\bar{\Delta}^{2}_{\rm pg}({\bf k})}\right ],\\
E^{-}_{{\rm h}{\bf k}}={1\over 2}\left [\xi_{{\bf k}}-M_{\bf k}-\sqrt{(\xi_{{\bf k}}+M_{\bf k})^{2}+ 16\bar{\Delta}^{2}_{\rm pg}({\bf k})}\right ],
\end{eqnarray}
\end{subequations}
while $\alpha^{\rm (n)}_{1{\bf k}}=(E^{+}_{{\rm h}{\bf k}}+M_{\bf k})/(E^{+}_{{\rm h}{\bf k}}-E^{-}_{{\rm h}{\bf k}})$ and $\alpha^{\rm (n)}_{2{\bf k}}=-(E^{-}_{{\rm h}{\bf k}} +M_{\bf k})/(E^{+}_{{\rm h}{\bf k}}-E^{-}_{{\rm h}{\bf k}})$ satisfy the sum rule $\alpha^{({\rm n})}_{1{\bf k}}+\alpha^{({\rm n})}_{2{\bf k}}=1$ for any wave vector ${\bf k}$. In this case, the conductivity in the normal-sate is obtained in terms of the full charge-carrier Green's function (\ref{normal-state-HGF}) as,
\begin{eqnarray}\label{normal-state-conductivity}
\sigma (\omega)=\left ({Ze\over 2}\right )^{2}{1\over N}\sum_{\bf k}\gamma_{{\rm s}{\bf k}}^{2}\int^{\infty}_{-\infty}{{\rm d}\omega'\over 2\pi}A_{\rm h}({\bf k},\omega'+\omega) A_{\rm h}({\bf k},\omega'){n_{\rm F}(\omega'+\omega)-n_{\rm F}(\omega')\over\omega}, ~~~~~
\end{eqnarray}
with the charge-carrier spectral function $A_{\rm h}({\bf k},\omega)=-2{\rm Im}g({\bf k},\omega)$. In the fermion-spin theory, although both charge carriers and spins contribute to the charge dynamics, the results in Eqs. (\ref{current}) and (\ref{normal-state-conductivity}) show that the anomalous conductivity properties are mainly caused by charge carriers \cite{Feng97,Yuan03,Qin14,Qin14a}, which are strongly renormalized because of the strong interaction with the fluctuation of the surrounding spin excitations, and then the low-energy spectral weight of the conductivity spectrum is proportional to the charge-carrier doping concentration $\delta$ \cite{Orenstein90,Uchida91}.

\subsection{Doping dependence of conductivity in normal-state}\label{normal-conductivity}

\begin{figure}[h!]
\centering
\includegraphics[scale=0.4]{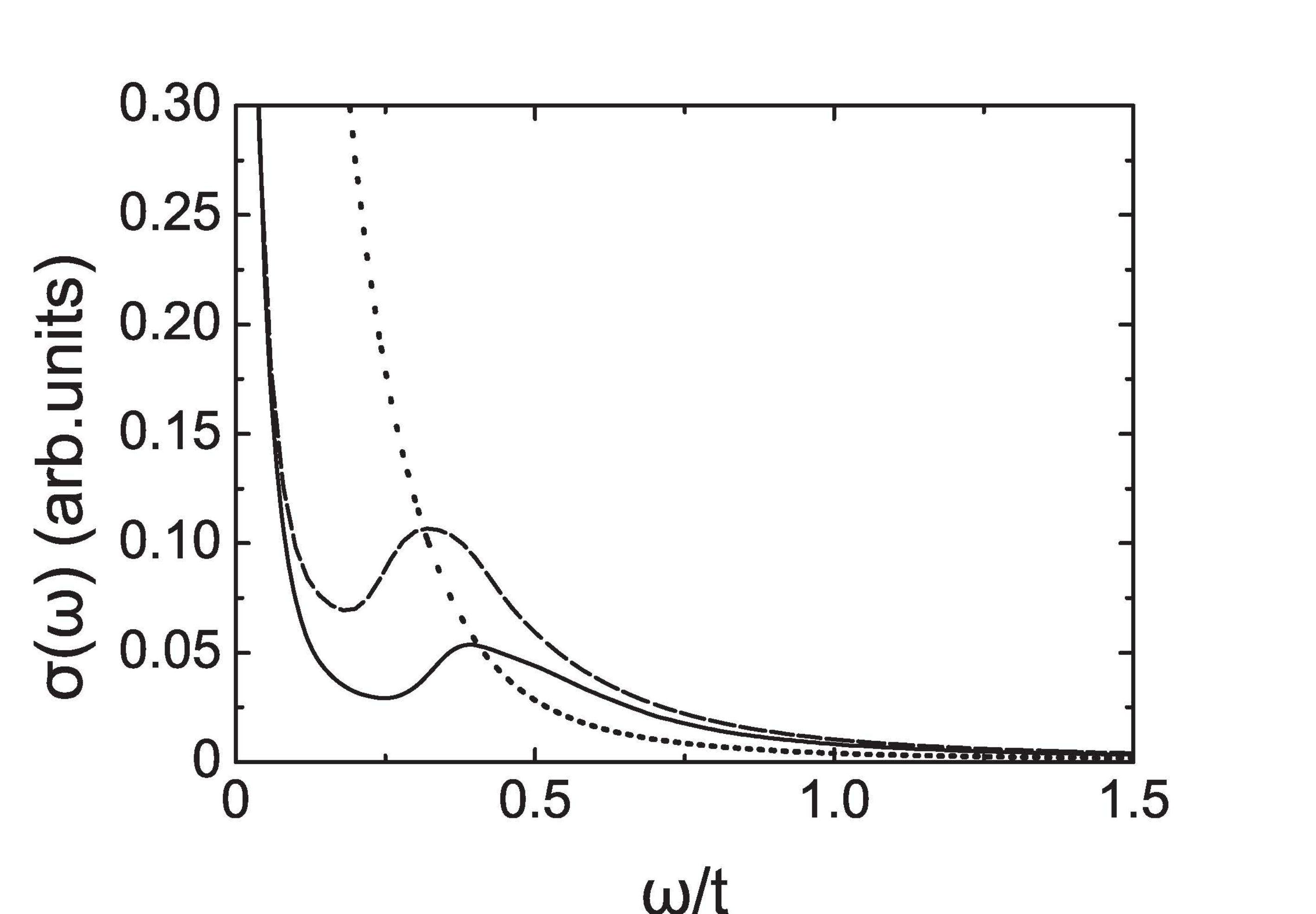}
\caption{The conductivity in the normal-sate as a function of energy at $\delta=0.09$ (solid line), $\delta=0.15$ (dashed line), and $\delta=0.25$ (dotted line) with $T=0.002J$ for $t/J=2.5$ and $t'/t=0.3$. [From Ref. \cite{Qin14}.] \label{norma-state-conductivity-doping}}
\end{figure}

\begin{figure}[h!]
\centering
\includegraphics[scale=0.4]{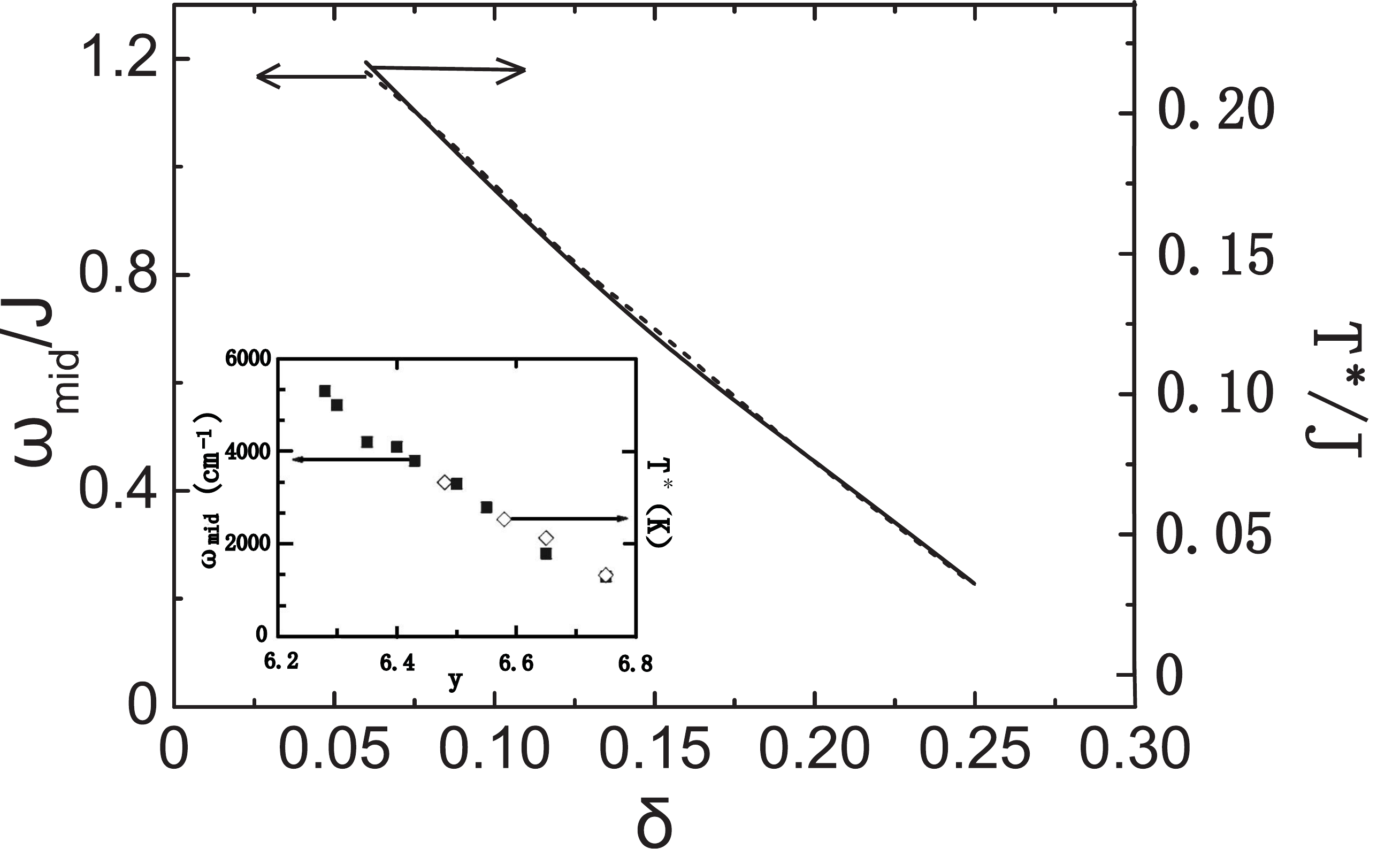}
\caption{The position of the midinfrared peak (dashed line) and $T^{*}$ (solid line) as a function of doping with $T=0.002J$ for $t/J=2.5$ and $t'/t=0.3$. Inset: the corresponding experimental data of YBa$_{2}$Cu$_{3}$O$_{y}$ taken from Ref. \cite{Lee05}.\label{midb-temp}}
\end{figure}

The conductivity $\sigma (\omega)$ in Eq. (\ref{normal-state-conductivity}) in the normal-state has been calculated \cite{Feng97,Yuan03,Qin14}, and the result \cite{Qin14} of  $\sigma (\omega)$ as a function of energy at $\delta=0.09$ (solid line), $\delta=0.15$ (dashed line), and $\delta=0.25$ (dotted line) with $T=0.002J$ for $t/J=2.5$ and $t'/t=0.3$ is shown in Fig. \ref{norma-state-conductivity-doping}, hereafter in this section we set the charge $e$ and reduced Planck constant $\hbar$ as the unity. Obviously, this calculated result captures all qualitative features of the doping dependence of the conductivity observed experimentally on cuprate superconductors in the normal-state \cite{Timusk99,Homes04,Schlesinger90,Orenstein90,Uchida91,Puchkov96,Puchkov96a,Basov96,Lee05,Basov05,Hwang07,Mirzaei13,Cooper09}. In the underdoped regime, there are two bands in the conductivity $\sigma(\omega)$ separated by a gap at $\omega\sim 0.2t$. The higher energy band, corresponding to the {\it midinfrared band}, shows a broad peak at $\omega\sim 0.38t$. In particular, the transferred weight of the low-energy band forms a sharp peak at $\omega\sim 0$, which can be described formally by the non-Drude formula, while the onset of the region to which the spectral weight is transferred, is always close to the normal-state pseudogap $\bar{\Delta}_{\rm pg}$, reflecting a fact that due to the presence of the normal-state pseudogap in cuprate superconductors, the part of the low-energy spectral weight of the conductivity spectrum in the normal-state in the underdoped regime is transferred to the higher energy region to form the unusual midinfrared band. In the other words, the appearance of the higher energy midinfrared band is closely related to the effect of the normal-state pseudogap on the infrared response in cuprate superconductors \cite{Lee05,Hwang07,Mirzaei13,Yu08,Hwang08}. In a given doping, the spectral weight is proportional to the area under the conductivity curve in Fig. \ref{norma-state-conductivity-doping}. However, the weight and position of the midinfrared band are strongly doping dependent. In particular, the result in Fig. \ref{norma-state-conductivity-doping} indicates that as the charge-carrier doping increases, although the overall conductivity increases, the magnitude of the gap in the conductivity spectrum decreases, and then the higher energy midinfrared band moves towards the low-energy non-Drude band. In the optimal doping, although two band features are still apparent, the positions of the gap in the conductivity spectrum and midinfrared peak appreciably shift towards the lower energies at $\omega\sim 0.16t$ and $\omega\sim 0.3t$, respectively, reflecting a tendency that with increasing doping, the magnitude of the gap in the conductivity spectrum decreases, while the midinfrared band moves towards the low-energy non-Drude band. However, as in the case in the underdoped regime, the low-energy peak in the optimal doping still shows the non-Drude formula. This follows from a fact that the result of the conductivity spectrum in the optimal doping $\delta=0.15$ has been fitted \cite{Qin14}, and the fitted result shows that the lower-energy peak decay perfectly as $\rightarrow A/\omega$,, with $A\sim 0.01$. On the other hand, the tendency of the decrease of the magnitude of the gap in the conductivity spectrum and the midinfrared band moving towards to the low-energy non-Drude band with increasing doping is particularly obvious in the overdoped regime. In particular, the low-energy non-Drude peak incorporates with the midinfrared band in the heavily overdoped regime, and then the midinfrared feature disappears, which leads to that the low-energy Drude type behavior of the conductivity recovers, and then in contrast to the case in the underdoped and optimally doped regimes, the lower-energy peak \cite{Qin14} decay as $\rightarrow A/(\omega^{2}+B)$ in the heavily overdoped regime, with $A\sim 0.01$ and $B\sim 0.0045$. In section \ref{SC-Pseudogap}, it has been shown that the magnitude of the normal-state pseudogap (then $T^{*}$) is particularly large in the underdoped regime, and then smoothly decreases upon increase of doping. The calculated result of the doping dependence of the conductivity in Fig. \ref{norma-state-conductivity-doping} also implies that the onset of the region to which the spectral weight is transferred (then the midinfrared peak) shows the same trend with doping in like manner of the doping dependence of the normal-state pseudogap (then $T^{*}$). To show this point clearly, we plot the position of the midinfrared peak (dashed line) and $T^{*}$ (solid line) as a function of doping with $T=0.002J$ for $t/J=2.5$ and $t'/t=0.3$ in Fig. \ref{midb-temp} in comparison with the corresponding experimental data \cite{Lee05} of YBa$_{2}$Cu$_{3}$O$_{y}$ (inset). This calculated result in Fig. \ref{midb-temp} is very well consistent with the experimental data observed from the conductivity measurements \cite{Timusk99,Lee05}, and therefore confirm the effect of the normal-state pseudogap on the infrared response.

\begin{figure}[h!]
\centering
\includegraphics[scale=0.4]{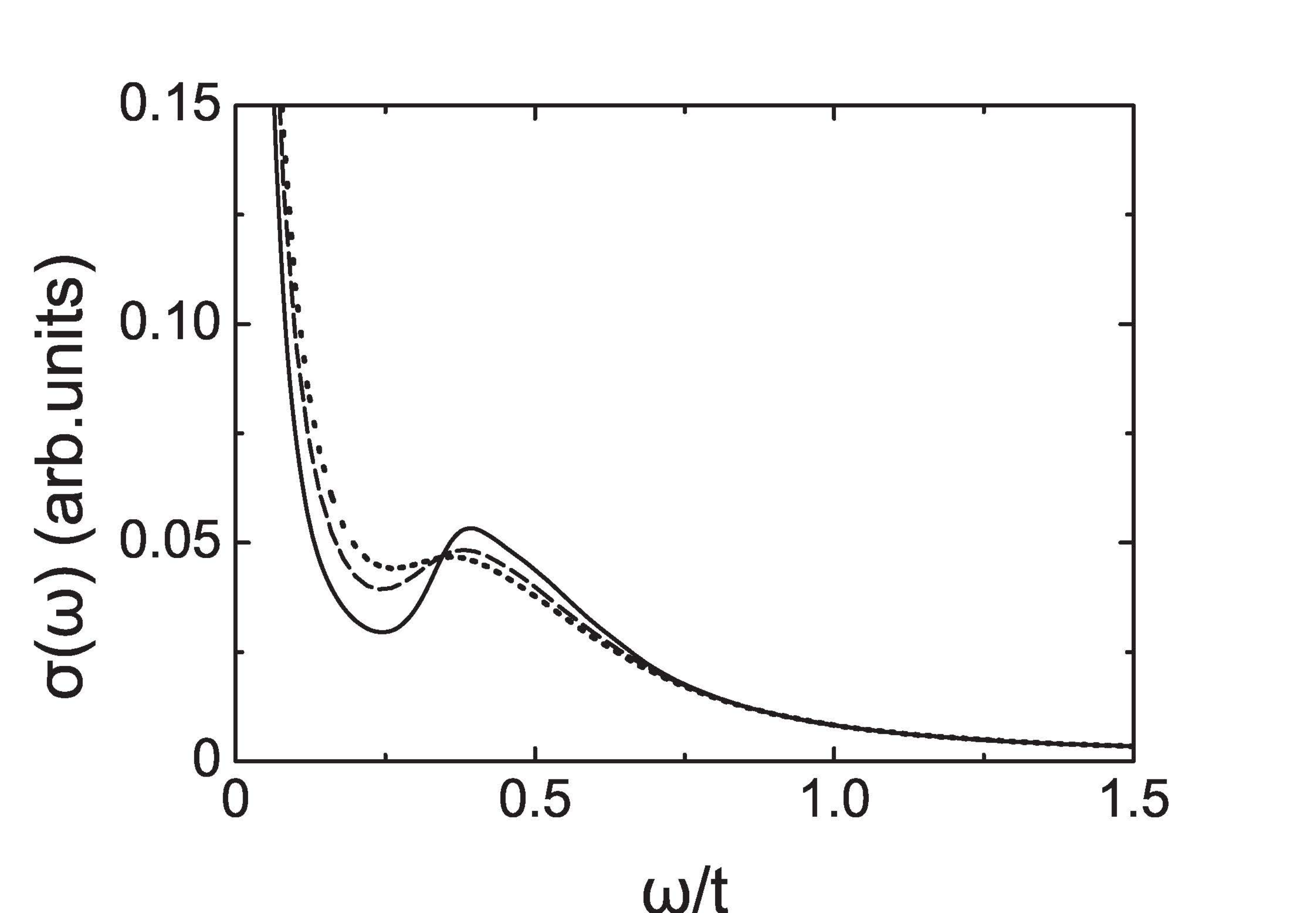}
\caption{The conductivity in the normal-sate as a function of energy at $\delta=0.09$ with $T=0.02J$ (solid line), $T=0.146J$ (dashed line), and $T=0.186J$ (dotted line) for $t/J=2.5$ and $t'/t=0.3$. [From Ref. \cite{Qin14}.] \label{norma-state-conductivity-temp}}
\end{figure}

The low-energy non-Drude peak and unusual midinfrared band of the conductivity spectrum in the normal-state in the underdoped and optimally doped regimes are also temperature dependence. In Fig. \ref{norma-state-conductivity-temp}, we \cite{Qin14} show the conductivity in the normal-state as a function of energy at $\delta=0.09$ with $T=0.02J$ (solid line), $T=0.146J$ (dashed line), and $T=0.186J$ (dotted line) for $t/J=2.5$ and $t'/t=0.3$. In section \ref{SC-Pseudogap}, the calculated normal-state pseudogap crossover temperature is $T^{*}\sim 0.19J$ at $\delta=0.09$. The result in Fig. \ref{norma-state-conductivity-temp} shows that the weight of the midinfrared band is severely suppressed with increasing temperatures. In particular, the weight of the midinfrared band vanishes above the temperatures $T>T^{*}$, and then the low-energy Drude type behavior of the conductivity is recovered, which is also qualitatively consistent with the experimental data observed in cuprate superconductors in the normal-state \cite{Timusk99,Schlesinger90,Puchkov96,Puchkov96a,Basov96,Orenstein90,Uchida91,Homes04}.

\subsection{Doping dependence of conductivity in superconducting-state}\label{SC-state-conductivity}

The discussions of the conductivity of cuprate superconductors in the normal-state in subsection \ref{normal-conductivity} can also be generalized to the SC-state in the condition of the charge-carrier pair gap $\bar{\Delta}_{\rm h}\neq 0$. In the SC-state, the electron current-current correlation function can be obtained in terms of the full charge-carrier diagonal and off-diagonal Green's functions (\ref{HGF-1}) as  \cite{Qin14a},
\begin{eqnarray}\label{correlation-1}
\Pi(i\omega_{n})&=&-{1\over 2}(Ze)^{2}{1\over N}\sum_{\bf k}\gamma^{2}_{{\rm s}{\bf k}}{1\over\beta}\sum_{i\omega_{n'}}[g({\bf k},i\omega_{n'}+i\omega_{n})g({\bf k},i\omega_{n'}) \nonumber\\
&+&\Gamma({\bf k},i\omega_{n'}+i\omega_{n})\Gamma^{\dagger}({\bf k},i\omega_{n'})],
\end{eqnarray}
and then the finite-frequency conductivity (\ref{conductivity-1}) of cuprate superconductors in the SC-state can be evaluated explicitly as \cite{Qin14a},
\begin{eqnarray}\label{conductivity}
\sigma(\omega) &=& \left ({Ze\over \hbar}\right )^{2}{1\over N}\sum_{\bf k}\gamma^{2}_{{\rm s}{\bf k}}\int^{\infty}_{-\infty}{{\rm d}\omega'\over 2\pi}[A_{g}({\bf k},\omega+\omega') A_{g}({\bf k},\omega')+A_{\Gamma}({\bf k},\omega+\omega')A_{\Gamma}({\bf k},\omega')]\nonumber\\
&\times&{n_{\rm F}(\omega')-n_{\rm F}(\omega+\omega')\over\omega},
\end{eqnarray}
where the spectral functions $A_{g}({\bf k},\omega)$ and $A_{\Gamma}({\bf k},\omega)$ are obtained in terms of the charge-carrier diagonal and off-diagonal Green's functions in Eq. (\ref{HGF-1}) as $A_{g}({\bf k},\omega)=-2{\rm Im}g({\bf k},\omega)$ and $A_{\Gamma}({\bf k},\omega) =-2{\rm Im}\Gamma^{\dagger}({\bf k},\omega)$, respectively.

\begin{figure}[h!]
\centering
\includegraphics[scale=0.4]{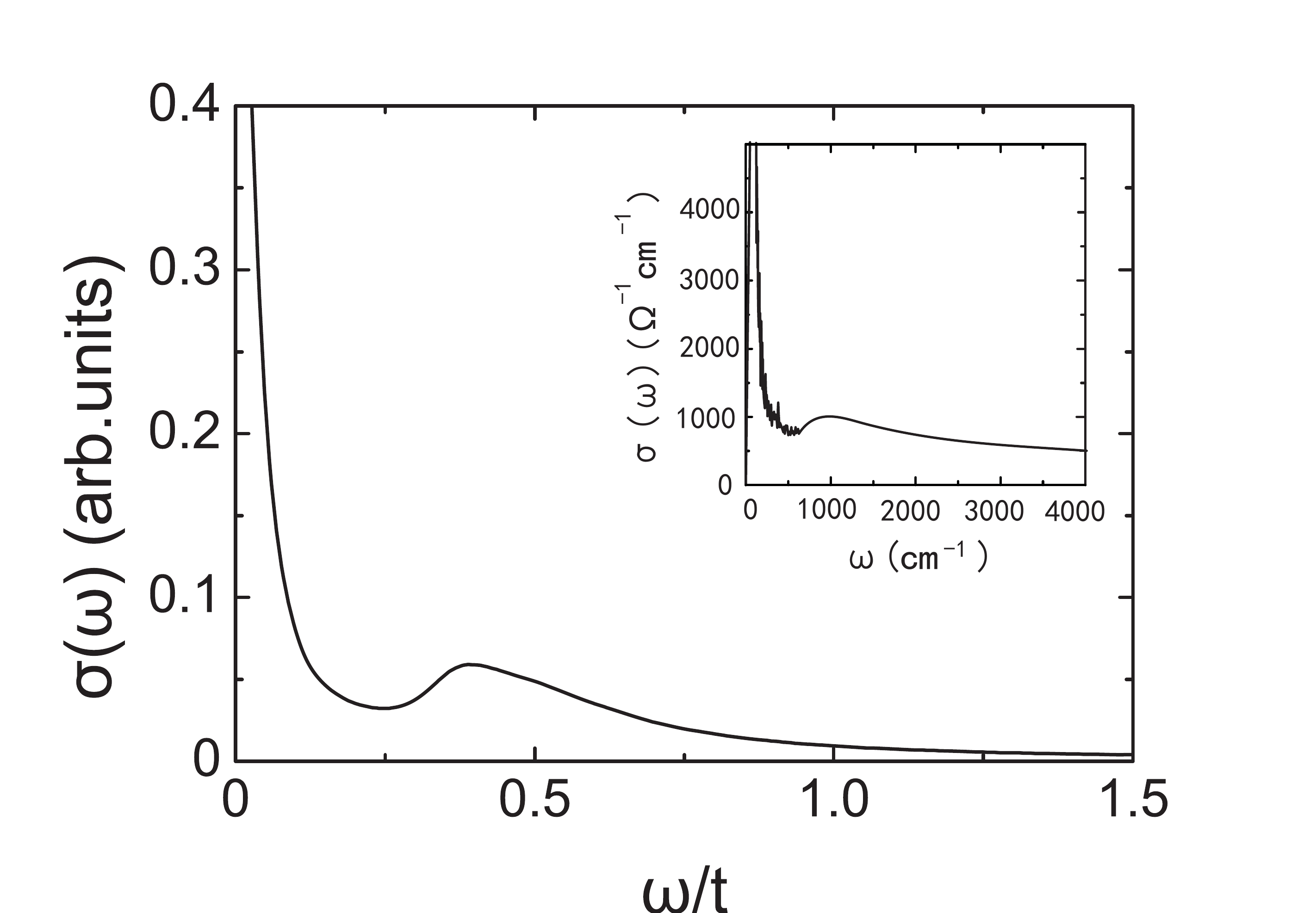}
\caption{The conductivity in the SC-state as a function of energy at $\delta=0.09$ with $T=0.002J$ for $t/J=2.5$ and $t'/t=0.3$. Inset: the corresponding experimental data of the underdoped Bi$_{2}$Sr$_{2}$CaCu$_{2}$O$_{8+\delta}$ taken from Ref. \cite{Hwang07}. [From Ref. \cite{Qin14a}.] \label{SC-conductivity}}
\end{figure}

In Fig. \ref{SC-conductivity}, we \cite{Qin14a} show the calculated result of the conductivity (\ref{conductivity}) in the SC-state as a function of energy at $\delta=0.09$ for $t/J=2.5$ and $t'/t=0.3$ with $T=0.002J$ in comparison with the corresponding experimental result \cite{Hwang07} of the underdoped Bi$_{2}$Sr$_{2}$CaCu$_{2}$O$_{8+\delta}$ (inset). The result in Fig. \ref{SC-conductivity} shows clearly that the two-component feature of the conductivity spectrum in the SC-state is the same as in the normal-state case. In particular, in comparison with the result of the conductivity in the normal-state in subsection \ref{normal-conductivity}, it is found that the spectral weight of the low-energy component of the conductivity in the SC-state is further suppressed by the SC gap, however, there is no depletion of the spectral weight of the higher energy midinfrared band of the conductivity in the SC-state, which is consistent with the experimental observation on cuprate superconductors \cite{Lee05}. In this case, although there is a coexistence of the SC gap and normal-state pseudogap, the onset of the region to which the spectral weight is transferred is also close to the normal-state pseudogap $\bar{\Delta}_{\rm pg}$, then in analogy to the evolution of the conductivity spectrum with doping in the normal-state, the positions of the gap in the conductivity spectrum and midinfrared peak gradually shift to the lower energies with increasing doping \cite{Qin14a}. All the calculated results \cite{Qin14a} of the conductivity spectrum in the SC-state are also qualitatively consistent with the corresponding experimental data of cuprate superconductors in the SC-state \cite{Timusk99,Schlesinger90,Puchkov96,Puchkov96a,Basov96,Orenstein90,Uchida91,Homes04}.

In the SC-state, there are two parts of the charge-carrier quasiparticle contribution to the redistribution of the spectral weight in the conductivity spectrum in the SC-state: the contribution from the first term of the right-hand side in Eq. (\ref{conductivity}) comes from the spectral function obtained in terms of the charge-carrier diagonal Green's function (\ref{HDGF-1}), and therefore is closely associated with the normal-state pseudogap $\bar{\Delta}_{\rm pg}$ in the particle-hole channel, while the additional contribution from the second term of the right-hand side in Eq. (\ref{conductivity}) originates from the spectral function obtained in terms of the charge-carrier off-diagonal Green's function (\ref{HODGF-1}), and is closely related to the charge-carrier pair gap $\bar{\Delta}_{\rm h}$ in the particle-particle channel. However, since $\bar{\Delta}_{\rm h}\ll \bar{\Delta}_{\rm pg}$ in the underdoped and optimally doped regimes as we have mentioned in section \ref{SC-Pseudogap}, the charge-carrier pair gap only suppresses the spectral weight of the low-energy component, while the normal-state pseudogap related shift of the spectral weight from the low-energy to the higher energy midinfrared band in the SC-state conductivity spectrum becomes arrested.

The effect of the normal-state pseudogap on the infrared response in cuprate superconductors has been also discussed based on the preformed pair theory \cite{Dan12} and the phenomenological theory of the normal-state pseudogap state \cite{Illes09}, and the results of the unusual two-component conductivity spectrum are qualitatively consistent with the above obtained result based on the kinetic-energy driven SC mechanism by considering the interplay between the SC gap and normal-state pseudogap. In particular, their results \cite{Dan12,Illes09} also indicate that in the underdoped and optimally doped regimes, the transfer of the part of the low-energy spectral weight of the conductivity spectrum to the higher energy region to form a midinfrared band is intrinsically associated with the presence of the normal-state pseudogap.

In a standard Landau Fermi-liquid, the shape of the conductivity spectrum $\sigma(\omega)$ is normally well accounted for by the low-energy Drude formula that describes the free charge carrier contribution to $\sigma (\omega)$, and then when the temperatures $T<T_{\rm c}$, the spectral weight of the condensate in the SC-state comes from low energies \cite{Schrieffer64}. However, in cuprate superconductors, the part of the low-energy spectral weight in the conductivity spectrum in the underdoped and optimally doped regimes is transferred to the higher energy region to form the unusual midinfrared band, and then the width of the low-energy band is narrowing, while the onset of the region to which the spectral weight is transferred is close to the normal-state pseudogap $\bar{\Delta}_{\rm pg}$. Moreover, since the unusual midinfrared band is taken from the low-energy band, so that both the low-energy non-Drude peak (the conductivity decays as $\rightarrow 1/\omega$ at low energies) and unusual midinfrared band describe the actual charge-carrier density. In the framework of the kinetic-energy driven SC mechanism, the normal-state pseudogap state is the result of the strong electron correlation, and therefore the transfer of the part of the low-energy spectral weight of the conductivity spectrum in the underdoped and optimally doped regimes to the higher energy region to form the unusual midinfrared band is a natural consequence of the strongly correlated nature in cuprate superconductors. In particular, this strong electron correlation which induces a shift of the spectral weight from the low-energy to the higher energy midinfrared band in the conductivity spectrum, has been confirmed by the early numerical simulations based on the $t$-$J$ model in the normal-state \cite{Stephan90,Moreo90,Jaklic94,Dagotto94} and in the SC-state \cite{Haule07}. In the normal-state, the $1/\omega$ perfect decay of the conductivity at low energies in the optimally doped regime is closely related with the linear temperature resistivity, since it reflects an anomalous frequency dependent scattering rate proportional to $\omega$ instead of $\omega^{2}$ as would be expected in the standard Landau Fermi-liquid. This linear temperature resistivity is one of the characteristically anomalous properties of cuprate superconductors in the normal-state, and has been also phenomenological discussed within the marginal Fermi-liquid theory \cite{Varma89}. In particular, based on the slave-boson gauge theory, it has been shown within the $t$-$J$ model that above the Bose-Einstein temperature, the boson inverse lifetime due to scattering by the gauge field is of order $T$, which suppresses the condensation temperature and leads to a linear $T$ resistivity \cite{Lee92,Lee06}. However, in the SC-state, the large normal-state pseudogap in the underdoped and optimally doped regimes heavily reduces the fraction of the charge carriers that condense in the SC-state \cite{Homes04}.

\subsection{Summary}

The calculated result of the conductivity spectrum summarized in this section shows very clearly that if the effect of the normal-state pseudogap is taken into account in the framework of the kinetic-energy driven SC mechanism, the conductivity of the $t$-$J$ model calculated based on the linear response approach per se can correctly reproduce the main features found in infrared response measurements on cuprate superconductor in both the normal- and SC-states. The conductivity spectrum in the underdoped and optimally doped regimes contains the low-energy non-Drude peak and unusual midinfrared band. However, the position of the midinfrared band shifts towards to the low-energy non-Drude peak with increasing doping. In particular, the low-energy non-Drude peak incorporates with the midinfrared band in the heavily overdoped regime, and then the low-energy Drude behavior recovers. The qualitative reproduction of the main features of infrared response measurements on cuprate superconductors also shows that the transfer of the part of the low-energy spectral weight in the conductivity spectrum in the underdoped and optimally doped regimes to the higher energy region to form the unusual midinfrared band can be attributed to the effect of the normal-state pseudogap on the infrared response in cuprate superconductors.

\section{Conclusion and discussions}\label{conclusion}

In this article, we have given a brief review of the kinetic-energy driven SC mechanism, where the main conclusions are summarized as:

(a) In the fermion-spin theory (\ref{CSS}), the constrained electron is decoupled as a product of a charge carrier and a localized spin, and the charge carrier represents the charge degree of freedom of the constrained electron together with some effects of the spin configuration rearrangements due to the presence of the doped charge carrier itself, while the spin operator represents the spin degree of freedom of the constrained electron. In the {\it decoupling scheme}, this fermion-spin representation (\ref{CSS}) is a natural representation of the constrained electron defined in a restricted Hilbert space without double electron occupancy. The main advantage of the fermion-spin theory (\ref{CSS}) is that the electron local constraint for single occupancy is satisfied in actual calculations. In particular, the charge carrier {\it or} spin {\it itself} is $U(1)$ gauge invariant, and in this sense, the collective modes for the charge carrier and spin are real and can be interpreted as the physical excitations of cuprate superconductors. Although both charge carriers and spins contribute to the charge and spin dynamics, the charge-carrier relaxation time is responsible to the charge transport, and the spin relaxation time is responsible to the dynamical spin response, while as a result of the charge-spin recombination, the the electronic properties are dominated by electron quasiparticles. This is an efficient calculation scheme which can provide very good results even at the MF level.

(b) In the framework of the kinetic-energy driven SC mechanism developed based on the fermion-spin theory (\ref{CSS}), the charge-carrier pairing interaction originates directly from the kinetic energy of the $t$-$J$ model (\ref{CSStJmodel}) by the exchange of spin excitations in the higher powers of the doping concentration, and then these charge-carrier pairs (then the electron Cooper pairs) condense to the d-wave SC-state, where the spin excitation in cuprate superconductors plays a similar role to that of the phonon in conventional superconductors. Although the physical properties of cuprate superconductors in the normal-state are fundamentally different from these in the standard Landau Fermi-liquid state, the kinetic-energy driven SC-state still is conventional BCS-like with the d-wave symmetry, and then the obtained formalism for the charge-carrier pairing can be used to compute $T_{\rm c}$ and the related SC coherence of the low-energy excitations in cuprate superconductors on the first-principles basis much as can be done for conventional superconductors. Moreover, the kinetic-energy driven charge-carrier pair state is controlled by both the charge-carrier pair gap and quasiparticle coherence, which leads to that the maximal $T_{\rm c}$ occurs around the optimal doping, and then decreases in both the underdoped and overdoped regimes. This kinetic-energy driven SC mechanism also indicates that the strong electron correlation favors superconductivity, since the main ingredient is identified into a charge-carrier pairing mechanism not from the external degree of freedom such as the phonon, but rather solely from the internal spin degree of freedom of the constrained electron.

(c) The same charge-carrier interaction arising through the exchange of spin excitations that induces the d-wave SC-state in the particle-particle channel also generates the normal-state pseudogap state in the particle-hole channel, therefore there is a coexistence of the SC gap and normal-state pseudogap in the whole SC dome. Consequently, this normal-state pseudogap is identified as being a region of the self-energy effect in the particle-hole channel in which the normal-state pseudogap suppresses the spectral weight of the low-energy excitation spectrum. This normal-state pseudogap vanishes at the normal-state pseudogap crossover temperature $T^{*}$, with $T^{*}$ that is much larger than $T_{\rm c}$ in the underdoped and optimally doped regimes, and monotonically decreases upon the increase of doping, eventually disappearing together with $T_{\rm c}$ at the end of the SC dome. In particular, the normal-state pseudogap is directly related to the quasiparticle coherence, and therefore antagonizes superconductivity. Moreover, both the normal-state pseudogap and charge-carrier pair gap are dominated by one energy scale, and they are the result of the strong electron correlation in cuprate superconductors, while the domelike shape of the doping dependence of $T_{\rm c}$, the monotonic decrease of $T^{*}$ with doping, and relatively anomalous normal-state properties are a natural consequence of the Mott physics in which double occupancy is suppressed by strongly Coulombic repulsion. The theory also indicates that the kinetic-energy driven SC mechanism provides a natural explanation of both the origin of the normal-state pseudogap state and the charge-carrier pairing mechanism for superconductivity.

(d) Within the framework of the kinetic-energy driven SC mechanism, a number of typical properties of cuprate superconductors have been studied. In this review article, the selected results are summarized, including the doping dependence of the electromagnetic response, the dynamical spin response from low-energy to high-energy, and the charge transport, and are qualitatively comparable to the corresponding experimental results observed in cuprate superconductors. Furthermore, this kinetic-energy driven BCS-type formalism gives an explanation of the Raman scattering spectra \cite{Geng10} obtained in terms of the electronic Raman response measurement technique \cite{Devereaux07}. It also gives a consistent description of the thermodynamic properties \cite{Zhao12} observed from heat capacity measurements \cite{Loram94,Wen09,Loram00,Luo00,Wang06,Wen08}. In particular, it has been used to successfully describe a number of the SC-state properties in the presence of impurities \cite{Wang09}, including the microwave conductivity \cite{Wang08}, the scanning tunneling microscopic measurements of the coherent Bogoliubov quasiparticle dispersion, and the related extinction of Bogoliubov quasiparticle scattering interference at low temperatures \cite{Wang10}. Establishing these agreements between the calculated results obtained based on the kinetic-energy driven SC mechanism and the corresponding experimental data observed from a wide variety of measurement techniques are important to confirm the nature of the SC phase of cuprate superconductors to be the kinetic-energy driven d-wave SC-state.

In this review article, we have restricted our attention to the hole-doped cuprate superconductors. However, superconductivity in cuprates also emerges when electrons are doped into Mott insulators \cite{Tokura89,Armitage10}. Both the hole-doped and electron-doped cuprate superconductors have the layered structure of the square lattice of the CuO$_{2}$ plane separated by insulating layers \cite{Bednorz86,Wu87,Schilling93,Tokura89,Armitage10}. Although the significantly different behaviors of the hole-doped and electron-doped cuprate superconductors are observed due to the electron-hole asymmetry, the symmetry of the SC order parameter is common in both case \cite{Tsuei00,Tsuei00a}, manifesting that two systems have similar underlying SC mechanism. In particular, the strong electron correlation is common for both the hole-doped and electron-doped cuprate superconductors, and then it is possible that superconductivity in the electron-doped cuprate superconductors is also driven by the kinetic energy as in the hole-doped case. In this case, the charge asymmetry \cite{Ma05} in superconductivity of the electron-doped and hole-doped cuprate superconductors and the electronic Raman response \cite{Geng11} in the electron-doped cuprate superconductors have been discussed based on the kinetic-energy driven SC mechanism, and the calculated results are in qualitative agreement with the experimental data observed on the electron-doped cuprate superconductors.

Besides the square lattice cuprate superconductors, some cuprate materials \cite{Dagotto96,Dagotto99,Uehara96}, such as Sr$_{14}$Cu$_{24}$O$_{41}$, do not contain CuO$_{2}$ planes common to cuprate superconductors but consist of two-leg Cu$_{2}$O$_{3}$ ladders and edge-sharing CuO$_{2}$ chains. In particular, the doped two-leg ladder cuprates are a system in which the SC state is realized by applying a high pressure in the highly charge-carrier doped regime \cite{Dagotto99,Uehara96}. These ladder cuprate materials also are natural extensions of the Cu-O chain compounds towards the CuO$_{2}$ sheet structures. Within the framework of the kinetic-energy driven SC mechanism, some typical properties of the two-leg ladder cuprate superconductors have been studied, including the pressure dependence of $T_{\rm c}$ \cite{Qin07,Qin06}, the charge dynamics \cite{Qin02}, and the spin dynamics \cite {He03,Qin11}, and the calculated results are also qualitatively consistent with the experimental data obtained from the experimental measurements on the two-leg ladder cuprate superconductors.

Finally, we should be noted that much remains to be done. In particular, for the normal-state pseudogap, which grows upon underdoping, it seems natural to seek a connection to the physics of the AF insulating parent compounds \cite{Timusk99,Norman05,Hufner08}. However, at half-filling, the $t$-$J$ model is reduced as the AF Heisenberg model with an AFLRO. Although a small density of charge carriers is sufficient to destroy AFLRO, this AFLRO remains until the extremely low-doped regime ($\delta <0.045$) \cite{Fujita12,Dean14,Lee88}. As we have mentioned in Eq. (\ref{ECpair}), the conduct is disrupted by AFLRO at the extremely low-doped regime, and then an important issue is how to extend the theory of the normal-state pseudogap state for the doped regime without AFLRO to the case at the extremely low-doped regime with AFLRO for a proper description of the connection between the finite doping normal-state pseudogap and the zero-doping quasiparticle dispersion.

\section*{Acknowledgements}

One of authors (SF) thanks Professor Z. B. Su and Professor L. Yu for the early collaborations, and he also thanks Li Cheng, Huaiming Guo, Zheyu Huang, Zhongbing Huang, Zhihao Geng, Mateusz Krzyzosiak, Ying Liang, Bin Liu, Tianxing Ma, Jihong Qin, Yun Song, Weifang Wang, Zhi Wang, Feng Yuan, and Jingge Zhang for the collaborations. This work was supported by the funds from the Ministry of Science and Technology of China under Grant Nos. 2011CB921700 and 2012CB821403, the National Natural Science Foundation of China under Grant Nos. 11274044 and 11447144, and the Science Foundation of Hengyang Normal University under Grant No. 13B44.


\appendix
\numberwithin{equation}{section}
\renewcommand\theequation{\Alph{section}\arabic{equation}}
\renewcommand\thesection{Appendix \Alph{section}}

\section{Matrix representation of projection operator}\label{matrix}

The charge-carrier operators $h^{\dagger}_{l\sigma}$ and $h_{l\sigma}$ in the basis \cite{Feng94},
\begin{eqnarray}
\left (\begin{array}{cc} {1}\\{0}\end{array}\right)_{h},~~~~~~\left (\begin{array}{cc} {0}\\{1}\end{array}\right)_{h},
\end{eqnarray}
of the charge-carrier states are given by,
\begin{eqnarray}
h^{\dagger}_{l\sigma}=\left(
\begin{array}{cc}
0 & 1 \\
0 & 0
\end{array} \right) \,,~~~~~~
h_{l\sigma}=\left(
\begin{array}{cc}
0 & 0 \\
1 & 0
\end{array} \right) \,,
\end{eqnarray}
while the spin raising and lowering operators $S^{+}_{l}$ and $S^{-}_{l}$ in the spin 1/2 space,
\begin{eqnarray}
\left (\begin{array}{cc} {1}\\{0}\end{array}\right)_{s},~~~~~~ \left (\begin{array}{cc} {0}\\{1}\end{array}\right)_{s},
\end{eqnarray}
are given by,
\begin{eqnarray}
S^{+}_{l}=\left(
\begin{array}{cc}
0 & 1 \\
0 & 0
\end{array} \right) \,,~~~
S^{-}_{l}=\left(
\begin{array}{cc}
0 & 0 \\
1 & 0
\end{array} \right) \,.
\end{eqnarray}
In the product space $|{\rm charge}\rangle\otimes |{\rm spin}\rangle$, the basis vectors are \cite{Feng94},
\begin{subequations}\label{CSSbasis}
\begin{eqnarray}
|1,\uparrow \rangle &=&\left (\begin{array}{cc}{1}\\{0}\end{array}\right)_{h}\otimes\left (\begin{array}{cc} {1}\\{0}
\end{array}\right)_{s}=\left (\begin{array}{cccc} {1}\\{0}\\{0}\\{0}\end{array}\right), \\
|1,\downarrow \rangle &=&\left (\begin{array}{cc}{1}\\{0}\end{array}\right)_{h}\otimes\left (\begin{array}{cc} {0}\\{1}
\end{array}\right)_{s}=\left (\begin{array}{cccc} {0}\\{1}\\{0}\\{0}\end{array}\right), \\
|0,\uparrow \rangle &=&\left (\begin{array}{cc}{0}\\{1}\end{array}\right)_{h}\otimes\left (\begin{array}{cc} {1}\\{0}
\end{array}\right)_{s}=\left (\begin{array}{cccc} {0}\\{0}\\{1}\\{0}\end{array}\right), \\
|0,\downarrow \rangle &=&\left (\begin{array}{cc}{0}\\{1}\end{array}\right)_{h}\otimes\left (\begin{array}{cc} {0}\\{1}
\end{array}\right)_{s}=\left (\begin{array}{cccc} {0}\\{0}\\{0}\\{1}\end{array}\right),
\end{eqnarray}
\end{subequations}
which form a complete set, then the fermion-spin transformation defined by Eq. (\ref{CSS}) in this basis gives the following matrix representation for the constrained electron operators \cite{Feng94},
\begin{subequations}\label{CSSoperator}
\begin{eqnarray}
C_{l\uparrow}&=&h^{\dagger}_{l\uparrow}S^{-}_{l}=\left(
\begin{array}{cc}
0 & 1 \\
0 & 0
\end{array}\right)_{h}\,
\otimes\left(
\begin{array}{cc}
0 & 0 \\
1 & 0
\end{array} \right) \,
=\left(
\begin{array}{cccc}
0 & 0 & 0 & 0 \\
0 & 0 & 1 & 0 \\
0 & 0 & 0 & 0 \\
0 & 0 & 0 & 0
\end{array} \right) \,, \\
C_{l\downarrow}&=&h^{\dagger}_{l\downarrow}S^{+}_{l}=\left(
\begin{array}{cc}
0 & 1 \\
0 & 0
\end{array}\right)_{h}\,
\otimes
\left(
\begin{array}{cc}
0 & 1 \\
0 & 0
\end{array} \right) \,
=\left(
\begin{array}{cccc}
0 & 0 & 0 & 1 \\
0 & 0 & 0 & 0 \\
0 & 0 & 0 & 0 \\
0 & 0 & 0 & 0
\end{array} \right) \,,\\
C^{\dagger}_{l\uparrow}&=&h_{l\uparrow}S^{+}_{l}=\left(
\begin{array}{cc}
0 & 0 \\
1 & 0
\end{array}\right)_{h}\,
\otimes
\left(
\begin{array}{cc}
0 & 1 \\
0 & 0
\end{array} \right) \,
=\left(
\begin{array}{cccc}
0 & 0 & 0 & 0 \\
0 & 0 & 0 & 0 \\
0 & 1 & 0 & 0 \\
0 & 0 & 0 & 0
\end{array} \right) \,,\\
C^{\dagger}_{l\downarrow}&=&h_{l\downarrow}S^{-}_{l}=\left(
\begin{array}{cc}
0 & 0 \\
1 & 0
\end{array}\right)_{h}\,
\otimes
\left(
\begin{array}{cc}
0 & 0 \\
1 & 0
\end{array} \right) \,
=\left(
\begin{array}{cccc}
0 & 0 & 0 & 0 \\
0 & 0 & 0 & 0 \\
0 & 0 & 0 & 0 \\
1 & 0 & 0 & 0
\end{array} \right) \,.
\end{eqnarray}
\end{subequations}
However, as we have mentioned in Eq. (\ref{RHspace}), the restricted Hilbert space without double electron occupancy in the $t$-$J$ model (\ref{tJmodel}) consists of three states, $|0\rangle$, $|\uparrow \rangle$, $|\downarrow \rangle$, namely,
\begin{eqnarray}\label{physics-basis}
|0\rangle =\left (\begin{array}{cccc} {1}\\{0}\\{0}
\end{array}\right),~~~~
|\uparrow \rangle =\left (\begin{array}{cccc} {0}\\{1}\\{0}
\end{array}\right),~~~~
|\downarrow \rangle =\left (\begin{array}{cccc} {0}\\{0}\\{1}
\end{array}\right).
\end{eqnarray}
To remove the extra degrees of freedom in the $|{\rm charge}\rangle\otimes |{\rm spin}\rangle$ space, we \cite{Feng94} introduce a projection operator $P$. By requiring
$P|1,\uparrow\rangle=P|1,\downarrow\rangle=|0\rangle$, $P|0,\uparrow\rangle=|\uparrow\rangle$, and $P|0,\downarrow\rangle=|\downarrow\rangle$, we can easily obtain its matrix representation,
\begin{eqnarray}
P=\{P_{\kappa \alpha}\}=\left(
\begin{array}{cccc}
1 & 1 & 0 & 0 \\
0 & 0 & 1 & 0 \\
0 & 0 & 0 & 1
\end{array} \right) \,,
\end{eqnarray}
and its hermitian conjugation,
\begin{eqnarray}
P^{\dagger}=\left(
\begin{array}{ccc}
1 & 0 & 0 \\
1 & 0 & 0 \\
0 & 1 & 0 \\
0 & 0 & 1
\end{array} \right) \,.
\end{eqnarray}
Using this projection operator, the electron operators in the restricted Hilbert space without double electron occupancy are given by \cite{Feng94},
\begin{subequations}
\begin{eqnarray}
\tilde{C}_{l\uparrow}=P_{l}h^{\dagger}_{l\uparrow}S^{-}_{l}P^{\dagger}_{l}=\left(
\begin{array}{ccc}
0 & 1 & 0 \\
0 & 0 & 0  \\
0 & 0 & 0
\end{array} \right) \,,~~~
\tilde{C}^{\dagger}_{l\uparrow}=P_{l}h_{l\uparrow}S^{+}_{l}P^{\dagger}_{l}=\left(
\begin{array}{ccc}
0 & 0 & 0 \\
1 & 0 & 0  \\
0 & 0 & 0
\end{array} \right) \,,~~~\\
\tilde{C}_{l\downarrow}=P_{l}h^{\dagger}_{l\downarrow}S^{+}_{l}P^{\dagger}_{l}=\left(
\begin{array}{ccc}
0 & 0 & 1 \\
0 & 0 & 0  \\
0 & 0 & 0
\end{array} \right) \,,~~~
\tilde{C}^{\dagger}_{l\downarrow}=P_{l}h_{l\downarrow}S^{-}_{l}P^{\dagger}_{l}=\left(
\begin{array}{ccc}
0 & 0 & 0 \\
0 & 0 & 0  \\
1 & 0 & 0
\end{array} \right) \,,~~~
\end{eqnarray}
\end{subequations}
as quoted in Eq. (\ref{CEP2}). It is then straightforward to verify the operator relations quoted in Eq. (\ref{SR3}). In particular, the charge-carrier number operator,
\begin{eqnarray}\label{hole-number}
n^{(h)}_{i}=\left(
\begin{array}{ccc}
1 & 0 & 0 \\
0 & 0 & 0 \\
0 & 0 & 0
\end{array} \right) \,=|0\rangle\langle 0|
={1\over 2}P(|1\uparrow\rangle\langle 1\uparrow |+|1\downarrow\rangle\langle 1\downarrow |)P^{\dagger}.
\end{eqnarray}
The physical meaning of Eq. (\ref{hole-number}) is transparent: the empty state should be counted only once, not twice. Since the MF treatment of the constraint on average doping concentration $\delta$ is imposed directly on $h^{\dagger}_{l\sigma}h_{l\sigma}=h^{\dagger}_{l}h_{l}$, the sum rule for the constrained electron is satisfied.

\end{document}